\documentclass[a4paper,11pt]{article}
\pdfoutput=1 

\usepackage{jheppub} 
\usepackage{caption}
\usepackage{xcolor}
\usepackage{bm}
\usepackage{ tipa }
\usepackage{float}
\usepackage{pifont}
\usepackage{bbding}
\usepackage{amssymb}
\usepackage{subcaption}
\usepackage[T1]{fontenc} 
\newcommand{\comment}[1]{}
\usepackage{booktabs}
\usepackage{siunitx}
\usepackage[symbol]{footmisc}

\newcommand{\com}[1]{(*{\textbf{#1}}*)}
\newcommand{\KG}[1]{\com{KG: #1}}

\newcommand{\ba}{\begin{equation}\begin{aligned}}
\newcommand{\ea}{\end{aligned}\end{equation}}

\newcommand{\be}{\begin{equation}}
\newcommand{\ee}{\end{equation}}

\newcommand{\bea}{\begin{equation}\begin{aligned}}
\newcommand{\eea}{\end{aligned}\end{equation}}

\title{\boldmath Numerical Conformal bootstrap with Analytic Functionals and Outer Approximation}

\author{Kausik Ghosh}
\author{and Zechuan Zheng}

\affiliation{Laboratoire de Physique Th\'eorique, \\ de l'\'Ecole Normale Sup\'erieure, PSL University,\\ CNRS, Sorbonne Universit\'es, UPMC Univ. Paris 06\\ 24 rue Lhomond, 75231 Paris Cedex 05, France}

\emailAdd{kau.rock91@gmail.com}
\emailAdd{zechuan.zheng.phy@gmail.com}

\abstract{This paper explores the numerical conformal bootstrap in general spacetime dimensions through the lens of a distinct category of analytic functionals, previously employed in two-dimensional studies. We extend the application of these functionals to a more comprehensive backdrop, demonstrating their adaptability and efficacy across a wider range of dimensionalities. \KG{rephrase a wider range of dimensionalities ...somehow it doesn't sound right to me. I understand what it means but it is not a nice read.. may be say across a wider regime of spacetime dimensions} The bootstrap is implemented using the outer approximation methodology, with computations conducted in double precision. The crux of our study lies in comparing the performance of this category of analytic functionals with conventional derivatives at crossing symmetric points. Our numerical analysis indicates that these functionals offer a superior performance, thereby revealing potential alternative paradigm in the application of conformal bootstrap methodology \KG{Remove the word methodology. I think it's better without it}. }

\begin{document} 

\addtocontents{toc}{\protect\setcounter{tocdepth}{2}}

\begingroup\parindent0pt
\begin{flushright}\footnotesize
\end{flushright}
\centering
\begingroup\LARGE
\bf
Numerical Conformal bootstrap with Analytic Functionals and Outer Approximation
\par\endgroup
\vspace{3.5em}
\begingroup\large
{\bf Kausik Ghosh}\footnote{kau.rock91@gmail.com}$\,$ \it{and} $\,$
{\bf Zechuan Zheng}\footnote{zechuan.zheng.phy@gmail.com}
\par\endgroup
\vspace{2em}
\begingroup\sffamily\footnotesize
Laboratoire de Physique de l'\'Ecole Normale Sup\'erieure, ENS,\\ PSL Research University, CNRS, Sorbonne Universit\'e, Universit\'e de Paris,\\ 24 rue Lhomond, 75231 Paris Cedex 05, France
\vspace{1em}
\par\endgroup
\vspace{2em}

\endgroup

\begin{abstract}

This paper explores the numerical conformal bootstrap in general spacetime dimensions through the lens of a distinct category of analytic functionals, previously employed in two-dimensional studies. We extend the application of these functionals to a more comprehensive backdrop, demonstrating their adaptability and efficacy in general spacetime dimensions above two. The bootstrap is implemented using the outer approximation methodology, with computations conducted in double precision. The crux of our study lies in comparing the performance of this category of analytic functionals with conventional derivatives at crossing symmetric points. It is worth highlighting that in our study, we identified some novel kinks in the scalar channel during the maximization of the gap in two-dimensional conformal field theory. Our numerical analysis indicates that these analytic functionals offer a superior performance, thereby revealing a potential alternative paradigm in the application of conformal bootstrap.  
\end{abstract}

\begin{figure}[ht]
    \centering
    \includegraphics[width=10cm]{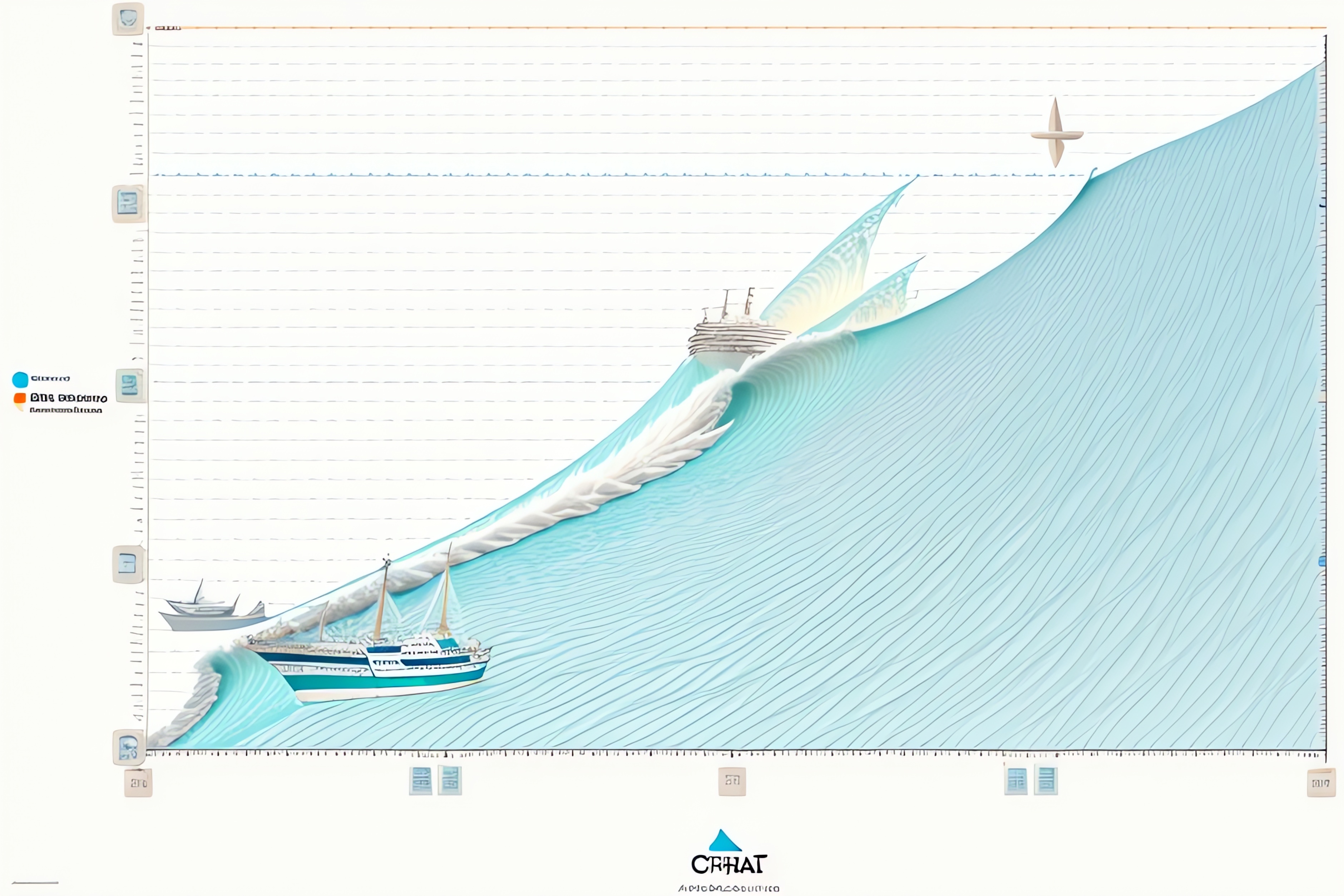}

\end{figure}

\thispagestyle{empty}

\newpage
\tableofcontents
\emailAdd{kau.rock91@gmail.com}
\emailAdd{zechuan.zheng.phy@gmail.com}
\flushbottom
\renewcommand*{\thefootnote}{\arabic{footnote}}
\section{Introduction}\label{sec: intro}

In the past decade, the understanding of conformal field theories (CFTs) across various spacetime dimensions has been revolutionized, with the analytical and numerical bootstrap program expanding into new territories, as initiated by the foundational work of \cite{Rattazzi:2008pe} \footnote{Comprehensive reviews of these advancements can be found in \cite{Poland:2018epd, Bissi:2022mrs}. See also \cite{Poland:2022qrs, Hartman:2022zik} for more recent updates. }. The core premise revolves around studying the constraints enclosed within the crossing symmetric equation:

\begin{equation}
    \sum_{\Delta,\ell} a_{\Delta,\ell} F_{\Delta,\ell}(z,\bar{z})=0,
\end{equation}

This equation is typically discretized by considering its derivative at the crossing-symmetric point $z=\bar{z}=\frac{1}{2}$. A suite of innovative techniques are employed to compute these derivatives efficiently, as proposed by \cite{Kravchuk:2016qvl, Erramilli:2019njx, Kravchuk:2017dzd, Karateev:2018oml, Erramilli:2020rlr}. Following discretization, polynomial approximation transforms the equation into a standard SDP constraint condition. Subsequently, the \textit{SDPB} solver \cite{Simmons-Duffin:2015qma} offers the solution of the corresponding SDP to arbitrary precision. With the constraints in place, the `navigitor' technique \cite{Reehorst:2021ykw} is deployed to locate physically permitted theories within the CFT parameter space. 

The predominant usage of derivative bases at the crossing-symmetric point remained unchallenged until Maz\'{a}\v{c}'s seminal work \cite{Mazac:2016qev}, which introduced an explicit functional for 1D CFT gap-maximization\footnote{See also \cite{El-Showk:2012vjm, El-Showk:2016mxr} for an initial exploration of the properties associated with the extremal functionals.}. This spurred a spate of studies, broadening the application of this functional and setting a mathematically rigorous foundation for its validity \cite{Qiao:2017lkv, Mazac:2018mdx, Mazac:2018ycv}. Demonstrating immense versatility, these functionals have since been expanded to boundary CFT \cite{Kaviraj:2018tfd, Mazac:2018biw}, CFT with global symmetry and large $N$ \cite{Ghosh:2021ruh, Li:2023whn}, CFT in higher dimensions \cite{Paulos:2019gtx, Mazac:2019shk, Caron-Huot:2020adz}, and CFT with mixed correlators \cite{Trinh:2021mll, Ghosh:2023lwe}. These advancements have led to a broad relationship with the Mellin amplitude \cite{Gopakumar:2016cpb, Gopakumar:2016wkt, Dey:2016mcs, Dey:2017fab, Ferrero:2019luz}, dispersion relations \cite{Caron-Huot:2020adz, Paulos:2020zxx,Gopakumar:2021dvg, Kaviraj:2021cvq}, sphere packing \cite{Hartman:2019pcd, Afkhami-Jeddi:2020hde} and real projective space\cite{Giombi:2020xah}, thereby highlighting the broad applicability and potential of this class of functionals.

However, the application of analytic functionals within numerical bootstrap remains less explored. Studies have indicated that 1D analytic functionals present a superior convergence behavior than derivative bases \cite{Paulos:2019fkw,Ghosh:2021ruh}, and have been successfully applied to bootstrap integrable theories \cite{Caron-Huot:2022sdy}. The limited progress, from our perspective, lies in the difficulties associated with approximating analytic functionals by polynomials due to their transcendental nature, thereby inhibiting the utilization of \textit{SDPB}. Furthermore, efficient numerical evaluation methods for most analytic functional frameworks are lacking. Additionally, analytic functionals in dimensions greater than 1 are often devoid of assurances for positivity or completeness conditions.

\begin{figure}[ht]
    \centering
    \includegraphics[width=15cm]{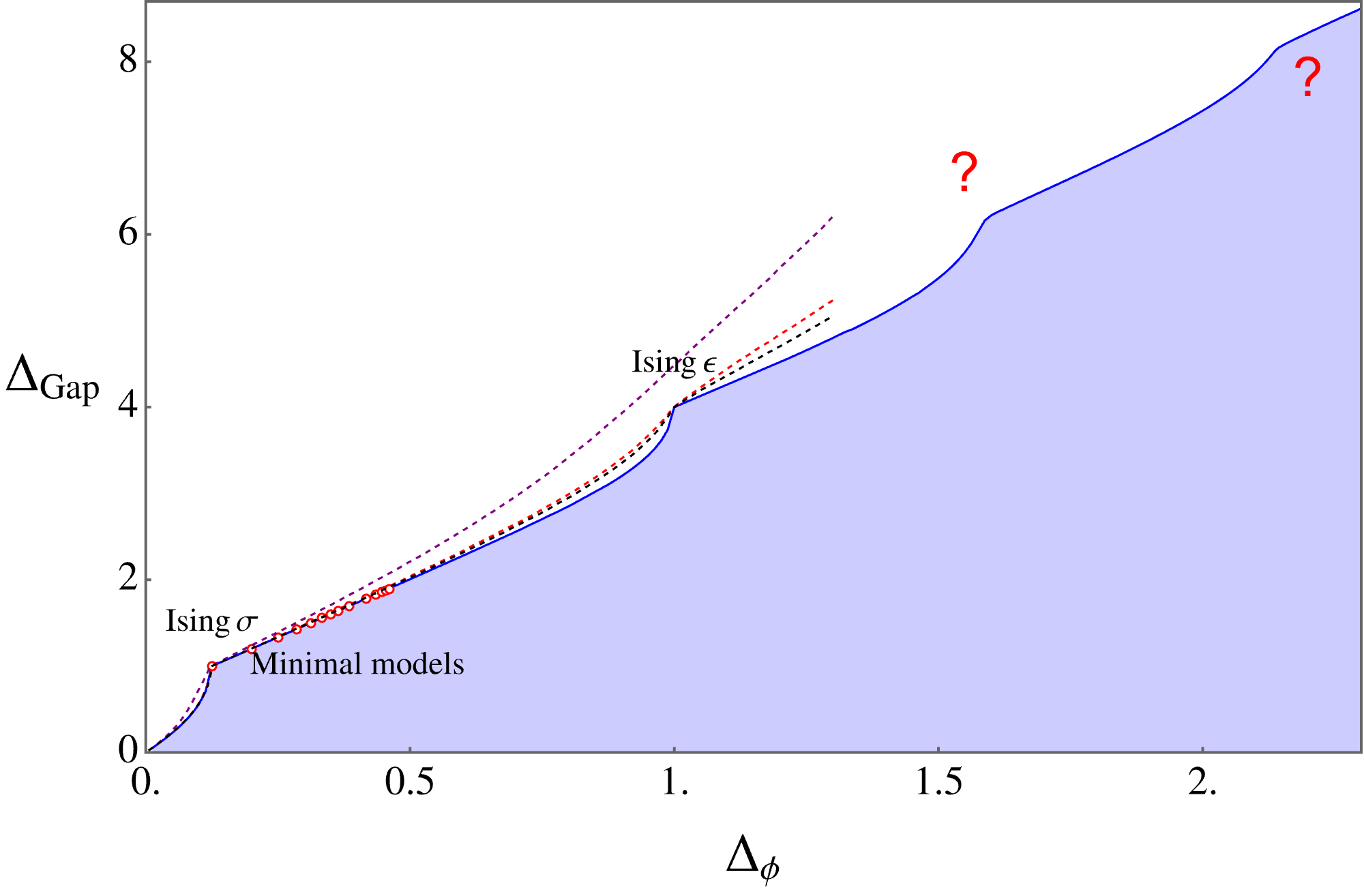}
    \caption{\label{fig: spin0full} Spin $\ell=0$ gap maximization. The purple, red, and black dashed lines represent the gap maximization with 15, 91, and 171 derivatives, respectively. The dotted circles indicate the positions of several selected minimal models. The shaded blue region is the allowed region by the analytic functional basis.}
\end{figure}

We should note that the optimization problems associated with the conformal bootstrap, regardless of whether they utilize a derivative basis or analytic functionals, are encompassed within the framework of Semi-Infinite Programming (SIP) \cite{hettich1993semi, reemtsen1998semi}. The outer approximation method, used for solving SIP, involves generating a sequence of discretized problems which are expected to converge to the original SIP problem. This discretization approach has been implemented in early numerical conformal bootstrap work, and an initial application of a self-refining grid was seen in the extremal functional method \cite{El-Showk:2012vjm}. Subsequent discussions regarding the feasibility of outer approximation for conformal bootstrap began on the Bootstrap Collaboration Slack Workspace\footnote{We also acknowledge the preliminary application of outer approximation in \textit{JuliBoots}\cite{Paulos:2014vya} for 1D CFT bootstrap, and 
\href{https://github.com/davidsd/sdpb/blob/master/docs/Outer_Limits.pdf}{SDPB Github} for an application of outer approximation in \textit{SDPB} after the implementation of outer approximation in \cite{Caron-Huot:2021rmr}.}. We posit that the outer approximation method is particularly well-suited to the analytic functionals selected for this study\footnote{It's worth mentioning the possibility of completely bypassing the discretization process, as was originally implemented by \cite{El-Showk:2014dwa}. This method is implemented in \cite{ElShowkRychkov}, \href{https://gitlab.com/bootstrapcollaboration/SIPSolver/-/tree/master/}{SIPsolver} and \textit{JuliBoots}\cite{2014arXiv1412.4127P}. The question of whether this approach or the discretization method is more efficient for analytic functionals, however, remains open.
}.

Our primary focus in this study is on product functionals, initially introduced by Maz\'{a}\v{c} \cite{Mazac:2016qev} and later expanded by \cite{Paulos:2019gtx, Kaviraj:2021cvq}. These functionals are designed for CFTs in even dimensions and they exhibit sign-definiteness asymptotically, while also being conjectured to be complete. Importantly, the product functional can be evaluated efficiently, with methods detailed in previous works \cite{Ghosh:2021ruh, Faller:2017hyt, Paulos:2019gtx, Kaviraj:2018tfd, Zhou:2018sfz}. Our numerical bootstrap study employs an outer approximation technique for the computation using these product functionals, thereby circumventing the limitations of conventional polynomial approximation that is otherwise inapplicable for analytic functionals\footnote{In order to apply the usual semidefinite programming techniques as in \textit{SDPB}, it
is not strictly necessary to have a polynomial approximation. A set of functions $\chi_i(\Delta)$
such that bilinears $\chi_i(\Delta)\chi_j(\Delta)$  span the space of functionals is enough. We thank David Simmons-Duffin for pointing this out.}. Our algorithm is implemented in the Julia package \textit{FunBoot} and performs all computations in double precision, which significantly enhances the efficiency of numerical evaluation. Through the dimension reduction of the conformal block\cite{Hogervorst:2016hal}, we generalize the action of the product functional to odd dimensions. We aim to demonstrate the superiority of our proposed basis over traditional ones through practical examples, highlighting the enhancements brought about by our novel methodology and underscoring the potential and effectiveness of this framework. One of the interesting results is summarized in Fig.\ref{fig: spin0full}, where we discovered new kinks in the $2d$ CFT gap maximization, hinting at potentially interesting unknown theories.

The rest of the paper is organized as follows. In Section 2, we discuss fundamental 1D functionals and their bootstrap applications. Section 3 introduces the product functionals designed for general dimensions. Section 4 outlines the numerical implementation of our framework. In Section 5, we present our numerical results, providing numerical evidence for the effectiveness of our approach.  We conclude with future directions and open questions arising from this work. Technical details and supplementary information are provided in the appendices.

\section{Fundamentals of 1D analytic Functionals}
\label{sec:1d-func}

In this section, we review the theoretical background and properties of $1d$ analytical functionals acting on (anti) crossing symmetric vectors. These functionals are integral to the study of four-point functions of identical scalars with dimension $\Delta_\phi$ in one-dimensional conformal field theories (CFTs). 

\subsection{Formal Representation of Four-point Functions}
The four-point function in one-dimensional CFTs can be represented as a single cross-ratio $z$:

\begin{equation}
\langle\phi(x_1)\phi(x_2)\phi(x_3)\phi(x_4) \rangle= \frac{1}{x_{12}^{2\Delta_{\phi}} x_{34}^{2\Delta_{\phi}}} \mathcal{G}(z), \quad z=\frac{x_{12}x_{34}}{x_{13}x_{24}},\,\quad x_{ij}:=x_i-x_j\,
\end{equation}

The function $\mathcal{G}(z)$ adheres to the s-t crossing relation:

\begin{equation}\label{1dcrossing}
    z^{-2\Delta_{\phi}}\mathcal{G}(z)-(1-z)^{-2\Delta_{\phi}}\mathcal{G}(1-z)=0.
\end{equation}

By virtue of the state-operator correspondence, we can represent the correlator in terms of the sum of $SL(2,\mathbb R)$ conformal blocks with the square of OPE coefficients, $a_{\Delta}\geq 0$:

\begin{equation}\label{eq: OPE}
\begin{split}
    \mathcal{G}(z) & =\sum_{\Delta} a_{\Delta} G^{d=1}_{\Delta}(z),\qquad G^{d=1}_{\Delta}(z) := z^{\Delta}\,{ }_2 F_1\left(\Delta, \Delta, 2\Delta ; z\right).
\end{split}
\end{equation}

With the conformal block expansion, we can recapitulate the crossing equation in terms of the crossing vectors:
\begin{equation}
    F^{d=1}_{\Delta}(z|\Delta_\phi):=\left(z^{-2\Delta_\phi}G^{d=1}_{\Delta}(z)-(1-z)^{-2\Delta_\phi}G^{d=1}_{\Delta}(1-z))\right),
\end{equation}
\begin{equation}
    \sum_{\Delta} a_{\Delta}F^{d=1}_{\Delta}(z|\Delta_\phi)=0.
\end{equation}

\subsection{Decomposition of the Crossing Vector}
It was demonstrated in the study \cite{Mazac:2018ycv} that the crossing vector can be decomposed into a complete basis set:

\begin{equation}
    F^{d=1}_{\Delta}(z|\Delta_\phi)=\sum_{n} (\alpha^-_n(\Delta)  F^{d=1}_{\Delta_n}(z|\Delta_\phi)+\beta^-_n(\Delta)  \partial F^{d=1}_{\Delta_n}(z|\Delta_\phi)).
\end{equation}

Here, $\Delta_n$ denotes the dimensions of operators in Generalized Free Field Theory (GFF) correlators. We employ two sets of GFF: one corresponding to a free bosonic field and the other to a free fermion in $AdS_2$. The correlators for these fields are given by:

\begin{equation}
\mathcal{G}(z)=1+\eta z^{2\Delta_{\phi}}+(\frac{z}{1-z})^{2\Delta_{\phi}},
\end{equation}

where $\eta=1$ or $-1$ for boson and fermion respectively. These correlators encapsulate operators with dimensions:

\begin{equation}
\begin{split}
    & \Delta^B_n=2\Delta_{\phi}+2n,\\
   &  \Delta^F_n=2\Delta_{\phi}+2n+1.
\end{split}
\end{equation}

where $B,F$ stand for boson and fermion respectively. For this spectrum, the corresponding OPE in Eq.\ref{eq: OPE}  is given by:
\begin{equation}
    a_{\Delta}^{\mathrm{gff}}=\frac{2 \Gamma(\Delta)^2}{\Gamma(2 \Delta-1)} \frac{\Gamma\left(\Delta+2 \Delta_\phi-1\right)}{\Gamma\left(2 \Delta_\phi\right)^2 \Gamma\left(\Delta-2 \Delta_\phi+1\right)}.
\end{equation}

We also introduce the concept of anti-crossing ($``+"$ type) vectors\cite{Hartman:2019pcd,Paulos:2019gtx} for our study:

\begin{equation}
    H^{d=1}_{\Delta}(z|\Delta_\phi):= \left(z^{-2\Delta_\phi}G^{d=1}_{\Delta}(z)+(1-z)^{-2\Delta_\phi}G^{d=1}_{\Delta}(1-z))\right).
\end{equation}

These vectors also admit a similar basis decomposition:

\begin{equation}
       H^{d=1}_{\Delta}(z|\Delta_\phi)= \sum_{n} (\alpha^+_n(\Delta)  H^{d=1}_{\Delta_n}(z|\Delta_\phi)+\beta^+_n(\Delta)  \partial H^{d=1}_{\Delta_n}(z|\Delta_\phi)).
\end{equation}

Considering the two potential choices for $\Delta_n$ (bosonic or fermionic), we have the following set of functionals:

\begin{equation}
    \alpha^{-,B/F}_n,\,\beta^{-,B/F}_n, \,\alpha^{+,B/F}_n,\,\beta^{+,B/F}_n.
\end{equation}

The functionals under consideration in this study adhere to the duality relations stated above. For the fermionic basis, the functionals meet the following conditions:
\begin{equation}
\begin{split}
   &  \alpha^{\pm,F}_n(\Delta^F_m)=\delta_{n,m},\,\,\qquad \partial \alpha^{\pm,F}_n(\Delta^F_m)=0,\,\,\\
    & \beta^{\pm,F}_n(\Delta^F_m)=0,\,\,\,\,\,\,\,\,\,\,\,\qquad
    \partial \beta^{\pm,F}_n(\Delta^F_m)=\delta_{n,m}
\end{split}
\end{equation}

Analogously, the bosonic basis abides by the subsequent relations:

\begin{equation}
\begin{split}
   &  \qquad\,\,\alpha^{\pm,B}_n(\Delta^B_m)=\delta_{n,m},\,\,\qquad \partial \alpha^{\pm,B}_n(\Delta^B_m)=-c_n^{\pm}\delta_{m,0}\\
    & \qquad\,\,\beta^{\pm,B}_n(\Delta^B_m)=0,\,\,\,\,\,\,\,\,\,\,\,\qquad
    \partial \beta^{\pm,B}_n(\Delta^B_m)=\delta_{n,m}-d_n^{\pm} \delta_{m,0},
\end{split}
\end{equation}

In these relations, the coefficients $c_n^\pm$ and $d_n^\pm$ have been suitably selected to enhance the Regge behavior of the functionals. Detailed calculations can be performed to ascertain these coefficients explicitly. For further computational elaboration pertaining to these functionals, the reader is directed to Appendix \ref{funcwittenrelation}.

The concrete manifestation of a one-dimensional functional is realized via contour integration, as expressed in the following equation:
\begin{equation}\label{eq: contour}
    \omega\left(\Delta \mid \Delta_\phi\right)=\int_1^{\infty} \frac{\mathrm{d} z}{\pi} h(z) \mathcal{I}_z F_{ \Delta}\left(z \mid \Delta_\phi\right)
\end{equation}
where
\begin{equation}
    \mathcal{I}_z F(z):=\lim _{\epsilon \rightarrow 0^{+}} \frac{F(z+i \epsilon)-F(z-i \epsilon)}{2 i}
\end{equation}

We can do the transformation of the integration contour. This process effectively captures the majority of the double zeros of the functional action:

\begin{equation}
    \omega_n(\Delta) \equiv \omega_n\left[F_{\Delta}\right]=2 \sin^2(\frac{\pi}{2}(\Delta-\Delta^{B/F}_n)) \int_0^1 \mathrm{~d} z g_n(z) \frac{G_{\Delta}(z)}{z^{2 \Delta_\phi}}
\end{equation}
where
\begin{equation}
    g_n(z)=-\frac{\operatorname{Disc}[h_n(z)]}{2 \pi i} \quad \text { for } z \in(0,1)
\end{equation}
\subsection{Asymptotic Behavior of 1D Functionals}

In this section, we examine the asymptotic behavior of one-dimensional functionals. 

\begin{figure}[ht]
    \centering
    \begin{subfigure}{0.495\textwidth}
        \centering
        \includegraphics[width=\textwidth]{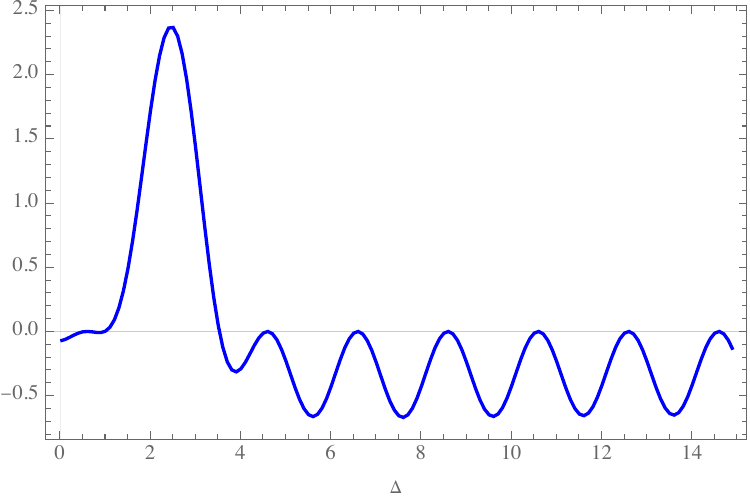}
        \caption{Asymptotic behavior of $\alpha_1^{-B}$}
        
    \end{subfigure}
    \hfill
    \begin{subfigure}{0.495\textwidth}
        \centering
        \includegraphics[width=\textwidth]{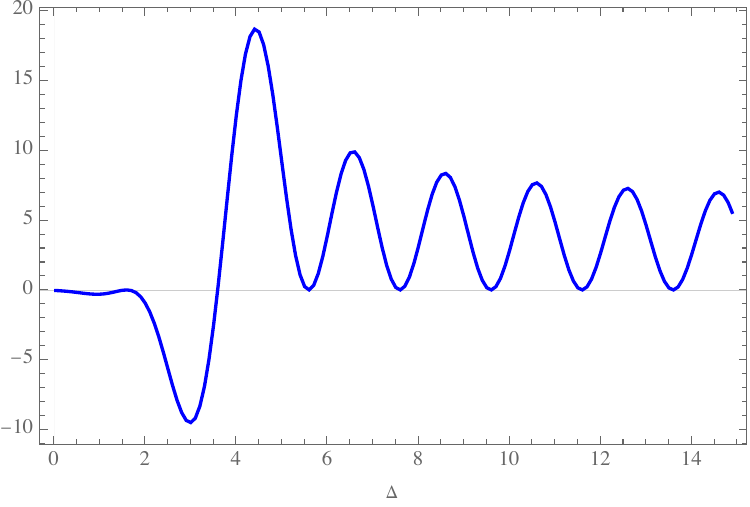}
        \caption{Asymptotic behavior of $\beta_1^{+F}$}
        
    \end{subfigure}
    \caption{Depiction of the asymptotic behavior of one-dimensional functionals.}
    \label{fig: 1dplot}
\end{figure}

It is noteworthy that the functional family of plus type and minus type exhibit the following universal asymptotics in large $\Delta$\cite{Mazac:2018ycv}\cite{Paulos:2019gtx}:
\begin{equation}\label{eq:asymp}
    \omega_n^{B, F}\left(\Delta \mid \Delta_\phi\right)\rightarrow \frac{4 \sin ^2\left[\frac{\pi}{2}\left(\Delta-\Delta_n^{B, F}\right)\right]}{\pi^2}\left(\frac{a_{\Delta_n^{B, F}}^{\mathrm{gff}}}{a_{\Delta}^{\mathrm{gff}}}\right) R_{\omega_n} \left(\Delta \mid \Delta_\phi\right)
\end{equation}
For both plus type functional and minus type functional, the $R_\omega\left(\Delta \mid \Delta_\phi\right)$ takes the form:
\begin{equation}
R_{\omega_{-}}^{B, F}\left(\Delta \mid \Delta_\phi\right) \underset{\Delta \rightarrow \infty}{\sim} \frac{C_{\omega-}}{\Delta^3}, \quad R_{\omega_{+}}^{B, F}\left(\Delta \mid \Delta_\phi\right) \underset{\Delta \rightarrow \infty}{\sim} \frac{C_{\omega+}}{\Delta^5}
\end{equation}
The constants $C_{\omega+}$ and $C_{\omega-}$ are, in essence, functions of the external dimension $\Delta_\phi$ and the functional itself. While a closed-form formula for the bosonic and fermionic bases is currently unknown, it could, in principle, be derived using the formula provided in \cite{Mazac:2018ycv}. By utilizing this asymptotic information, we normalize the functional by a universal factor for both the plus and minus types\footnote{This is important for the numerical stability.}. Further details of the normalization procedure are delineated in Appendix.\ref{Normalization}. Figure \ref{fig: 1dplot} presents some of the normalized functionals for reference.

\section{Product functionals and its action in general dimensions}

 In this section, we will embark on a concise introduction to two-dimensional CFTs under the global conformal symmetry, specifically eschewing the Virasoro algebra. Subsequently, we will reexplore the definition and underlying concepts of the product functional, detailing several of its cardinal properties that render it an optimal candidate for the numerical conformal bootstrap. Finally, we will delve into its implications in the context of general dimensional CFTs and draw several immediate inferences from this discussion.



\subsection{Two-Dimensional Conformal Field Theory}

When we consider the four-point correlator in the $2d$ CFT, the conformal group $SL(2,\mathbb{C})$ groups the contribution of descendants of a given primary operator in terms of the $2d$ conformal block, it takes a factorized sum of $SL(2, \mathbb{R})$:

\begin{equation}
    G^{d=2}_{\Delta,\ell}(z,\bar{z})=\frac{1}{2} \big(k_\tau(z)k_\rho(\bar{z})+k_\tau(\bar{z})k_\rho(z)\big),\quad \tau=\Delta+\ell,\, \rho=\Delta-\ell
\end{equation}
Here $k_\beta(z)$ is the $SL(2, \mathbb{R})$ block:

\begin{equation}
    k_h\left(z \right)  :=z^{\frac{h}{2}}{ }_2 F_1\left(\frac{h}{2}, \frac{h}{2}, h ; z\right) 
\end{equation}
The reader could easily recongnize its relationship between $1d$ conformal block as $k_\beta(z)$ =$G^{d=1}_{\beta/2}(z)$. With this definition in mind, we can expand the $2d$ crossing vector and anticrossing vector in terms of the $1d$ counterpart \cite{Mazac:2016qev}:
\begin{equation}\label{2dF}
    F^{d=2}_{\Delta,\ell}(z,\bar{z}|\Delta_\phi)=\frac{1}{4} (F_{\tau}(z|\Delta_\phi)H_{\rho}(\bar{z}|\Delta_\phi)+F_{\rho}(z|\Delta_\phi)H_{\tau}(\bar{z}|\Delta_\phi)+(z\leftrightarrow\bar{z})),
\end{equation}
\begin{equation}
    H^{d=2}_{\Delta,\ell}(z,\bar{z})=\frac{1}{4} (F_{\tau}(z|\Delta_\phi)F_{\rho}(\bar{z}|\Delta_\phi)+H_{\rho}(z|\Delta_\phi)H_{\tau}(\bar{z}|\Delta_\phi)+(z\leftrightarrow\bar{z}))
\end{equation}

Here we used the notation:
\begin{equation}
    \begin{aligned}
k_h\left(z | \Delta_\phi\right) & :=z^{\frac{h}{2}-\Delta_\phi}{ }_2 F_1\left(\frac{h}{2}, \frac{h}{2}, h ; z\right) \\
F_h\left(z | \Delta_\phi\right) & :=k_h\left(z | \Delta_\phi\right)-k_h\left(1-z | \Delta_\phi\right) \\
H_h\left(z | \Delta_\phi\right) & :=k_h\left(z | \Delta_\phi\right)+k_h\left(1-z | \Delta_\phi\right)
\end{aligned}
\end{equation}

We also remind the reader about the following identity in terms of 1d (anti) crossing vector,
\begin{equation}
\begin{split}
   F_h\left(z | \Delta_\phi\right)= F^{d=1}_{h/2}\left(z | \frac{\Delta_\phi}{2}\right),\,\,\,  H_h\left(z | \Delta_\phi\right)= H^{d=1}_{h/2}\left(z | \frac{\Delta_\phi}{2}\right).
   \end{split}
\end{equation}

\subsection{Product Functional Action on Crossing and Anti-crossing Vectors}

Let us introduce the product functional action on the crossing vector. Its definition is as follows:

\begin{equation}
    \left(\omega^{-} \otimes \omega^{+}\right)\left(F_{\Delta, \ell}(z, \bar{z})\right):=2 \int_{++} \frac{\mathrm{d} z \mathrm{~d} \bar{z}}{\pi^2} h_{-}(z) h_{+}(\bar{z})\left[\mathcal{I}_z \mathcal{I}_{\bar{z}} F_{\Delta, \ell}(z, \bar{z})+\mathcal{I}_z \mathcal{I}_{\bar{z}} F_{\Delta, \ell}(z, 1-\bar{z})\right]
\end{equation}

The corresponding $1d$ functional and integration contour $++$ are defined as follows:

\begin{equation}
    \omega_{ \pm}\left(\Delta \mid \Delta_\phi\right)=\int_1^{\infty} \frac{\mathrm{d} z}{\pi} h_{ \pm}(z) \mathcal{I}_z F_{ \pm, \Delta}\left(z \mid \Delta_\phi\right)
\end{equation}
\begin{equation}
    \mathcal{I}_z F(z):=\lim _{\epsilon \rightarrow 0^{+}} \frac{F(z+i \epsilon)-F(z-i \epsilon)}{2 i} .
\end{equation}

Given the $2d$ crossing vector defined in Eq.\ref{2dF}, the product functional's action can be determined in an instant as a sum of factorized $1d$ functional evaluations, but with half of the external dimension:
\begin{equation}\label{eq: productdef}
\left(\omega_1^{-} \otimes \omega^{+}_2\right)\left(F_{\Delta, \ell}(z, \bar{z}|\Delta_\phi)\right):=\omega_1^- \otimes \omega_2^+(\Delta,\ell| \Delta_{\phi})=\frac{1}{2} \left(\omega_1^-(\tau)\omega_2^+(\rho)+\omega_1^-(\rho)\omega_2^+(\tau)\right)
\end{equation}

This equation uses the shorthand substitution:

\begin{equation}
\omega\left(h \right) = \omega\left(\left. \frac{h}{2} \right\vert \frac{\Delta_\phi}{2}\right) , \quad \tau=\Delta-l,\quad \rho=\Delta+l
\end{equation}

The functionals $\omega_1^-$ and $\omega_2^+$ are not limited to a specific form in our discussion, as long as they have the correct plus-minus type. As we will demonstrate, these functionals can belong to both the bosonic and fermionic basis, or even be the master functional that maximizes the four-point correlator at a certain point\cite{Paulos:2020zxx}. Since our functionals can be either $\alpha$ or $\beta$, our construction suggests that we have the following expansion of crossing vector $F^{d=2}_{\Delta,\ell}(z,\bar{z})$,

    \begin{equation} \label{2dFexp}
\begin{split} 
   & F^{d=2}_{\Delta,\ell}(z,\bar{z})=\mathcal{F}_{\Delta,\ell}(z,\bar{z})+\mathcal{F}_{\Delta,\ell}(\bar{z},z),\\
    &\mathcal{F}_{\Delta,\ell}(z,\bar{z})= \sum_{m,n} \bigg[ \bigg(\alpha^-_m(\tau|\Delta_{\phi})\alpha^+_n(\rho|\Delta_{\phi}) +\alpha^-_m(\rho|\Delta_{\phi})\alpha^+_n(\tau|\Delta_{\phi})\bigg) F_{\Delta_m}H_{\Delta_n}\\
   & +\bigg(\beta^-_m(\tau|\Delta_{\phi})\alpha^+_n(\rho|\Delta_{\phi}) +\beta^-_m(\rho|\Delta_{\phi})\alpha^+_n(\tau|\Delta_{\phi})\bigg) \partial F_{\Delta_m}H_{\Delta_n} \\
    &  +\bigg(\alpha^-_m(\tau|\Delta_{\phi})\beta^+_n(\rho|\Delta_{\phi}) +\alpha^-_m(\rho|\Delta_{\phi})\beta^+_n(\tau|\Delta_{\phi})\bigg) F_{\Delta_m}\partial H_{\Delta_n}\\
    & \bigg(\beta^-_m(\tau|\Delta_{\phi})\beta^+_n(\rho|\Delta_{\phi}) +\beta^-_m(\rho|\Delta_{\phi})\beta^-_n(\tau|\Delta_{\phi})\bigg) \partial F_{\Delta_m}
    \partial H_{\Delta_n}
    \bigg].
    \end{split}
\end{equation}

Considering the crossing symmetry of the vector $F_{\Delta, \ell}(z, \bar{z}|\Delta_\phi)$, the action of the product functional $\omega_1^{+} \otimes \omega^{-}_2$ does not offer new insights. However, the product functional $\omega_1^{+} \otimes \omega^{+}_2$ and $\omega_1^{-} \otimes \omega^{-}_2$ present potential candidates for the anti-crossing vector:

\begin{equation}
    \left(\omega_1^{+} \otimes \omega_2^{+}\right)\left(H_{\Delta, \ell}(z, \bar{z})\right):=2 \int_{++} \frac{\mathrm{d} z \mathrm{~d} \bar{z}}{\pi^2} h_{1+}(z) h_{2+}(\bar{z})\left[\mathcal{I}_z \mathcal{I}_{\bar{z}} H_{\Delta, \ell}(z, \bar{z})+\mathcal{I}_z \mathcal{I}_{\bar{z}} H_{\Delta, \ell}(z, 1-\bar{z})\right]
\end{equation}

\begin{equation}
    \left(\omega_1^{-} \otimes \omega_2^{-}\right)\left(H_{\Delta, \ell}(z, \bar{z})\right):=2 \int_{++} \frac{\mathrm{d} z \mathrm{~d} \bar{z}}{\pi^2} h_{1-}(z) h_{2-}(\bar{z})\left[\mathcal{I}_z \mathcal{I}_{\bar{z}} H_{\Delta, \ell}(z, \bar{z})-\mathcal{I}_z \mathcal{I}_{\bar{z}} H_{\Delta, \ell}(z, 1-\bar{z})\right]
\end{equation}

The functional action on $2d$ anti-crossing vectors is expressed as:

\begin{equation}
\left(\omega_1^{+} \otimes \omega^{+}_2\right)\left(H_{\Delta, \ell}(z, \bar{z}|\Delta_\phi)\right):=\omega_1^+ \otimes \omega_2^+(\Delta,\ell| \Delta_{\phi})=\frac{1}{2} \left(\omega_1^+(\tau)\omega_2^+(\rho)+\omega_1^+(\rho)\omega_2^+(\tau)\right)
\end{equation}

\begin{equation}
\left(\omega_1^{-} \otimes \omega^{-}_2\right)\left(H_{\Delta, \ell}(z, \bar{z}|\Delta_\phi)\right):=\omega_1^- \otimes \omega_2^-(\Delta,\ell| \Delta_{\phi})=\frac{1}{2} \left(\omega_1^-(\tau)\omega_2^-(\rho)+\omega_1^-(\rho)\omega_2^-(\tau)\right)
\end{equation}

These equations will prove instrumental in discussions involving CFTs with global symmetry and in the context of mixed correlators.

\subsection{Product functional as a numerical bootstrap basis}\label{Section: criteria}

In the realm of numerical bootstrap, a valid basis, denoted as $\Omega_i$, is generally accepted to satisfy the following necessary conditions:

\begin{enumerate}
    \item \textbf{Finiteness:} Above the unitarity bound, the functional action on the crossing vector must always remain finite. This condition ensures that our numerical computations do not encounter any issues of diverging results.
    \item \textbf{Swapping:} It is required that for a physical crossing equation, the functional obeys: 
    \begin{equation}
        \Omega_i(\sum_{\Delta, L} a_{\Delta, L} F_{\Delta, L})=\sum_{\Delta, L}a_{\Delta, L} \Omega_i(F_{\Delta, L}).
    \end{equation} 
    This property underlines the linearity of the functional operation, thus facilitating its practical usage in algebraic computations.
    \item \textbf{Completeness:} Completeness is understood as the ability of a system of functionals to entirely capture the constraints imposed by crossing symmetry. Mathematically, this concept can be expressed by the following condition:
    \begin{equation}
    \sum_{\Delta, L} a_{\Delta, L} F_{\Delta, L}=0 \Leftrightarrow \sum_{\Delta, L} a_{\Delta, L} \Omega_i(F_{\Delta, L})=0, \forall i
    \end{equation}
    \item \textbf{Positivity:} Every functional $\Omega_i$ in the basis, for each spin channel $L$, should exhibit sign-definiteness, except at most within a compact region neighboring the unitarity bound. This condition is crucial for the stability and validity of the inequalities derived from the bootstrap equations\footnote{The previously mentioned requirement for positivity may indeed be more stringent than what is actually necessary. In principle, it would suffice to ensure the positivity of certain finite linear combinations of those functionals in the asymptotic region, including large $\Delta$ and large spin.
}.
    \item \textbf{Computability:} The requirement for an efficient algorithm that can precisely compute the value of each functional is a practical necessity, ensuring that our numerical computations are not only feasible but also reliable.
\end{enumerate}

The above-mentioned conditions form the bedrock principles that guide the construction and usage of numerical bootstrap problems, ensuring that the entire process is both mathematically robust and computationally feasible.  The swapping condition, initially discussed by \cite{Qiao:2017lkv}, is particularly specialized for conformal bootstrap among these considerations.

\begin{table}[h]
\centering
\begin{tabular}{|c|c|c|c|c|}
\hline
& Derivative & $\omega^F,\, \omega^B$($1d$) & $\nu_{i,j}\,\&\,\mu_{i,j}$\cite{Caron-Huot:2020adz} & Product Functional \\
\hline
Finiteness & \checkmark& \checkmark & \checkmark & \checkmark \\
\hline
Swapping & \checkmark& \checkmark & \checkmark & \checkmark \\
\hline
Completeness & \checkmark& \checkmark & \checkmark ? & \checkmark ? \\
\hline
Positivity & \checkmark& \checkmark & \ding{55} & \checkmark \\
\hline
Computability & \checkmark& \checkmark & Not available & \checkmark \\
\hline
\end{tabular}
\caption{Candidates of higher dimensional functionals which are believed to be complete.}
\label{tab:ComparisonTable}
\end{table}

In this discussion, we start off by elucidating the manner in which the derivative basis upholds the outlined properties. It is evident that the derivative basis complies with the finiteness and positivity conditions. This compliance is rooted in the rational expansion of the conformal block, which has been thoroughly explored in earlier studies \cite{Kos:2013tga, Kos:2014bka}. Furthermore, it is important to note that the said expansion exponentially converges when the derivative at the crossing symmetric point is computed. Moving on to the completeness and swapping properties of the derivative basis, they are affirmed by the absolute convergence of the Operator Product Expansion (OPE) \cite{Pappadopulo:2012jk}. Notably, the absolute convergence of the OPE is instrumental in guaranteeing the analyticity of the system.

In the context of 1-dimensional analytic functionals, the initial three conditions have been comprehensively demonstrated to hold as shown by Maz\'{a}\v{c} and Paulos \cite{Mazac:2018ycv}\footnote{The finiteness and swapping conditions of $\beta_0$ were originally proven by \cite{Qiao:2017lkv}.}. While the proof of asymptotic definiteness remains elusive for a broad array of situations, empirical investigations have provided substantial evidence endorsing its validity. As for the computability condition, we reviewed this in Appendix \ref{funcwittenrelation} how we can efficiently compute these functionals.

The functionals $\nu_{i,j}$ and $\mu_{i,j}$, as introduced by Caron-Huot \textit{et al.} \cite{Caron-Huot:2020adz}, are subtracted versions of the functionals associated with the generalized free theory. This implies that the unsubtracted basis, denoted as $\alpha_{n,\ell}$ and $\beta_{n,\ell}$, directly corresponds to the operators within the generalized free theory. It is widely accepted that this functional basis satisfies the finiteness, swapping, and completeness conditions. Nevertheless, as demonstrated by Caron-Huot \textit{et al.} \cite{Caron-Huot:2020adz}, it has been revealed that it is not feasible to construct a basis adhering to the positivity condition using a finite linear combination of the original basis. Additionally, an effective method for evaluating the functional value remains to be determined\footnote{In the context of large-scale calculations, computations by definition---specifically those involving multivariable integration against a kernel---tend to converge at a pace that is prohibitively slow for practical implementation.
}.

Amongst all known $d>1$ analytic functionals, the product functionals discussed in this study stand unique as they adhere to all the aforementioned criteria. The study by Paulos \cite{Paulos:2019gtx} affirmed that product functionals satisfy finiteness and swapping conditions. The essence of the proof is that if $1d$ functionals meet these conditions, then their product functional counterparts would inherently comply as well. The same logic holds for the asymptotic definiteness condition (which is strongly conjectured for the functionals considered in this study). The completeness of product functionals, however, remains an open question. Nonetheless, our numerical study implies that product functionals likely satisfy this condition, as they reproduce all expected known bounds.

In our study, we posit four immediate and conjectured complete families of product functions, as expressed below:

\begin{equation}\label{eq: four}
    \omega^{+F}\otimes\omega^{-F},\, \omega^{+F}\otimes\omega^{-B},\, \omega^{+B}\otimes\omega^{-B},\, \omega^{+B}\otimes\omega^{-F}
\end{equation}

Here, the functional $\omega$ can be any $\alpha_m$ or $\beta_n$. Our research primarily zeroes in on a numerical exploration of the first two categories, $\omega^{+F}\otimes\omega^{-F}$ and $\omega^{+F}\otimes\omega^{-B}$. For brevity, these will be referred to as FF and FB respectively, in the ensuing sections. 

Although the current study focused on the exploration of the first two categories of functional, the methodology for numerical investigation applied to the last two would remain congruent. It is an open avenue for future studies to further explore these product functional families and their potential applications.

\begin{figure}[ht]
    \centering
    \begin{subfigure}{0.495\textwidth}
        \centering
        \includegraphics[width=\textwidth]{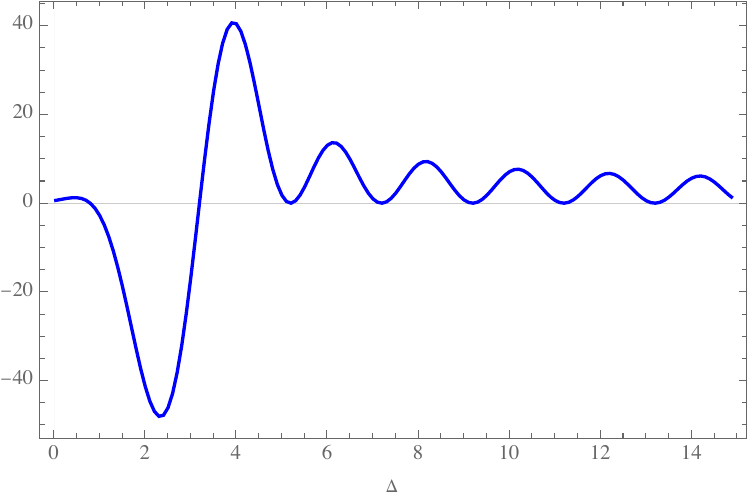}
        \caption{$\alpha_0^{-B}\otimes\beta_0^{+F}$}
        
    \end{subfigure}
    \hfill
    \begin{subfigure}{0.495\textwidth}
        \centering
    \includegraphics[width=\textwidth]{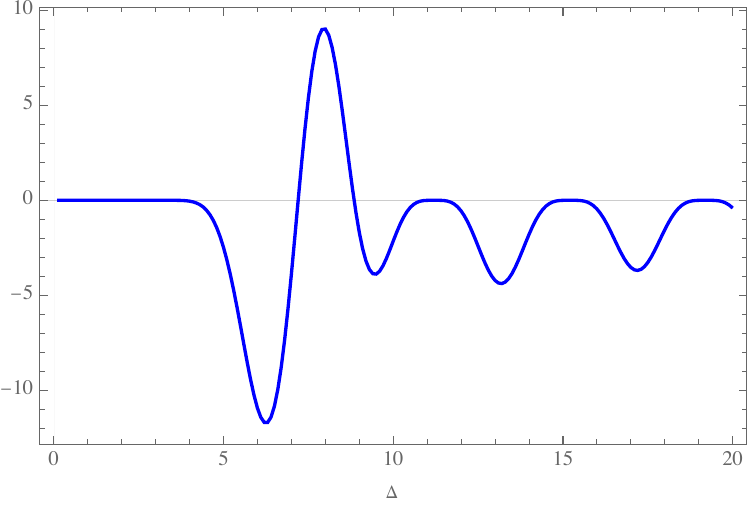}
        \caption{$\alpha_1^{-F}\otimes\beta_1^{+F}$}
        
    \end{subfigure}
    \caption{The action of the product functional $\alpha_0^{-B}\otimes\beta_0^{+F}$ and $\alpha_1^{-F}\otimes\beta_1^{+F}$ on the scalar channel in a $2d$ CFT. The functionals here are normalized by the factor in Appendix.\ref{Normalization}}
    \label{fig: 2dplot}
\end{figure}

We want to draw the reader's attention to the fact that although the numerical evidence provided in this study demonstrates that the four categories of product functionals depicted in equation \eqref{eq: four} can each independently form a complete basis, they exhibit distinct asymptotic zeros in two dimensions. Specifically, the functionals $\omega^{+F}\otimes\omega^{-B},\, \omega^{+B}\otimes\omega^{-F}$ display double zeros in the asymptotic region, which aligns with the correct spectrum of the generalized free theory. Contrastingly, the functionals $\omega^{+F}\otimes\omega^{-F},\, \omega^{+B}\otimes\omega^{-B}$ present fourth-order zeros in steps of four, a behavior that coincides with the asymptotics of $2d$ Ising model. The asymptotic zeros are more explicitly expressed as\footnote{These are the zeros for the $2d$ functional value, we will discuss the higher dimensional zeros later.}:
\begin{equation}
    \begin{cases}
    2\Delta_\phi +4n+l,       & \text{for } \omega^{+B}\otimes\omega^{-B}\\
    2\Delta_\phi +4n+l+2,       & \text{for }\omega^{+F}\otimes\omega^{-F}\\
    2\Delta_\phi +2n+l,       & \text{for }\omega^{+F}\otimes\omega^{-B},\, \omega^{+B}\otimes\omega^{-F}
\end{cases},\quad n\in \mathbb{Z}
\end{equation}
These patterns can be easily verified using the definition of functional action and the properties of $1d$ functional, as presented in Appendix.\ref{sec:1d-func}. Figure.\ref{fig: 2dplot} provides a graphical representation of this behavior.

In terms of numerical implementation, the product functionals confer several additional advantages. First and foremost, they can be normalized to a limit that does not diverge to infinity, both for large values of $\Delta$ and large spin $\ell$. Our numerical investigations have underscored that achieving an optimal normalization for these functionals is crucial for maintaining computational stability. Our specific choices for normalization are elaborated in Appendix \ref{Normalization}. Secondly, the product functional exhibits sign-definiteness exclusively within a compact neighborhood of the unitarity bound, a point further clarified in Section \ref{Section: instant}. This characteristic significantly streamlines the discretization implementation process. Specifically, in the sign-definite region, a minimal number of points is sufficient to ensure positivity in the asymptotic region.\footnote{Regarding the positivity, there is a special feature worth noting. For the product functional extensively studied in this work, specifically the FF and FB types, we observed numerically that $\alpha_0^{+F}\otimes\alpha_0^{-B}$, $\alpha_0^{+F}\otimes\alpha_0^{-F}$, and $\alpha_0^{+F}\otimes\beta_0^{-F}$ are positive-definite for $\ell \geq 2$ and for $\ell = 0$, only negative in a compact region in the vicinity of the unitarity bound. A more systematic understanding of this positivity is anticipated in future work.}

To conclude this section, we wish to discuss the relationship between the product functionals and the $\nu_{i,j}\,\&\,\mu_{i,j}$ functionals presented in the work of Caron-Huot et al. \cite{Caron-Huot:2020adz}. Both bases adhere to the swapping condition. For the functional categories $\omega^{+F}\otimes\omega^{-B}$ and $\omega^{+B}\otimes\omega^{-F}$, they even share the same asymptotic zeros as $\nu_{i,j}\,\&\,\mu_{i,j}$. Despite these similarities, they are inherently different bases. The $\nu_{i,j}\,\&\,\mu_{i,j}$ basis corresponds one-to-one with the Generalized Free Field (GFF) operators, albeit one to finite due to the subtraction, while the product functionals exhibit a one-to-infinite correspondence. This distinction is evident from the definition in Eq.\ref{eq: productdef}. Assuming both bases are complete functionals on the space of crossing vectors, the product functionals essentially represent infinite linear combinations of the $\nu_{i,j}\,\&\,\mu_{i,j}$ functionals.

\subsection{Functional action on Higher Dimensional Crossing Vectors} 

In this section, we explore the action of the product functional $\Omega^ {\otimes}$ constructed in the previous section on higher dimensional crossing vectors. Rather than directly performing the multi-complex-variable integration, we utilize an insightful observation \cite{Hogervorst:2016hal}:

\begin{equation}\label{eq: dd-1}
     G^{d}_{\Delta,\ell}(z,\bar{z})=\sum_{n=0}^{\infty} \sum_{j=\ell,\ell-2,....,\ell\, \text{mod}\, 2} \mathcal{A}_{n, j}(\Delta,\ell)   G^{d-1}_{\Delta+2n,j}(z,\bar{z}),
\end{equation}

where the detailed definition of $\mathcal{A}_{n, j}(\Delta,\ell)$ can be found in the appendix, and $d$ is the space-time dimension. We further stress that $\mathcal{A}_{n, j}(\Delta,\ell)$ is always positive. Notably, the conformal crossing vector $F^-(\Delta, \ell|\Delta_\phi)$ adheres to the same formula.

This insight naturally leads us to propose a corresponding functional formula:

\begin{equation}\label{functional3}
     \Omega^ {\otimes,\,d=3}(\Delta,\ell)=\sum_{n=0}^{\infty} \sum_{j=\ell,\ell-2,....,\ell\, \text{mod}\, 2} \mathcal{A}_{n, j}(\Delta,\ell)   \Omega^ {\otimes,\,d=2}(\Delta+2n,j),
\end{equation}

which can be rigorously justified. By utilizing the asymptotic limit given by Appendix.\ref{Normalization}, we find that the nth-term contribution behaves as:

\begin{equation}\label{3dconverge}
    \sum_{j=\ell,\ell-2,....,\ell\, \text{mod}\, 2} \mathcal{A}_{n, j}(\Delta,\ell)   \Omega^ {\otimes,\,d=2}(\Delta+2n,j)= \mathcal{O}\left(\frac{1}{n^{4\Delta_\phi +6}}\right),\, n\rightarrow\infty
\end{equation}

Thus, the series on the right-hand side of the functional formula is absolutely convergent for any $\Delta,\, \ell$, and $\Delta_\phi$ that satisfy the unitarity bound. Invoking Fubini's theorem, we find that the absolute convergence not only guarantees the existence of the functional action but also justifies the interchange of the infinite sum with the functional action. 

The formula described in Eq.\ref{functional3} is the key methodology employed to numerically compute the $3d$ functional in this study. The applicability of the $4d$ action becomes immediate. It is interesting to write done the $d$ dimension conformal block to $2$ dimension conformal block version of Eq.\ref{eq: dd-1}. This is due to the fact that the $4d$ block can be expressed as a resummation of the $2d$ block, multiplied by the $2d$ OPE.\footnote{We would like to highlight for the reader that some descendant fields in $4d$ can be interpreted as primary fields in $2d$, owing to the underlying group theoretical considerations.} Furthermore, under the same set of assumptions, the conditions of completeness and swapping are ensured for $d\geq 3$. 

We must emphasize that our approach towards dealing with $4d$ CFTs(and $6d$ CFT), as discussed in the preceding paragraph, diverges from the methodologies proposed in \cite{Paulos:2019gtx} and \cite{Kaviraj:2021cvq}. The latter studies utilize a product functional of shifted $\Delta_\phi$. One possible pitfall associated with their proposed functional is the potential lack of assured positivity. We refer the readers to appendix \ref{prodgend} for a discussion of product functionals in general dimensions, which is slightly different from \cite{Paulos:2019gtx} and \cite{Kaviraj:2021cvq}. Future research is required to establish whether this shifted functional approach can be made effective.

\subsubsection{Technical issue: slow convergence}\label{Sec: slow}

Equation \eqref{3dconverge} represents a promising outcome due to its strong power-law suppression. In practical terms, we observed that retaining a reasonable number of terms on the right-hand side allows for an accurate evaluation of the functional for small $\Delta$ in the scalar channel. Unfortunately, further investigation reveals that this strong power-law suppression is associated with a divergent factor:

\begin{equation}\label{3dconvergeslow}
    \sum_{j=\ell,\ell-2,....,\ell\, \text{mod}\, 2} \mathcal{A}_{n, j}(\Delta,\ell)   \Omega^ {\otimes,\,d=2}(\Delta+2n,j)=\mathcal{O}\left( \frac{\Delta^{4\Delta_\phi +11/2}}{n^{4\Delta_\phi +6}}\right)
\end{equation}

 in the limit where $n \gg 1,\, \Delta\gg 1,\, \mathrm{and}\, n\gg \Delta$. A notable impediment to extending our approach to higher spin channels in $3d$ is the current inability to compute a closed form of the leading coefficient $C_{\omega_\pm}$ in the asymptotic formula, as given by equation \eqref{eq:asymp}. A potential avenue for improvement could involve computing this expression explicitly, by which we can sum analytically to improve the convergence by one order in $n$. On a more speculative note, it may be feasible to express the $3d$ functional action as a rational expansion, just as the derivative basis. This conjecture is predicated on the possibility of recasting the evaluation of the $1d$ functional in terms of a rational expansion. As demonstrated in Appendix.\ref{funcwittenrelation}, the computation of the $1d$ functional hinges on the evaluation of the generalized hypergeometric function ${}_7F_6$, which potentially can admit a well-established rational expansion\cite{willis2012acceleration}\footnote{We would like to acknowledge Miguel Paulos to bring this paper to our attention and for confirming that this can be implemented in some cases.}.

\subsection{Instant bootstrap results}\label{Section: instant}

\begin{figure}[ht]
    \centering
    \begin{subfigure}{0.495\textwidth}
        \centering
        \includegraphics[width=\textwidth]{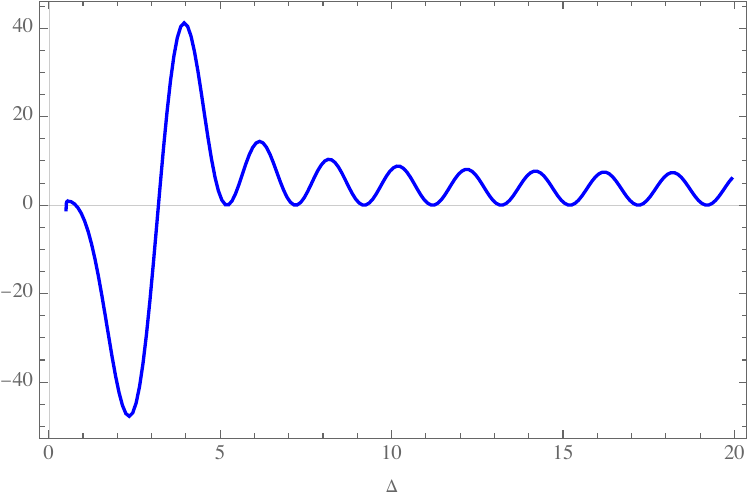}
        \caption{$\alpha_0^{-B}\otimes\beta_0^{+F}$}
        
    \end{subfigure}
    \hfill
    \begin{subfigure}{0.495\textwidth}
        \centering
    \includegraphics[width=\textwidth]{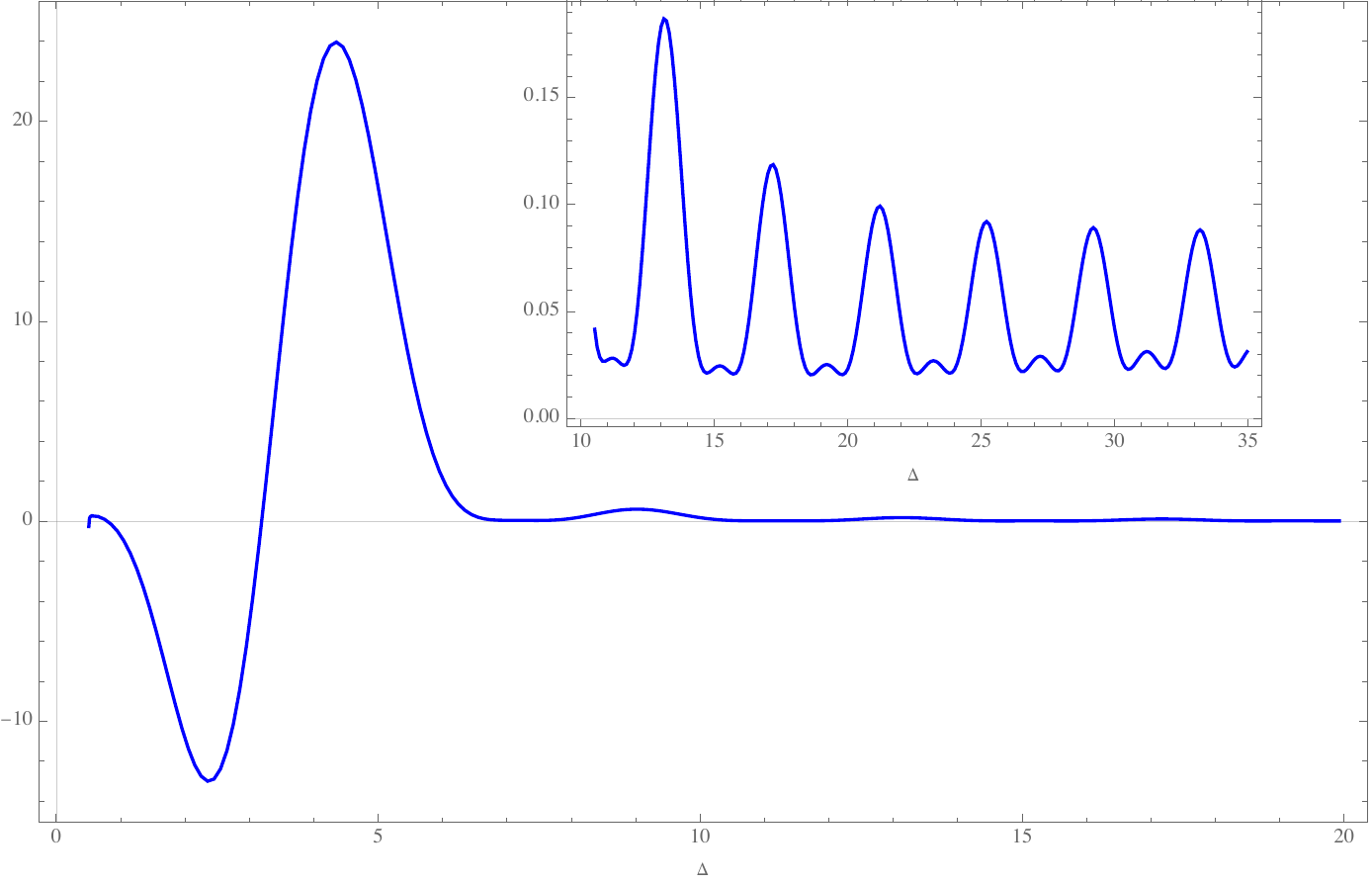}
        \caption{$\alpha_0^{-F}\otimes\beta_0^{+F}$}
        
    \end{subfigure}
    \caption{The action of the product functional $\alpha_0^{-B}\otimes\beta_0^{+F}$ and $\alpha_0^{-F}\otimes\beta_0^{+F}$ on the scalar channel in a $3d$ CFT. As the inset plot shows, $\alpha_0^{-F}\otimes\beta_0^{+F}$ has no double zeros asymptotically, but positive definite. The functionals here are normalized by the factor in Appendix.\ref{Normalization}}
    \label{fig: 3dplot}
\end{figure}

In this section, we delve deeper into the potency of product functionals by discussing several bootstrap results obtained using a single element of the product functional. The discussion is further facilitated by noting some key properties of the $2d$ and $3d$ functionals. These properties can be inferred from the characteristics of the $1d$ functional, as detailed in Appendix \ref{sec:1d-func}. 

\begin{enumerate}
    \item Consider the product functional $\omega_m^{+}\otimes\omega_n^{-}(\Delta, \ell)$, where $\omega$ represents $\alpha$ or $\beta$ of either fermionic or bosonic type. To ensure the functional is sign definite beyond a particular point, we require that $\Delta-\ell \geq 2\Delta_\phi+4\max (m, n)+6$. This bound holds valid in $2d$ as well as in higher spacetime dimensions since the coefficient in eq.\eqref{functional3} are all positive.
    
    \item In the $2d$ scenario, the FF and BB functionals exhibit fourth-order zeros in steps of 4 (reminiscent of minimal model type), while the FB and BF exhibit double zeros in steps of 2 (typical of GFF type). This is depicted in Figure \ref{fig: 2dplot}.
    
    \item In contrast, the $3d$ FF and BB type functionals do not have asymptotic zeros, as eq.\eqref{functional3} proceeds in steps of 2, contradicting the pattern of the zeros observed for FF and BB in $2d$. However, FB and BF still manifest the expected GFF type double zeros in the asymptotic region, as shown in Figure \ref{fig: 3dplot}.
\end{enumerate}

\subsubsection{An optimal functional in 2d}
Here we provide an example of an optimal analytical functional  where the gap in the scalar sector is saturated by a physical solution for the four-point crossing equation in 2d. Specifically, we focus on the four-point energy correlator of the 2D Ising model. Initially presented by \cite{Mazac:2016qev} and further analyzed in detail in \cite{Paulos:2019gtx}, we briefly mention it here for the sake of completeness. Let us consider the following product functional   $(\beta_0^{-,F} \otimes \alpha_0^{+,F})^{d=2} (\Delta,\ell| \Delta_{\phi})$. We set $\Delta_\phi=1$. We notice the following properties are satisfied by this functional,
\begin{enumerate}
    \item $(\beta_0^{-,F} \otimes \alpha_0^{+,F})^{d=2} (0,0| \Delta_{\phi})= 0$ \text{and} $(\beta_0^{-,F} \otimes \alpha_0^{+,F})^{d=2} (\Delta,0| \Delta_{\phi})\geq 0$, \text{for} $\Delta\geq 4$,
    \item $(\beta_0^{-,F} \otimes \alpha_0^{+,F})^{d=2} (\Delta,\ell| \Delta_{\phi})\geq 0$,  \text{for} $\Delta>\ell$ and $\ell>0$.
\end{enumerate}
We also note that the functional is negative in the scalar channel for $1.215<\Delta<4$. From the consistency of the bootstrap equation, we can deduce that the existence of an operator with a dimension lying within the range of 1.215 and 4 is necessary. We know that the energy correlator of the 2d ising model has gap 4 in the scalar channel and therefore the above functional gives an optimal bound for $\Delta_\phi=1$.

\subsubsection{Upper bound on OPE data of 3d Ising model}

\begin{figure}[ht] 
	\centering
	\includegraphics[width=8cm]{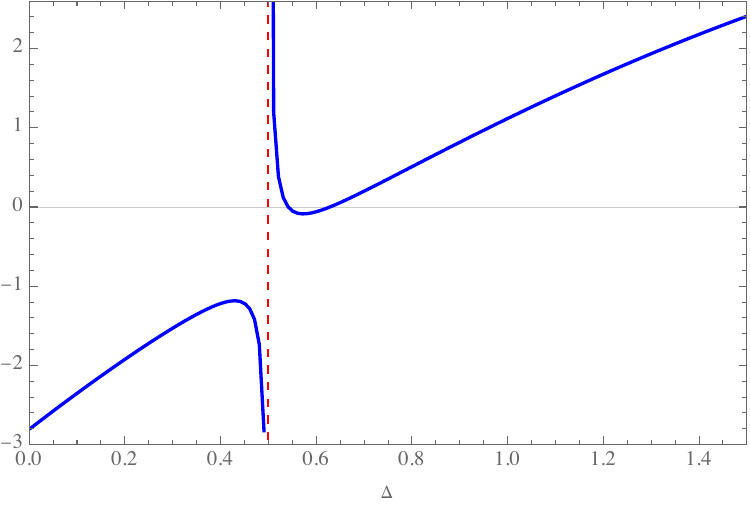}
	\caption{\label{aa}The action of the product functional $\alpha_0^{-B}\otimes\alpha_0^{+F}$ on the scalar channel in a $3d$ CFT. The red-dashed line represents the unitary bound at $\Delta=0.5$, at which the functional action tends to diverge.}

\end{figure}

In this section, we provide an illustrative example of an analytical functional $(\omega^- \otimes \omega^+)^{d=3} (\Delta,\ell| \Delta_{\phi})$ which provides an upper bound on the OPE data of 3d Ising model. Here we have chosen $\omega^-=\alpha^{-,B}_0$ and  $\omega^+=\alpha^{+,F}_0$. We set the dimension of the external scalar to $\Delta_{\phi}=0.518149$.This functional exhibits the following properties (see figure \ref{aa}),
\begin{enumerate}
    \item $(\alpha_0^{-,B} \otimes \alpha_0^{+,F})^{d=3} (0,0| \Delta_{\phi})<0$ and $(\alpha^{-,B} \otimes \alpha^{+,F})^{d=3} (\Delta,0| \Delta_{\phi})>0$ for $\Delta>0.7$.\\
    \item $(\alpha_0^{-,B} \otimes \alpha_0^{+,F})^{d=3} (\Delta,\ell| \Delta_{\phi})>0$ for $\Delta>1+\ell$ and $\ell>0$.
\end{enumerate}
  We find that,
$\alpha_0^{-B}\otimes\alpha_0^{+F}(0,0)=-2.80772$ and $\alpha_0^{-B}\otimes\alpha_0^{+F}(1.412625,0)=2.12498$. Provided the first non-trivial scalar has dimension greater than 0.7, our functional gives an upper bound on the ope coefficient of that operator with the external scalar field. Choosing the first non-trivial scalar operator, $\epsilon$, to have dimension 1.412625 we get the following upper bound,
\begin{equation}
C^2_{\phi \phi \epsilon}< -\frac{(\alpha_0^{-,B} \otimes \alpha_0^{+,F})^{d=3} (0,0| \Delta_{\phi})}{(\alpha_0^{-,B} \otimes \alpha_0^{+,F})^{d=3} (1.412625,0| \Delta_{\phi})}=1.32129.
\end{equation}

Since we know that the first non-trivial scalar operator in the 3d Ising model has a dimension greater than 1, the above bound is also valid for its OPE coefficient.The best value for the ope coefficient of $\sigma$ with $\epsilon$ operator, $C^2_{\sigma\sigma\epsilon}$, found from numerical bootstrap is 1.1064. The value obtained from our single functional is fairly close to the optimum value. This result gives us confidence that our basis will converge very fast to its optimum value with less number of functionals.

\subsection{Infinite Spin Channel}

The implementation of product functionals introduces a subtlety in comparison to the use of a derivative basis. In numerical bootstrap with a derivative basis, positivity for the remainder of the higher spin channels is secured by the inclusion of a sufficient number of spin channels, with the higher spin channel contributions exponentially suppressed. However, this scenario transforms when conducting a numerical bootstrap using functionals. For instance, with the Gaussian Free Field (GFF) spectrum, the contribution of the higher spin channel in the crossing equation of $\alpha_{0}^{B-}\otimes \beta_{0}^{F+}$ can be readily seen to behave as $l^{-5}$. This situation is quite universal for analytic functionals: the exponential decay of the Operator Product Expansion (OPE) is counterbalanced by the exponential growth of the functionals, resulting in a series converging in power-law. This observation suggests that the high spin channels might have an increased significance in the crossing equation of analytic functionals. For the numerical implementation of the $2d$ conformal bootstrap in this study, we included a single spin channel at $256$ to ensure positivity in the asymptotic region (as illustrated in Appendix \ref{Appendix: parameter}). However, we propose an intriguing alternative: considering the infinite spin channel. Given the asymptotic behavior listed in Eq. \ref{eq:asymp} or the normalized functional action Eq.\ref{eq: 2dnorm}, we have:

\begin{equation}
    \omega^{-} \otimes \omega^{+}(\tau+\infty,\infty )\sim C_{\omega_-}\sin^2(\frac{\pi}{2}(\tau-\Delta_{\omega_-})) \omega^{+}(\tau)
\end{equation}
This is due to the first term in Eq.\ref{eq: 2dnorm} is suppressed in the large $\ell$ limit. As a result, the positivity for the asymptotically large spin is equivalent to the positivity condition among some plus type functional.

We can further incorporate the $\sin$ square factor into the normalization of OPE. This suggests that the positivity in the large spin limit is primarily captured by the plus-type functional. This approach merits exploration in future studies.  The potential physical implications of the contribution of the infinite spin channel to the crossing equation remain unclear. We further suggest that this setup can also be employed in higher dimensions (e.g., $d=3$). In the large spin limit, a meticulous examination of the functional action on Eq. \ref{functional3} reveals that the functional action is predominantly influenced by the neighborhood of infinite spin, with a strong power-law convergence. This observation allows us to define the functional action on the infinite spin channel in $d=3$.

\section{Numerical Implementation}

To address the constraints imposed by the Conformal Field Theory (CFT) analytic functionals, we introduce the \textbf{FunBoot} package, implemented in Julia, which includes an outer approximation solver for semi-infinite programming. The outer approximation method tackles semi-infinite programming problems by solving a successive series of linear programs. In this section, we first provide a succinct introduction to semi-infinite programming, linear programming, and the simplex algorithm. Subsequently, we delineate the numerical implementation employed in our study.

\subsection{Conformal Bootstrap and Semi-Infinite Programming}

The Conformal Bootstrap method, particularly when applied to a single correlator, can be viewed as a Semi-Infinite Programming (SIP) problem\cite{reemtsen1998semi}. The bootstrap conditions serve as an infinite set of constraints on the space of Conformal Field Theories (CFTs), and the aim is to find CFT data that satisfy all these constraints.

A typical bootstrap problem can be formulated as follows\footnote{This is over-simplified, i.e. we make the parameter including spin to be dummy.}:

\begin{equation}
\begin{aligned}
& \text{maximize}
& & \Delta \\
& \text{subject to}
& & F(\Delta, \lambda; \Delta_i, \lambda_i) \geq 0, \; \forall \; \Delta_i, \lambda_i \in \mathbb{R},
\end{aligned}
\end{equation}


where $\Delta$ and $\lambda$ are scaling dimensions and OPE coefficients of the CFT, and $F$ is the bootstrap function that encodes the constraints from the bootstrap conditions.

Solving this kind of bootstrap problem is challenging due to the infinite number of constraints. A common technique to handle these challenges is the method of \textbf{outer approximation}.

The idea of outer approximation is to approximate the infinite set of constraints with a finite subset that captures the essential features of the problem. This is done by iteratively solving a series of finite sub-problems, where each sub-problem is a finite approximation of the original semi-infinite problem. The solution to each sub-problem is then used to generate a new constraint for the next sub-problem. This process is repeated until a suitable convergence criterion is met.

At each iteration $k$, we solve the following sub-problem:

\begin{equation}
\begin{aligned}
& \text{maximize}
& & \Delta \\
& \text{subject to}
& & F(\Delta, \lambda; \Delta_i, \lambda_i) \geq 0, \; \forall \; i \in \{1, 2, \ldots, k\},
\end{aligned}
\end{equation}

where $\{\Delta_1, \Delta_2, ..., \Delta_k\}$ and $\{\lambda_1, \lambda_2, ..., \lambda_k\}$ are finite subsets of the scaling dimensions and OPE coefficients chosen based on the solutions to the previous sub-problems. Therefore, we address a series of successive optimization problems that are characterized by finite degrees of freedom. Specifically, within the context of our single-correlator conformal bootstrap study, these problems translate into a series of linear programming.

\subsection{Brief Introduction to Linear Programming and the Simplex Algorithm}

In this section, we offer a concise introduction to the simplex algorithm as it pertains to linear programming. While not exhaustive, the main objective of this discussion is to establish the necessary notation and lay the groundwork for the subsequent examination of the outer approximation methodology in the sections that follow. For further reading and a deeper understanding of these concepts, the reader can consult the introductory book \cite{matouvsek2007understanding}.

The simplex algorithm is a pivotal method in linear programming designed to solve optimization problems. Linear programming pertains to the optimization of a linear objective function, dictated by linear equality and inequality constraints.

We begin with a formal definition of a linear programming problem in its standard form:

\begin{align}
\text{Minimize} \quad & c^T x \\
\text{subject to} \quad & Ax = b, \\
& x \geq 0,
\end{align}

In this formulation, $c$ denotes the cost vector, $x$ is the vector of variables to be optimized, $A$ is the coefficient matrix corresponding to the constraints, and $b$ is the vector representing available resources. The simplex algorithm is applied to find an optimal solution to this problem.

This general formulation can be reinterpreted in the context of the functional conformal bootstrap. Here, the elements of the $x$ vector represent the squares of the OPEs for primary operators. The columns of the $A$ matrix constitute vectors of functional values at the corresponding primary operators, while the vector $b$ corresponds to the negative functional value on the identity operators, i.e., $-\Vec{w}(\Delta=0,\, L=0)$. The choice of the objective function, given by the cost vector $c$, depends on the particular bootstrap problem under consideration. For instance, it may correspond to a single OPE for OPE maximization, or an exponentially decreasing vector for gap maximization, as we will discuss in Section.\ref{gap}.

The simplex algorithm, propounded by George Dantzig in 1947\footnote{Please note that the original paper by George Dantzig, where he proposed the simplex method in 1947, is believed to be an unpublished manuscript titled "Prospectus for the AAF electronic computer". The exact details and the full text of this manuscript are not publicly available.}, systematically inspects the vertices of the feasible region, defined as the set of all solutions satisfying the constraints. The algorithm commences with an initial feasible solution and proceeds iteratively, at each step transitioning to a better feasible solution, until it attains the optimal solution or concludes that no finite optimal solution exists.

A basic feasible solution (BFS) is a solution that corresponds to a basis of the system of linear equations. Let $A$ be a $m \times n$ matrix, $b$ be a $m$-dimensional vector, and $x$ be a $n$-dimensional vector. We can write the system of equations as $Ax = b$.

A basis of this system is a set of $m$ linearly independent columns of $A$. Without loss of generality, let's reorder the columns of $A$ such that the first $m$ columns form a basis. We can write $A$ as $[B \, | \, N]$, where $B$ is a $m \times m$ matrix corresponding to the basis, and $N$ is a $m \times (n-m)$ matrix corresponding to the non-basis. Similarly, we can partition the vector $x$ into $x = [x_B^T \, | \, x_N^T]^T$, where $x_B$ is the vector of basic variables and $x_N$ is the vector of non-basic variables.

An BFS is then obtained by setting the non-basic variables $x_N$ to zero and solving the system $Bx_B = b$ for the basic variables $x_B$. If all entries of $x_B$ are non-negative, then the solution is a BFS.

The simplex algorithm works by starting from an BFS, and iteratively moving to adjacent BFSs (by swapping one basic variable with one non-basic variable) in a way that improves the objective function. The selection of a candidate for the new BFS is guided by a pivot rule. The pivot rule determines the entering variable (a non-basic variable to be included in the basis) and the exiting variable (a basic variable to be removed from the basis).

Let's assume that we are at a feasible solution with basis $B$ and non-basic variables $x_N = 0$. The current objective function value is given by $c_B^T x_B$ where $c_B$ are the costs associated with the basic variables.

The reduced or equivalent cost $c_j'$ of a non-basic variable $x_j$ is computed as:

\begin{equation}\label{ReducedCost}
c_j' = c_j - c_B^T B^{-1}A_j,
\end{equation}

where $A_j$ is the column of $A$ corresponding to the non-basic variable $x_j$.

If all the reduced costs $c_j'$ are nonnegative, then the current solution is optimal. If there exists a $j$ such that $c_j' < 0$, then increasing $x_j$ from zero would improve the objective function (i.e., make it smaller since we're minimizing). Hence, such a $j$ can be selected as the entering variable for the next iteration of the simplex method. In the case of multiple negative $c_j'$, a common pivot rule is to select the most negative one (this rule is known as Dantzig's rule), but other selection rules can be applied as well.

Once the entering variable is chosen, the exiting variable is determined by the minimum ratio rule, which ensures that the solution remains feasible as we move to the new BFS.

The pivot rule is a crucial component of the simplex algorithm. Different pivot rules can affect the efficiency of the algorithm, and the choice of pivot rule can depend on the specific characteristics of the problem at hand. In the appendix of this document, we will provide a detailed discussion of the pivot rule implemented in our version of the simplex algorithm.

Please note that the simplex method is a fundamental algorithm for linear programming, but its implementation can be quite complex, especially for large-scale problems. We will discuss some details of our implementation in Section.\ref{pivot}.

\subsection{Two-phase Outer Approximation Method}

Our method of outer approximation involves resolving several sub-problems related to linear programming, each of which is addressed via the simplex algorithm. 

\subsubsection{Phase 1: Comprehensive Grid}

In the initial phase, we discretize the $\Delta$ axis into non-uniform grids for each spin channel. We allocate a denser grid to low twist operators and a relatively sparser grid for higher twist operators. For regions above the twist bound as delineated in Section.\ref{Section: instant}, where all functionals are sign-definite, we adopt an even sparser grid to accurately represent the asymptotic behavior. The exact implementation of the grids is described in detail in the appendix. We will refer to this grid as the \textit{comprehensive grid}. 

Subsequently, we proceed to solve this discretized linear programming problem, with the goal of achieving an optimal (or near-optimal) solution. As a basic feasible solution, the optimal solution will present a full spectrum of all operators where the reduced cost, computed as:
\begin{equation}
    c_j' = c_j - c_B^T B^{-1}A_j,
\end{equation}
equals zero. This behavior is illustrated in Fig.\ref{outer}.

It can be noted that there may still be operators that satisfy the unitarity condition but have negative reduced costs (but absent from the comprehensive grid). The next step involves integrating these operators into the linear programming framework and re-optimizing.

 \subsubsection{Phase 2: Focused Grid}

\begin{figure}[ht]
	\centering
	\includegraphics[width=8cm]{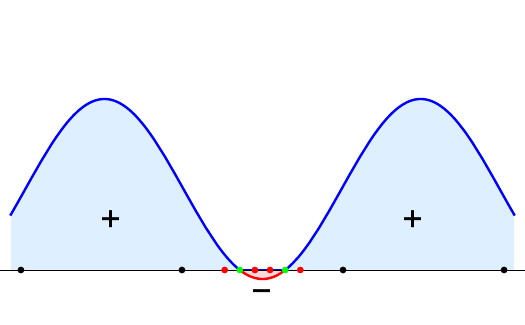}
	\caption{\label{outer} This diagram illustrates the construction of the focused grid. The black points denote the operators drawn from the comprehensive grid, while the green points represent the operators in the optimal solution derived from the last optimization on the comprehensive grid. The red points indicate the new operators that will be incorporated into the focused grid. The distance between a green point and its corresponding red points is one-third of the distance from the green point to the nearest point, which in this case is another green point.}
\end{figure}

An initial approach could be to locate the local minima of the interpolated reduced cost exhibiting negative values and incorporate these points into the Comprehensive Grid. However, a more efficient strategy is available. For the operators present in the optimal solution from the first phase, there is a guaranteed zero reduced cost, and a region of negative reduced cost can be found nearby. 

To capitalize on this, we construct a new grid, dubbed the \textit{Focused Grid}, using the following methodology: for each operator from the optimal solution, we add two new operators on both the right-hand side and the left-hand side. The spacing is set to be one-third of the distance from the operator to the remaining grid (or equivalently the distance from the closest operator), which may be the Comprehensive Grid or the Focused Grid from a previous iteration. In addition, we meticulously avoid the inclusion of operators that contravene the unitarity bound. 

This process is visually represented in Fig.\ref{outer}. The resulting Focused Grid has a size approximately double that of the number of functionals (after deleting the new operators violating the unitarity bound), thus permitting immediate optimization.

\subsubsection{Iteration: Revisiting the Comprehensive Grid}

The optimization process on the Focused Grid can lead to non-perturbative changes in the functional, potentially generating new regions of negative reduced cost on the Comprehensive Grid. Even though such changes are less likely during the later iterations, they can indeed occur during the initial iterations.

Our approach in such a scenario is straightforward: we place the optimal solution obtained from the Focused Grid back onto the Comprehensive Grid and re-optimize it. This allows us to verify whether the optimization process from the previous step has induced new regions of negative reduced cost. 

As the iteration progresses, we anticipate that the effective grid distance becomes three times smaller in each iteration compared to its previous state. Typically, we can obtain an extremely precise solution within approximately 20 iterations. Provided there are no issues with numerical stability, there is generally no need for optimization on the comprehensive grid beyond the initial few iterations.

The algorithm delineated above fundamentally relies on a simplex solver at each step. While we have not yet examined its applicability to mixed correlator situations, we maintain a confident stance that the approach should continue to hold true with minimal necessary adjustments. In this context, our course of action would involve identifying operators where positivity is saturated. A major challenge would then be leveraging the optimal solution derived from the preceding iteration to initialize the interior point algorithm, commonly referred to as 'hot-starting.'

\subsection{Technical Issue: Pivot Rule and Numerical Stability}\label{pivot}

The maximum cost variation (MCV) pivot rule appears to be the most efficient for our bootstrap problem, for reasons that are not yet entirely clear \cite{Paulos:2014vya}. This is to the best of our knowledge, and we must stress that this pivot rule may not be optimal for a general class of linear programming problems. The potential success of the MCV pivot rule might be because our linear programming derives from a smooth function (analytic functional or derivative value at crossing symmetric point). This pivot rule selects a new operator with the maximum cost variation in the objective function. Each iteration requires substantial computational effort to determine the final cost variation of each new candidate element of the BFS, but it results in fewer iterative steps. However, a naive implementation of MCV could encounter insurmountable numerical instability around 80 functionals, unlike the arbitrary precision implementation in \cite{Paulos:2014vya}. Therefore, it is beneficial to identify the source of the numerical instability before proposing our implemented pivot rule.

The simplex algorithm is a pivotal method that is numerically stable, with numerical errors that do not accumulate across iterations. In our current problem, there is evidence that after the normalization introduced in Appendix \ref{Normalization}, the numerical instability originates from the ill-conditioned matrix $B$ (the matrix of column vectors of BFS). The best way to avoid numerical instability is to evade such ill-conditioned situations.

To achieve this, we introduced an additional check to the iteration, alongside the MCV pivot rule:
\begin{enumerate}
    \item We ensure all coefficients of the solution are positive. Although this is theoretically guaranteed, it may be violated if our BFS matrix $B$ is ill-conditioned.
    \item We calculate the cost variation in two different ways: first, by multiplying the reduced cost Eq.\ref{ReducedCost} with the corresponding max variation operator, and second, by subtracting the old cost from the new cost. These two quantities must coincide in the absence of rounding error. Therefore, if we observe a significant discrepancy in their values (for example, 10\%), we interpret this as a strong indication of ill-conditionedness. 
\end{enumerate}
The advantage of this check is that it does not introduce additional computational cost, as we would calculate all necessary quantities when applying the MCV pivot rule. If the new BFS matrix $B$ turns out to be ill-conditioned, we will attempt the next candidate with the second-highest cost variation.

In practice, we have tested this pivot rule with up to 144 functionals and it has demonstrated excellent numerical stability. The area of concern is the selection of BFS candidates from variables with negative cost variation, which has become the most computationally demanding part of the process, if the ill-conditioness begins to appear. Encountering ill-conditionedness across all BFS candidates (all variables with negative reduced cost) would be problematic. In such cases, we would need to proceed with higher precision (extended precision, such as double-double precision). While it would be intriguing to implement such a dynamic precision algorithm in the future, all the numerical results in this article have been produced with machine precision (double precision with 64 bits). Possible improvements to the current solver might further enhance its efficacy. Specifically, incorporating a stabilizer for LU factorization, akin to the stabilizer for Cholesky factorization employed in \textit{SDPB}\cite{Simmons-Duffin:2015qma}, could be highly beneficial. This adjustment could potentially counteract the issue of ill-conditionedness.
 In anticipation of addressing extremely large-scale problems in the future, we recommend the implementation of \textit{CAMPARY} \cite{joldes2016campary}. \textit{CAMPARY} offers CUDA acceleration for extended precision floating point numbers, significantly improving computational efficiency and facilitating the handling of larger and more complex datasets.\footnote{We would like to highlight the history of evolution in the precision used in conformal bootstrap studies. Initial papers, including the seminal study solving the $3d$ Ising problem \cite{El-Showk:2012cjh}, utilized double precision calculations. It wasn't until \cite{Poland:2011ey} that multiple precision was employed for the first time in numerical bootstrap. This high-precision approach became more prevalent after its successful application in the subsequent $3d$ Ising study \cite{El-Showk:2014dwa}.}

\section{Numerical Results}
In this section, we illustrate the efficacy of the methods detailed in previous sections by presenting a series of numerical results. These results revolve around gap maximizations performed under a variety of scenarios. 

The section is structured as follows: We initially describe an exponential decaying cost method that facilitates the expedited execution of gap maximization procedures. Subsequently, we extend the application of the numerical functional bootstrap approach to an assortment of gap maximization problems in both two- and three-dimensional settings. This provides an opportunity to evaluate and compare the convergence rates and benchmarks resulting from our numerical implementations. The parameters of the implementations are summarized in Appendix.\ref{Appendix: parameter}.

\subsection{Gap Maximization and Exponential Decaying Cost}\label{gap}

All the numerical results in this section fall within the realm of gap maximization, the initial application considered in modern conformal bootstrap. In the case of single-correlator bootstrap, the bisection method is a conventional tool used to pinpoint the exact bound. However, the outer approximation framework provides a more efficient alternative, which is directly linked to the correlator minimization discussed in the study \cite{Paulos:2021jxx}\cite{Paulos:2020zxx}. The authors in that study considered both maximization and minimization of the four-point correlator. Their results demonstrated that the minimization of the spin $L$ correlator at some given points (generally $z\ll 1$ and $\bar{z} \ll 1$) reproduced the gap maximization in the corresponding spin channel.

In this paper, we extend this technique to a more general background. This method allows us to obtain the optimal bound in a single run, avoiding the bisection method, which is an order of magnitude more computationally demanding.

For the spin channel $\ell$, we consider the following optimization problem:

    \begin{align}\label{exp}
        \text{Minimize} \quad & \sum_\Delta \epsilon^\Delta a_{\Delta, \ell} \\
        \text{subject to} \quad & \sum_{\Delta, L} a_{\Delta, L} F_{\Delta, L}(z, \bar{z})= 0, \\
        & a_{\Delta, L} \geq 0,\\
    \end{align}
Where the spectrum is above the unitarity bound:
\begin{equation}
    \begin{cases}
        \Delta\geq d-2+L,       & \text{for } L\geq 2\\
    \Delta\geq \frac{d-2}{2},       & \text{for }L=0.\\
    \end{cases}
\end{equation}

Based on our limited numerical tests, the results suggest that for sufficiently small $\epsilon< \epsilon_0$, the optimal solution to this optimization problem is the same as the gap maximization in the respective spin channel $\ell$. 

The rest of this subsection provides a non-rigorous argument to explain this finding. We denote the gap in the corresponding spin channel as $\Delta_{\mathrm{Gap}, \ell}$, and generally, there is a unique solution to saturate the gap. For sufficiently small $\epsilon$, the cost is dominated by the lowest operator in the spin channel $\ell$, with other terms being exponentially suppressed. In essence, we are minimizing $\epsilon^\Delta_{\mathrm{gap}} a_{\Delta_{\mathrm{gap},\ell}}$. This is a combination of gap maximization and OPE minimization at the lowest operator. 

To clarify this concept, we consider the auxiliary OPE minimization problem:

\begin{align}\label{expg}
    \text{Minimize} \quad &  a_{\Delta_g, \ell} \\
    \text{subject to} \quad & \sum_{\Delta, L} a_{\Delta, L} F_{\Delta, L}(z, \bar{z})= 0, \\
    & a_{\Delta, L} \geq 0,\, \Delta_g\leq \Delta_{\mathrm{Gap},\ell} \\
    & a_{\Delta, \ell}= 0\, \text{for } \Delta<\Delta_{\mathrm{Gap},\ell} \text{ and } \Delta\neq \Delta_g 
\end{align}

In this auxiliary bootstrap problem, $\Delta_g$ is the only operator under the spin $\ell$ gap, so it must have a minimal OPE value, otherwise the gap is violated. Furthermore, when $\Delta_g=\Delta_{\mathrm{Gap},\ell}$, the we expect a unique possible value of the gap maximization solution. We denote the optimal value as $a(\Delta_g)$. As a result if we have:

\begin{equation}\label{eq: expbound}
    \log (\epsilon)<-\frac{1}{a(\Delta_{\mathrm{Gap},\ell})} \left. \frac{\partial}{\partial \Delta_{g}} a(\Delta_g)\right|_{\Delta_{g}=\Delta_{\mathrm{Gap},\ell}}\equiv \log(\epsilon_0)
\end{equation} 

The gap maximization is at least a local solution to the bootstrap problem with exponential decaying cost Eq.\ref{exp}. But for our numerical implement after descretization, they are always convex optimization problems. For convex optimization problem, local minimization guarantees the global optimization. This finish our argument that the bootstrap problem Eq.\ref{exp} is equivalent to the gap maximization in the corresponding spin channel. 

The argument presented above fails when the derivative diverges:

\begin{equation}\label{eq: divder}
    \left. \frac{\partial}{\partial \Delta_{g}} a(\Delta_g)\right|_{\Delta_{g}=\Delta_{\mathrm{Gap},\ell}}\rightarrow-\infty
\end{equation} 

We observe that when the external dimension $\Delta_\phi$ approaches zero, the corresponding derivative could be very large. However, there is no obstacle to setting $\epsilon$ to be exponentially small. In our numerical implementation, we set $\epsilon=10^{-50}$. \footnote{We reassure the readers that this doesn't bring us numerical instability, as discussed in Section.\ref{pivot}} 

In a sense, this exponential decaying cost is an automatic navigator\cite{Reehorst:2021ykw} for the numerical bootstrap, and it would be interesting to consider $\epsilon^\mathrm{Navigator}$ in future studies. We are not aware of a way to implement this method in the framework of polynomial approximation (SDPB), as generally, it is not advisable to approximate exponential functions with polynomials.

In closing, it is worth noting a significant aspect of gap minimization. One might be inclined to attempt the minimization of $\epsilon_{\mathrm{big}}^\Delta a_{\Delta, \ell}$, with $\epsilon_{\mathrm{big}}$ representing a significantly large, exponentially growing cost. However, this strategy does not perform as effectively as gap maximization for several reasons. Firstly, gap minimization typically involves additional constraints, such as the premise of only a single relevant operator. Our existing framework struggles to preclude the possible presence of double relevant operators. A potential suggestion might be to apply the exponentially growing cost within a small region suspected to harbor the minimized gap, while disregarding all regions beneath the spacetime dimension. Secondly, and most crucially, the divergence described in Eq.\ref{eq: divder} appears to manifest during gap maximization. In such circumstances, using an excessively large exponential value (for instance, $10^{2000}$) will yield a gap minimization accurate up to the fourth decimal place\footnote{One could also consider other, more assertive functions, for example, by trying to minimize $\exp (\exp(\Delta+10)-\exp(10))a_{\Delta, \ell}$. However, implementing such aggressive proposals( including $10^{2000}$) can be challenging due to the overflow of double-precision floating-point numbers.}.

\subsection{Gap Maximization in Two-Dimensional CFTs}

\begin{figure}[ht]
    \centering
    \includegraphics[width=15cm]{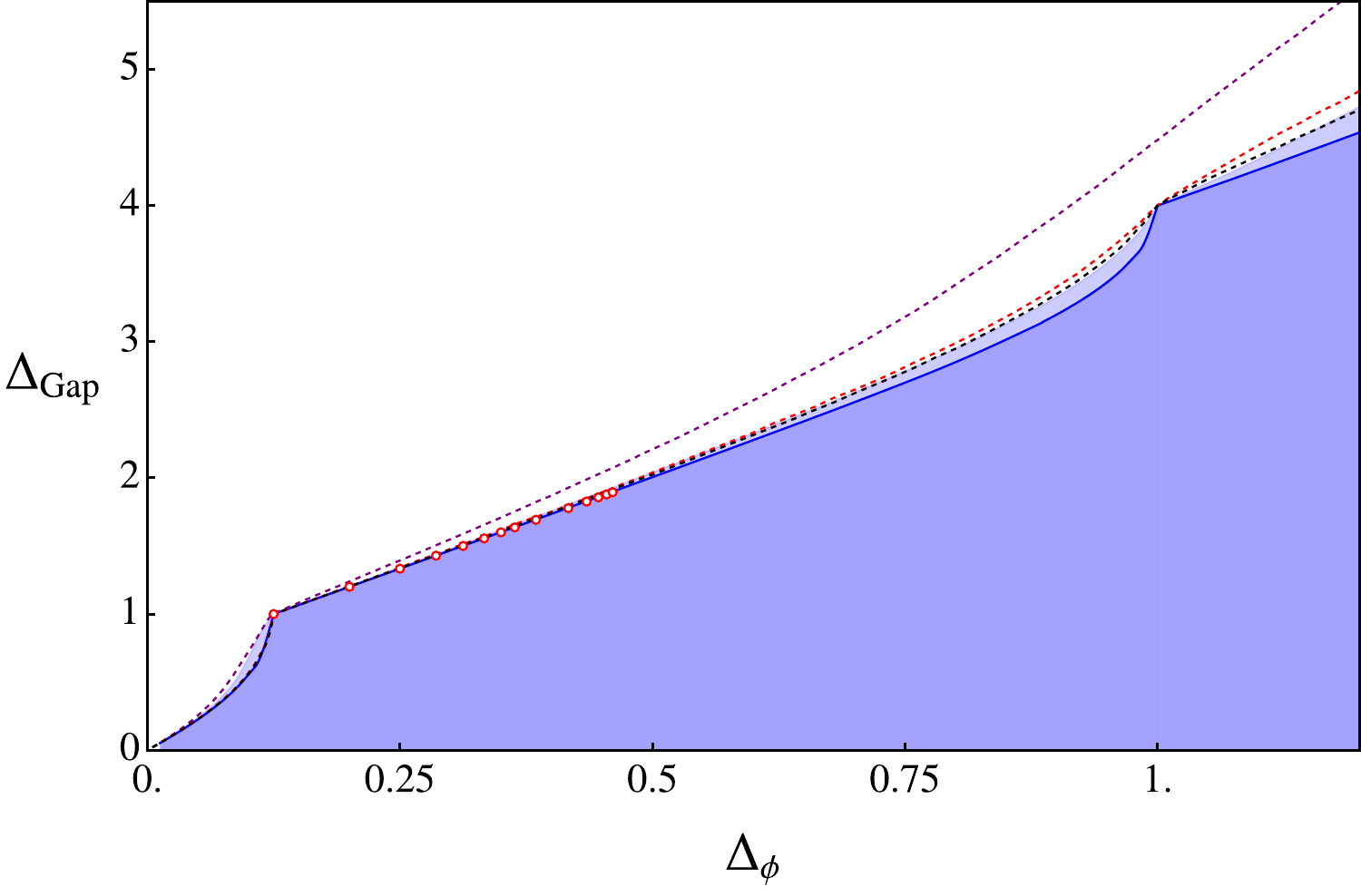}
    \caption{\label{spin0} Spin $\ell=0$ gap maximization. The gap for $60$ $F^+F^-$ functionals is depicted by the darker blue region, while the gap for $12$ $F^+F^-$ functionals is depicted by the lighter blue region. The purple, red, and black dashed lines represent the gap maximization with 15, 91, and 171 derivatives, respectively. The dotted circles indicate the positions of several selected minimal models.}
\end{figure}

\begin{figure}[ht]
	\centering
	\includegraphics[width=12cm]{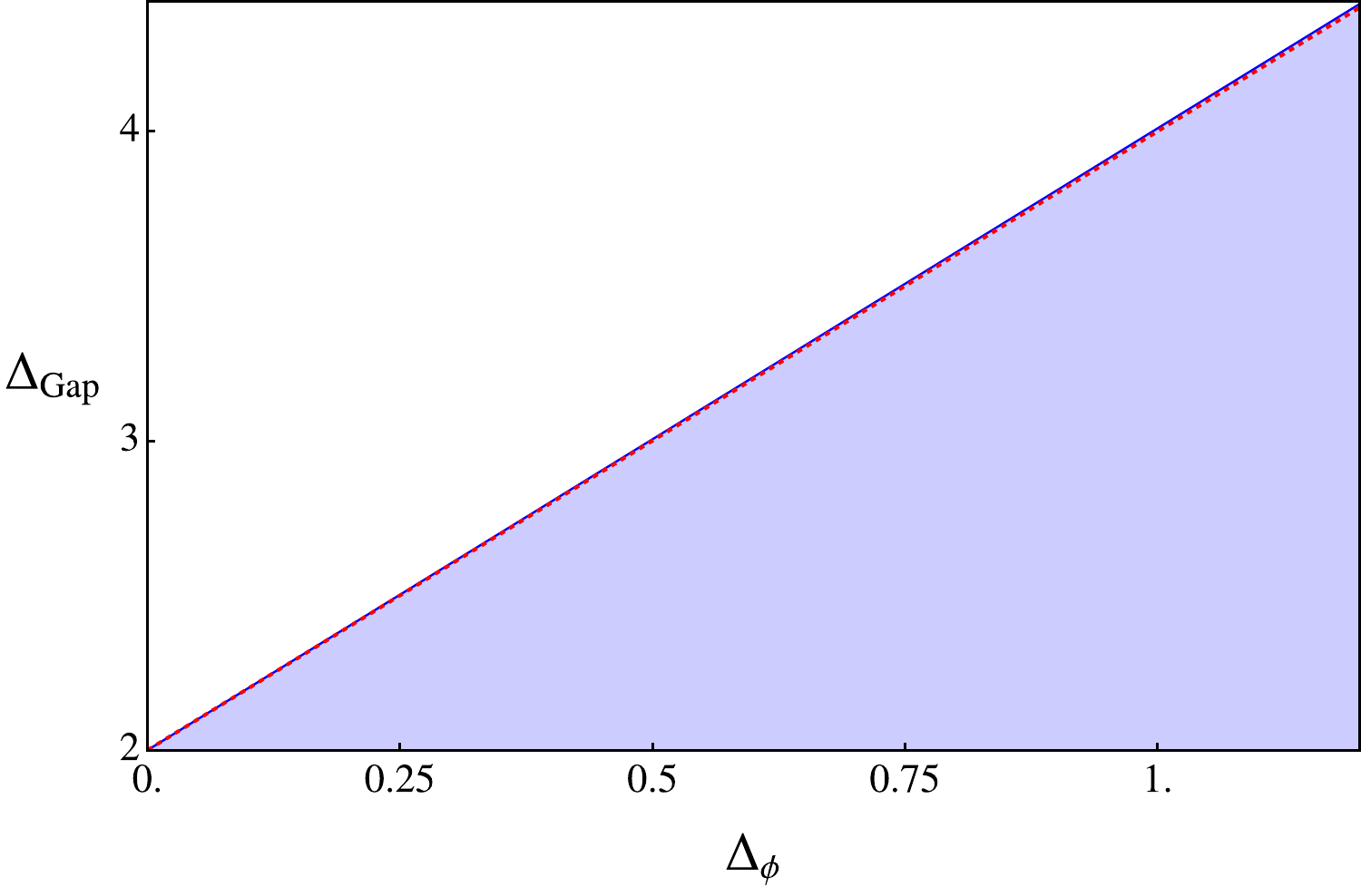}
	\caption{\label{spin2} Spin $\ell=2$ gap maximization, which is expected to reproduce the Generalized free field solution with the corresponding $\Delta_\phi$ (red dashed line on the plot).}
\end{figure}

In this subsection, we present the results of the scalar channel and $L=2$ channel gap maximization, as derived using the functional bootstrap. We juxtapose these results with established exact results at several points, as well as those obtained via derivative bases. Similar bootstrap problems have been studied in \cite{El-Showk:2012vjm, Behan:2017rca, Gowdigere:2018lxz, Nakayama:2019jvm, Li:2021uki,  Paulos:2019gtx}.

A significant result of our investigation is depicted in Fig.\ref{spin0}. Here, we exhibit the scalar channel gap maximization up to $\Delta_\phi=1.2$. A comparison with the derivative method distinctly shows that the functional method's convergence rate significantly outperforms the latter's\footnote{The derivative result is basically from the example of the SDPB package \textbf{Bootstrap2dExample.m}, with the corresponding $\Lambda= 9, 25, 35.$}. 

It is worth noting that the plot displayed in Fig.\ref{spin0} was drawn utilizing the $F^+ F^-$ (FF) functional, as introduced in Section.\ref{Section: criteria}. However, we also conducted a similar experiment using the $F^+ B^-$ (FB) functional. The key insights gained from the results can be succinctly summarized as follows: the FB functional demonstrates faster convergence around the first kink, while the FF functional outpaces it around the second kink\footnote{It is pertinent to highlight that at $\Delta_\phi=1$, the $\alpha^{+F}_0 \otimes \beta^{-F}_0$ is the extremal functional, so this observed behaviour in terms of convergence is fully anticipated.}.

For high enough number of functionals, both categories of functionals - FF and FB - converge to the expected solution. This observation corroborates the completeness of both categories of basis sets.

A widely recognized challenge in this area is the derivative basis' slow convergence when $\Delta_\phi$ deviates substantially from the unitarity bound. This issue is clearly observable in Fig.\ref{spin0}. However, the use of the FF-type functional significantly mitigates this problem. In particular, the second kink becomes notably sharper compared to the results derived using the derivative bases. Given the sluggish convergence rate of the derivative basis, we posit that it is unlikely to achieve acceptable numerical precision within a reasonable allocation of computational resources.

Then we delve into the maximization of the spin 2 gap and its association with the generalized free field (GFF). The GFF is expected to saturate the gap maximization bound, represented as $2\Delta_\phi+l$. Our findings, as presented in Fig.\ref{spin2}, show an almost perfect agreement with this expectation, offering further evidence to support the completeness of our product functional family.

Interestingly, one might anticipate that the "fermionic+bosonic-" (FB) functional would converge more rapidly than the "fermionic+fermionic-" (FF) functional for this problem. This expectation is due to the fact that the FB functional possesses the appropriate asymptotic double zeros, in contrast to the FF functional which displays fourth-order zeros. Nevertheless, our empirical results reveal a different narrative, demonstrating that both categories of functionals perform comparably well, with negligible differences in their rates of convergence.

Lastly, we would like to discuss gap minimization. Existing evidence indicates that even in the case of a single correlator, highly constraining bounds can be obtained\cite{Li:2017kck}. When conducting a conformal bootstrap with a single correlator, a lower bound on the leading correlator is typically derived by applying additional assumptions. For instance, these assumptions could stipulate that there is only one relevant operator, or that a gap exists in the $L=2$ spin channel above the current stress tensor. Among these conditions, implementing the assumption of a singular relevant operator is not directly feasible using the exponential decay cost method. We conducted an investigation using the bisection method for gap minimization in 2-dimensional conformal field theory (2d CFT). The results align with the study conducted by Behan\cite{Behan:2017rca}, which suggests that the line for the minimal model cannot be disentangled into a series of islands for a single correlator.

To aid the reader, we have included in Appendix.\ref{App: spectrum} a list of the spectra of optimal solutions for a selection of $\Delta_\phi$ values. In the ensuing subsection, we showcase data for various $\Delta_\phi$ and $\Lambda$ values to further illustrate the convergence behaviour of our functionals.

\subsection{Comparison of the convergence and efficiency}\label{Sec: convergence}
\begin{table}[h]
\centering
\begin{tabular}{|c|c|c|c|c|c|c|}
\hline
($\Delta_\phi$, Gap channel)\textbf{\textbackslash} \#func & 12 & 24 & 40 & 60 & 84 & 112 \\
\hline
$(1.2,\, \ell=0)$ & 4.72796 & 4.5902 & 4.55348 & 4.53738 & 4.52769 & 4.52055\\
$(1.2,\, \ell=2)$ & 4.51436 & 4.45741 & 4.43341 & 4.42153 & 4.41558 & 4.41161\\
$(0.9,\, \ell=0)$ & 3.3305 & 3.25153 & 3.21836 & 3.1995 & 3.18765 & 3.18104\\
$(0.9,\, \ell=2)$ & 3.87072 & 3.83523 & 3.82041 & 3.81344 & 3.80954 & 3.80703\\
$(0.075,\, \ell=0)$ & 0.40518 & 0.37354 & 0.36761 & 0.36664 & 0.36624 & -\\
$(0.075,\, \ell=2)$ & 2.15409 & 2.15107 & 2.15045 & 2.15023 & 2.15014 & -\\
\hline
\end{tabular}
\caption{A comprehensive tabulation of the FF type functional bounds for an array of parameters. Each row represents a unique configuration characterized by a specific $\Delta_\phi$ and gap channel. The first row enumerates the number of functionals used in each case. This table facilitates an analytical comparison of functional bounds under varying conditions. For $\Delta_\phi=0.075$, we don't get the $112$ functionals result due to the numerical instability.}

\label{table:convergenceFF}
\end{table}

\begin{table}[h]
\centering
\begin{tabular}{|c|c|c|c|c|c|c|}
\hline
($\Delta_\phi$, Gap channel)\textbf{\textbackslash} \#func & 8 & 18 & 32 & 50 & 72 & 98\\
\hline
$(1.2,\, \ell=0)$ & 5.0865 & 4.79528 & 4.66345 & 4.6108 & 4.5778 & 4.55635\\
$(1.2,\, \ell=2)$ & 4.61361 & 4.5307 & 4.46287 & 4.43548 & 4.42435 & 4.4173\\
$(0.9,\, \ell=0)$ & 3.8498 & 3.45051 & 3.28872 & 3.24396 & 3.21153 & 3.19928\\
$(0.9,\, \ell=2)$ & 3.96285 & 3.89317 & 3.84109 & 3.82219 & 3.81484 & 3.8104\\
$(0.075,\, \ell=0)$ & 0.41212 & 0.36948 & 0.36729 & 0.36672 & 0.36618 & -\\
$(0.075,\, \ell=2)$ & 2.16484 & 2.15365 & 2.15089 & 2.15027 & 2.15019 & -\\
\hline
\end{tabular}
\caption{This table is parallel to the previous one, but for the FB type functional. Each row represents a unique configuration characterized by specific $\Delta_\phi$ and gap channel parameters. The first row enumerates the number of functionals used in each case. For $\Delta_\phi=0.075$, we don't get the $98$ functionals result due to the numerical instability.}

\label{table:convergenceFB}
\end{table}

\begin{table}[h]
\centering
\begin{tabular}{|c|c|c|c|}
\hline
($\Delta_\phi$, Gap channel) & 15 & 91 & 171\\
\hline
$(1.2,\, \ell=0)$ & 5.61683 & 4.8433 & 4.7034\\
$(0.9,\, \ell=0)$ & 3.92618 & 3.403122 & 3.35037\\
$(0.075,\, \ell=0)$ & 0.44723 & 0.371575 & 0.368788\\
\hline
\end{tabular}
\caption{This table presents the results of scalar channel gap maximization using a derivative basis, under the same $\Delta_\phi$ conditions as the prior two tables that used product functionals. It provides an opportunity to directly compare the performance of the derivative basis with that of the product functionals in the same scalar channel gap maximization scenarios.}

\label{table:convergenceDer}
\end{table}

In this section, we delve into a meticulous comparison of the convergence behavior exhibited by the functionals and their derivatives. This detailed exploration is facilitated by the selection of three diverse $\Delta_\phi$ values, chosen for their representative character within the region interspersed between the kinks, as depicted in Fig.\ref{spin0}.  It is important to note that for the bosonic functional, the functional $\beta_0$ is identically zero, which implies that at each level there are fewer functionals than for the FF type. This difference is reflected in the presented results. 

Tables \ref{table:convergenceFF} and \ref{table:convergenceFB} display the gap bound at each spin channel for $\Delta_\phi=0.075, 0.9, 1.2$ respectively. A significant point to observe here is the identical zero value for the functional $\beta_0$ under the bosonic functional, which implies fewer functionals at each level compared to the FF type. This discrepancy is mirrored in the results shown.

For a more comprehensive view of the spectrum at selected $\Delta_\phi$, we encourage readers to consult Appendix \ref{App: spectrum}. In Table.\ref{table:spectruml0FB0.125}, we provide a comparison of several leading operators with the $2d$ Ising model. Notably, there is significant agreement, extending even to higher twist operators. For comparison with the same problem using a $60$ derivative basis, readers are directed to Table.\ref{table:derspectruml0FB0.125} in Appendix \ref{App: spectrum}.

We now distill several critical observations concerning the convergence behavior from Tables \ref{table:convergenceFF}, \ref{table:convergenceFB}, and \ref{table:convergenceDer}:

\begin{enumerate}
    \item Both product functionals and the derivative exhibit faster convergence at the two kinks compared to nearby points.
    \item Given an equivalent number of functionals, the product functional basis consistently outperforms the derivative basis. In terms of scalar channel gap maximization, prior to the first kink, the FB basis, employing $18$ functionals, generates bounds nearly comparable to $171$ derivatives. Meanwhile, preceding the second kink, the FF basis with $24$ functionals yields bounds that significantly surpass $171$ derivatives.
    \item It becomes evident that the selection of an optimal functional can be more beneficial than merely increasing the number of functionals in different regions. The rapid convergence of the FF functional around the second kink is primarily attributed to its provision of the extremal functional exactly at the second kink. Consequently, future inclusion of more non-trivial functionals into our basis, such as the master functional presented in \cite{Paulos:2020zxx}, appears promising.
    \item While the FB functional is generally expected to converge quicker than the FF functional for spin $L=2$ gap maximization due to its correct asymptotic double zeros structure, our results do not corroborate this expectation conclusively. The only definite inference is that both functional bases converge to the precise solution as the number of functionals is increased.
    \item It is worth considering the hybrid FF\&FB basis. Though our numerical experience in this direction is limited, some evidence suggests faster convergence at certain $\Delta_\phi$ for the hybrid basis compared to pure bases at the same number of functionals.
    \item We observe that numerical instability issues are notably more severe for low $\Delta_\phi$ (before the first kink) than high $\Delta_\phi$ (after the second kink), although the reasons for this remain elusive.
    \item The task of encapsulating the convergence behavior of both the functional and derivative in explicit formulae does not seem feasible. Such convergence behavior is intimately tied to external dimensions and specific optimization problems at hand. For instance, a general trend is that convergence tends to be slower at high $\Delta_\phi$ values, while it accelerates at the kink. However, one can confidently assert that, irrespective of the particular bootstrap problem being addressed, the analytic functional consistently demonstrates superior convergence behavior as compared to the derivative basis\footnote{For a comprehensive understanding of the convergence behavior in the one-dimensional (1D) scenario, we encourage the reader to refer to the work \cite{Paulos:2019fkw}, which provides an in-depth examination of this topic.}. 
\end{enumerate}

\begin{table}[h]
\begin{equation*}
    \begin{array}{|c|c|c||c|c|c|c|}
\hline \text{Spin} & \Delta & \text {OPE}^2 & \Delta_{\text{num}} & \operatorname{Err}_{\Delta}(\%)& \text {OPE}_{\text{num}}^2 & \operatorname{Err}_{\mathrm{OPE}^2}(\%)     \\
\hline 0  & \,1\,  & \text{ 2.50000000e-1 } & \,1.00000009\, & \text{9.0e-6} & \text{ 2.49999985e-1 } & \text{6.0e-6}\\
          & \,4\,  & \text{ 2.44140625e-4 } & \,4.00001322\, & \text{3.3e-4} & \text{ 2.44137861e-4 } & \text{1.1e-3}\\
          & \,8\,  & \text{ 4.82812729e-8 } & \,7.99525353\, & \text{5.9e-2} & \text{ 4.78906432e-8 } & \text{8.1e-1}\\
\hline 2  & \,2\,  & \text{ 3.12500000e-2 } & \,2.00000000\, & \text{0}      & \text{ 3.12499999e-2 } & \text{3.2e-7}\\
          & \,6\,  & \text{ 6.86644375e-6 } & \,6.00080943\, & \text{1.3e-2} & \text{ 6.80179598e-6 } & \text{9.4e-1}\\
\hline 4  & \,4\,  & \text{ 4.39451562e-4 } & \,4.00000000\, & \text{0}      & \text{ 4.39455331e-4 } & \text{8.6e-4}\\
          & \,5\,  & \text{ 3.05175590e-5 } & \,5.00004158\, & \text{8.3e-4} & \text{ 3.05149040e-5 } & \text{8.7e-3}\\
          & \,8\,  & \text{ 2.12871504e-7 } & \,8.00476230\, & \text{6.0e-2} & \text{ 2.11291737e-7 } & \text{7.4e-1}\\
\hline 6  & \,6\,  & \text{ 1.36239239e-5 } & \,6.00022916\, & \text{3.8e-3} & \text{ 1.36305441e-5 } & \text{4.9e-2}\\
        & \,7\,  & \text{ 1.52586728e-6 } & \,7.00259564\, & \text{3.7e-2} & \text{ 1.51943252e-6 } & \text{4.2e-1}\\
\hline       8  & \,8\,  & \text{ 5.39324266e-7 } & \,8.00012269\, & \text{1.5e-3} & \text{ 5.39297105e-7 } & \text{5.0e-3}\\
\hline
\end{array}
\end{equation*}
\caption{Scalar channel gap maximization with $50$ FB functionals at $\Delta_\phi=0.125$. We compare the spectrum with the exact value from $2d$ critical Ising model. We show all the operators with $\Delta\leq 8$ and we discard the operator from numerical artifacts.}
\label{table:spectruml0FB0.125}
\end{table}

Turning to the computational efficiency of our implementation, all executions (barring the $1d$ functional evaluation) are carried out in double precision. As per the specific pivot rule described in Section \ref{pivot}, the evaluation is nearly instantaneous (under 10 seconds) when numerical stability is upheld. However, in cases of severe numerical instability, the pivot rule attempts to discover a singular new BFS vector that retains numerical stability from a pool of $100k$ candidates. This process could be considerably slower, spanning several orders of magnitude. A potential solution to this issue lies in the implementation of dynamic precision and possibly extensive parallelization facilitated by GPU acceleration, as mentioned in Section \ref{pivot}.

In conclusion, we wish to touch on the possible reasons why the analytic functional basis tends to converge faster than the derivative basis during numerical bootstrap. The most apparent reason is that the product functionals have well-controlled asymptotics, presenting both double and fourth-order zeros, thus approximating the exact extremal functionals more closely than the derivative does. Digging deeper, this could be attributed to the nature of the derivative crossing equation which exhibits exponential convergence, leading to a significant suppression of higher twist operators. Within the analytic functional basis, however, the crossing equation converges according to a power-law, and any exponential factors cancel out between the Operator Product Expansion (OPE) and the functional action. As a result, the contributions from higher twist operators become more pronounced within the equation for product functionals.

\subsection{Exploring the Open Sea of 2D CFT}\label{section: thirdkink}

\begin{table}[h]
\centering
\begin{subtable}{0.49\textwidth}
\centering
\begin{tabular}{|c|c|c|}
\hline
Dimension & Spin & OPE$^2$ \\
\hline
2.0798516 & 2 & 6.283760e+00 \\
4.0000000 & 4 & 1.328040e+00 \\
4.1249310 & 2 & 2.618865e+00 \\
4.5121104 & 4 & 7.135302e-02 \\
5.2704201 & 4 & 1.871181e-01 \\
6.0000000 & 6 & 1.793652e-01 \\
6.1856025 & 0 & 1.382652e+00 \\
6.3547912 & 4 & 3.225483e-01 \\
7.1748664 & 6 & 2.564648e-02 \\
7.3245433 & 0 & 5.008631e-03 \\
\hline
\end{tabular}
\caption{112 functionals}
\label{subtable:cutoff1}
\end{subtable}
\hfill
\begin{subtable}{0.49\textwidth}
\centering
\begin{tabular}{|c|c|c|}
\hline
Dim & Spin & OPE$^2$ \\
\hline
2.0846070 &   2 & 6.212455e+00 \\
4.0145347 &   4 & 1.346949e+00 \\
4.1258509 &   2 & 2.586199e+00 \\
5.2541847 &   4 & 2.240048e-01 \\
6.0000000 &   6 & 1.741817e-01 \\
6.1607411 &   0 & 1.379108e+00 \\
6.3828673 &   4 & 3.073846e-01 \\
7.0562103 &   0 & 1.971739e-02 \\
7.1805061 &   6 & 2.783298e-02 \\
8.0037135 &   8 & 1.845400e-02 \\
\hline
\end{tabular}
\caption{144 functionals}
\end{subtable}
\caption{Scalar channel gap maximization with FF functionals at $\Delta_\phi=1.5875$. This table highlights the leading terms in the spectrum.}
\label{table:spectrumlkink}
\end{table}

The previous two subsections were dedicated to substantiating the efficacy of our method against established results. In contrast, this section is aimed at boldly venturing into the uncharted realm of 2D conformal field theories (CFTs) — specifically, into regions where $\Delta_\phi$ significantly deviates from the unitary bound.

\begin{figure}[ht]
\centering
\includegraphics[width=15cm]{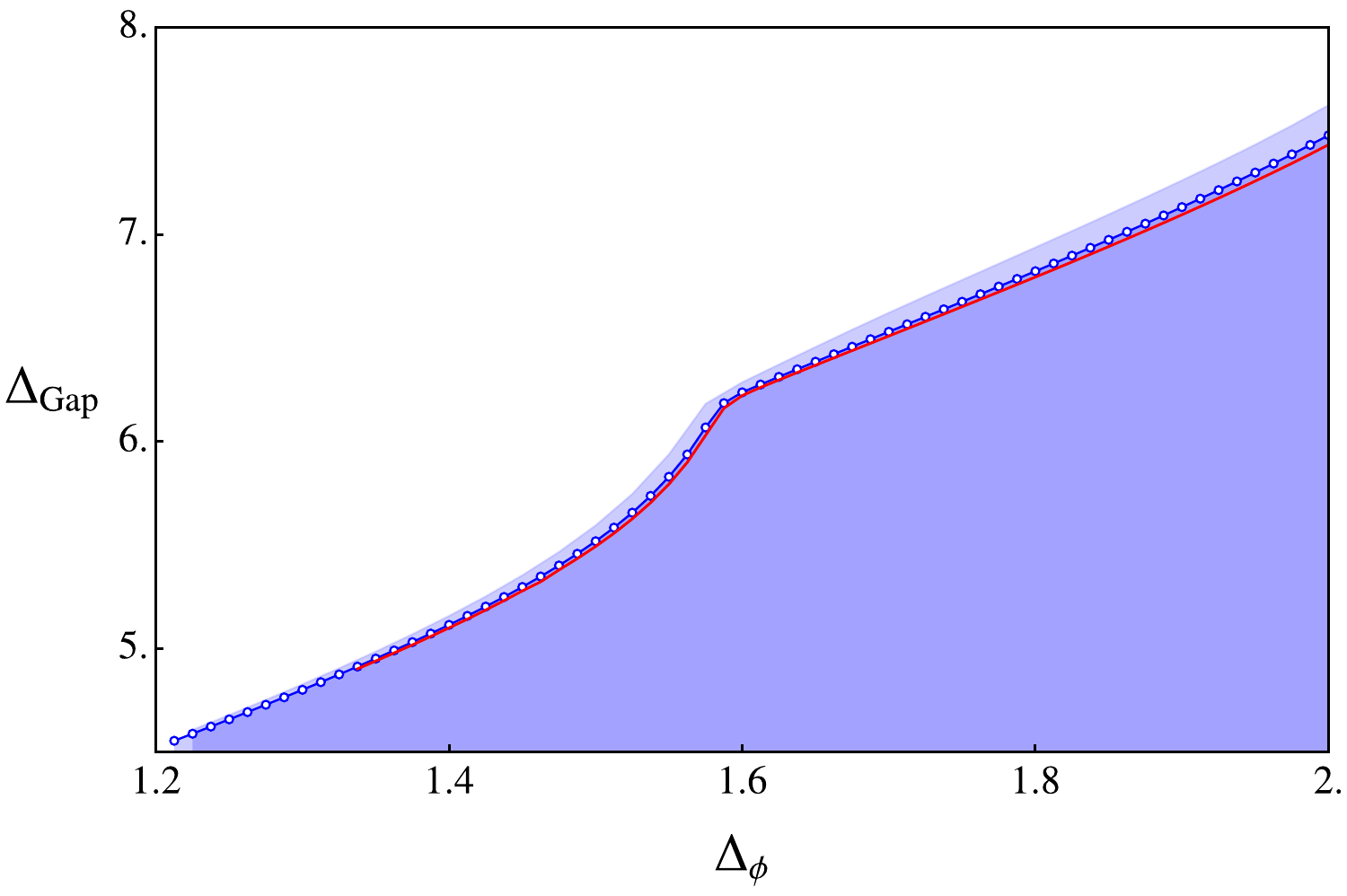}
\caption{Gap maximization for spin $\ell=0$, with $\Delta_\phi$ ranging from $1.2$ to $2$. The gap for $60$ $F^+F^-$ functionals is represented by the lighter blue region, while the darker blue region (marked with markers) represents the gap for $112$ $F^+F^-$ functionals. Note that this figure differs from Fig.\ref{fig: 2dplot} in that the darker blue region, corresponding to $60$ functionals in the previous figure, now corresponds to the lighter blue region. Finally, the red line represents the bound from $144$ $F^+F^-$ functionals, for a reference of the speed of convergence.}
\label{fig: thirdkink}
\end{figure}

Fig.\ref{fig: thirdkink} illustrates the scalar channel upper bound for $\Delta_\phi$ ranging from $1.2$ to $2$. A notable feature is a pronounced kink following the first two kinks at $\Delta_\phi=1$, which correlates with the 2D critical Ising model. The kink locates approximately at $(1.59, 6.15)$. The nature of this kink is, as of now, unclear. The most we can infer is that it signifies a strongly coupled and non-perturbative infrared (IR) fixed point.

\begin{figure}[ht]
\centering
\includegraphics[width=15cm]{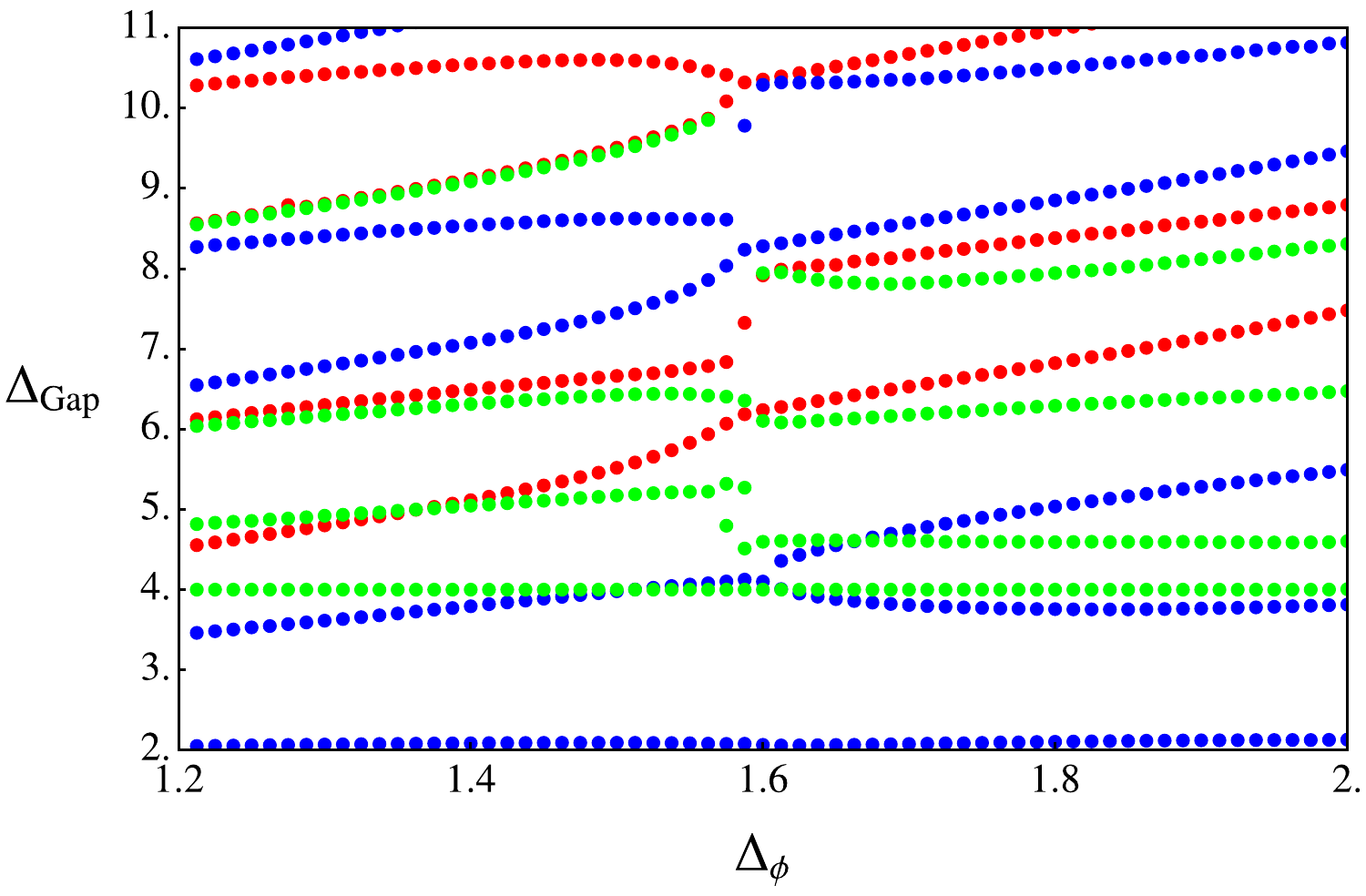}
\caption{The spectrum corresponding to the $112$ FF functional upper bound as depicted in Fig.\ref{fig: thirdkink}. The red dots denote scalars, the blue dots represent $L=2$ tensors, and the green dots signify $L=4$ tensors.}
\label{fig: Spectrum}
\end{figure}

Fig.\ref{fig: Spectrum} showcases the spectrum of the upper bound theory, harkening back to the spectrum across each well-established kink. Clear annihilation and creation of new operators across this kink are observed. An intriguing detail is the leading $L=2$ operator, which does not strictly adhere to the unitary bound, but appears slightly above in the range of $2.05$ to $2.10$ throughout the plot. We can compute the central charge $c$ of this CFT, suppose this is a numerical artifact and the leading $L=2$ tensor is a stress tensor. From \cite{El-Showk:2012cjh, Petkou:1994ad}, for the central charge of $2d$ CFT, we have the following relationship:
\begin{equation}
    c=\frac{\Delta_\sigma^2}{a_{\sigma\sigma T}}
\end{equation}

From the data from Table.\ref{table:spectrumlkink}, we can calculate $c\approx 0.41$. This is a very low value, beyond the unitarity condition of the minimal model. We come to the conclusion that if this kink corresponds to an interesting theory, it must not be a local theory(i.e. without stress tensor and Virasoro algebra). 

We have extended the analytic functional bootstrap beyond $\Delta_\phi=2$. Although the convergence rate diminishes significantly beyond this point, we have successfully pinpointed a distinct kink within the interval $\Delta_\phi=2$ to $\Delta_\phi=3$, specifically at $(2.14, 8.15)$. Furthermore, our analysis reveals slight upward shifts in the upper bounds, indicating the potential presence of an additional kink. However, our present methodology, utilizing 144 functionals, is insufficient to confirm this hypothesis conclusively. For a comprehensive discussion of this aspect, we direct the reader to Appendix.\ref{Appendix: beyond}. For a full plot containing the profile of all the kinks, the reader is invited to refer to Fig.\ref{fig: spin0full} in the Introduction Section.\ref{sec: intro}.

At this point\footnote{We would like to acknowledge Dalimil Mazac for bringing to our attention the observation \cite{Collier:2016cls}.} we would like to mention that while we still lack a comprehensive understanding of the third and fourth kinks and their associated theory, we have identified a captivating link between our plot and a comparable plot discovered in spinning modular bootstrap \cite{Collier:2016cls}. This resemblance can be justified by the observation from the study \cite{Hartman:2019pcd}. Here we can map the torus partition function of 2d CFT to a four-point function of twist operator of $\mathbb{Z}_2$ symmetric product orbifold of the CFT. If the CFT on torus has central charge $c$ then the symmetric product orbifold CFT contains a twist operator $\sigma_2$ of the dimension $\frac{c}{8}$. Also, an  operator with conformal dimension $(\frac{h}{2},\frac{\bar{h}}{2})$ gets mapped to a operator with dimension $(h,\bar{h})$. So the dimension of the first scalar operator after identity with dimension $\Delta_{\text{mod}}$ will  correspond to an operator with dimension $2\Delta_{\text{mod}}$ which appears in the OPE of $\sigma_2$. Further the conformal blocks of product orbifold CFT can be expanded in terms of two dimensional global conformal blocks. The identity block of product orbifold CFT contains $SL(2,R)\times SL(2,R)$ primaries with dimension $0,4,8,..$. If we find a functional which have zeroes on these locations then we see that the gap can be optimal and it will be same as what was observed in \cite{Collier:2016cls}. Following this argument, our first two kinks get mapped exactly to $c=1$ and $c=8$ kink from the modular bootstrap. These are spin and energy correlators of the 2D ising model. Therefore the extremal functional has zeroes at twist $4n$,\,$n\in \mathbb{Z}_{>0}$. Therefore the gaps are optimal. It would be interesting to compare other minimal models located in this range of dimension. Regarding the third and fourth kinks, they are closely positioned with a distinction that could be attributed to numerical error.

\subsection{Three-dimensional CFT gap maximization}

\begin{figure}[ht]
    \centering
    \begin{subfigure}{0.495\textwidth}
        \centering
        \includegraphics[width=\textwidth]{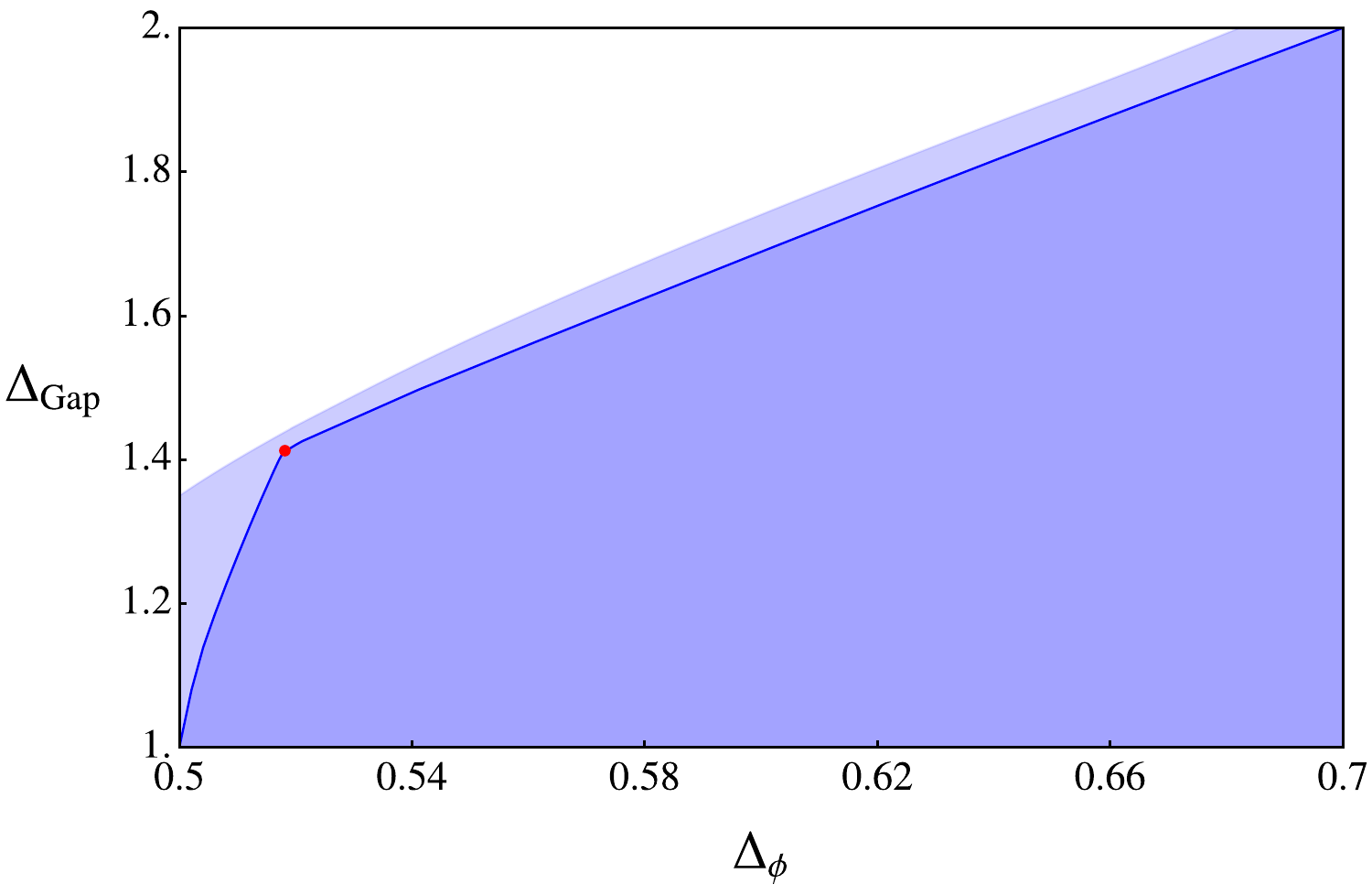}
        \caption{Scalar channel.}
        
    \end{subfigure}
    \hfill
    \begin{subfigure}{0.495\textwidth}
        \centering
    \includegraphics[width=\textwidth]{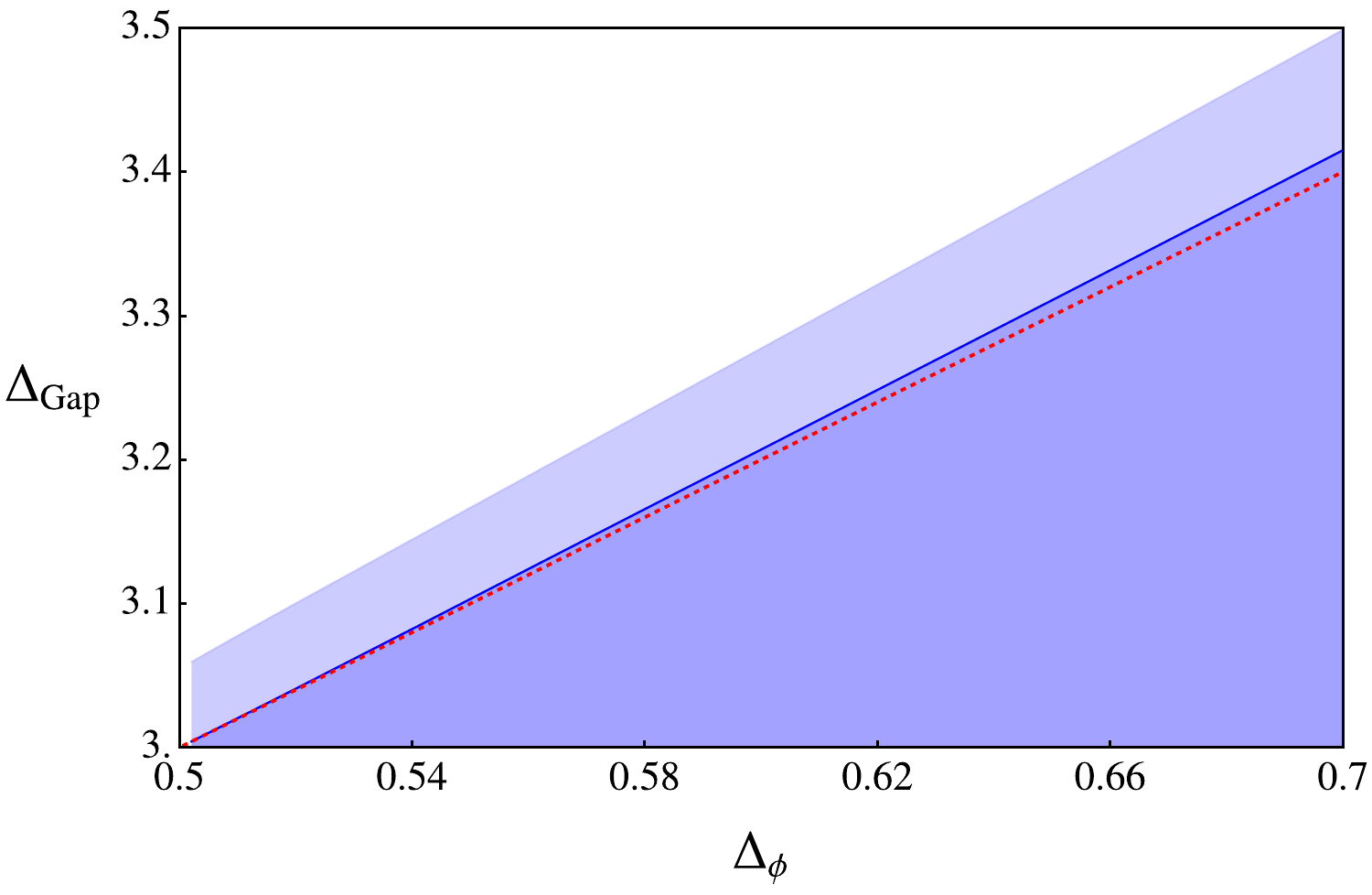}
        \caption{$L=2$ channel.}
        
    \end{subfigure}
    \caption{Gap maximization for $3d$ CFT by FB functional. The lighter blue region signifies the allowed space using 8 functionals, while the darker blue region illustrates the allowed space when employing 18 functionals. The left plot features the $3d$ Ising model, depicted as a red dot at the coordinates $(0.51815, 1.4126)$. In the right plot, the red dashed line represents the GFF for the corresponding $\Delta_\phi$.}
    \label{fig: 3d}
\end{figure}

In this subsection, we present the results obtained from the gap maximization using FB functional for three-dimensional conformal field theory (3D CFT). 

Figure \ref{fig: 3d} exhibits a rapid convergence of the bound, comparable to the numerical results in two dimensions. The maximization bounds for the scalar channel and the $L=2$ channel have become particularly pertinent in our analysis. These relevant visualizations first emerged in the landmark study which resolved the $3d$ Ising model\cite{El-Showk:2012cjh}. When we apply eight functionals, we observe that we are some distance away from the optimal bound. However, using 18 functionals, the results already align with the optimal value up to the third and fourth decimal places, and the profile almost coincides with the corresponding plot in \cite{El-Showk:2012cjh}.

Nevertheless, the numerical bootstrap in three dimensions is considerably hampered due to the issue discussed in Section \ref{Sec: slow}. Specifically, the evaluation of 3D product functional at high $\Delta$ is generally a challenging task. To incorporate a higher number of functionals, it becomes necessary to include higher spins, which inevitably leads to high $\Delta$ evaluation. Here for $18$ functionals, as shown in Appendix.\ref{Appendix: parameter}, we only incorporate spin up to $20$. We recognize this as a numerical challenge and defer its resolution to future studies.

Despite the challenges, it is noteworthy that the FF functional can still be employed to carry out numerical bootstrap in three dimensions. Our observations indicate that the convergence rate of the FF-type functional does not significantly trail that of the FB-type functional. However, the FF-type functional exhibits a higher degree of numerical instability. This instability could be attributed to the uncontrolled asymptotic behavior inherent in the FF-type functional.

\section{Bootstrapping: more general application}
In this section, we explain how our framework can be extended to a more general setup easily. We focus on explaining the framework we have developed while deferring the numerical exploration to future research.

\subsection{Global symmetry}

 Suppose we consider the scalar field is charged under some global symmetry and we denote it as $\phi_a$, such that 
\begin{equation}
    \phi_a \times \phi_b\sim \sum_r O_r,
\end{equation}
where $r$ denotes the number of irreps appearing in the tensor product of the scalar fields transforming in some representation $r$. For concreteness imagine the scalar field is a fundamental field of $O(N)$. In this case $r$ runs over the singlets, traceless symmetric and antisymmetric operators under $O(N)$. Analyzing the crossing equation of four-point functions of $\phi_a$ the bootstrap equations can be written down as,

 \begin{equation}\label{crosssonff}
 \begin{small}
     \sum_{\Delta} C^S_{\Delta,\ell} \begin{pmatrix}
0\\
F_{\Delta,\ell}(z, \bar{z}) \\
H_{\Delta,\ell}(z, \bar{z})
\end{pmatrix} + \sum_{\Delta} C^T_{\Delta,\ell} \begin{pmatrix}
F_{\Delta,\ell}(z, \bar{z})\\
(1-\frac{2}{N})F_{\Delta,\ell}(z, \bar{z}) \\
-(1+\frac{2}{N})H_{\Delta,\ell}(z, \bar{z})
\end{pmatrix} +\sum_{\Delta} C^A_{\Delta,\ell} \begin{pmatrix}
-F_{\Delta,\ell}(z, \bar{z})\\
F_{\Delta,\ell}(z, \bar{z}) \\
-H_{\Delta,\ell}(z, \bar{z})
\end{pmatrix}=0,
 \end{small}
\end{equation}
where we get (anti) symmetric combination of direct and crossed channel conformal blocks,
\begin{equation}
\begin{split}
   & F_{\Delta,\ell}(z,\bar{z})=(z \bar {z})^{-\Delta\phi} G_{\Delta,\ell}(z,\bar{z})-\big(z\rightarrow(1-z),\bar{z}\rightarrow(1-\bar{z})\big)\\
    &   H_{\Delta,\ell}(z,\bar{z})=(z \bar {z})^{-\Delta\phi} G_{\Delta,\ell}(z,\bar{z})+\big(z\rightarrow(1-z),\bar{z}\rightarrow(1-\bar{z})\big)
    \end{split}
\end{equation}

We have already described how the functionals act on $F_{\Delta,\ell}(z,\bar{z})$. Now we show how it acts on $H_{\Delta,\ell}(z,\bar{z})$ following \cite{Kaviraj:2021cvq}. $H^{d=2}_{\Delta,\ell}(z,\bar{z})$ can be expressed in terms of one dimensional crossing and anti crossing symmetric vectors as follows,
\begin{equation}
    H^{d=2}_{\Delta,\ell}(z,\bar{z})=\frac{1}{4} (F_{\tau}(z|\Delta_\phi)F_{\rho}(\bar{z}|\Delta_\phi)+H_{\rho}(z|\Delta_\phi)H_{\tau}(\bar{z}|\Delta_\phi)+(z\leftrightarrow\bar{z}))
\end{equation}

Now we can either symmetrize or antisymmetrize $H_{\Delta,\ell}(z,\bar{z})$ in $\bar{z}$ and $1-\bar{z}$ and act with product functionals for antisymmetric contribution,

\begin{equation}
    (\omega^{-} \otimes \omega^-)^{d=2} (\Delta,\ell| \Delta_{\phi})=\frac{1}{2} \Big(\omega^-(\tau|\Delta_{\phi})\omega^-(\rho|\Delta_{\phi})+\omega^-(\rho|\Delta_{\phi})\omega^-(\tau|\Delta_{\phi})\Big),
\end{equation}
$``-"$ should be replaced with $``+"$ type functionals for the symmetric combination of $H_{\Delta,\ell}(z,\bar{z})$ in $\bar{z}$ and $(1-\bar{z})$. So we get

\begin{equation} \label{2dHexp}
\begin{split} 
    & H^{d=2}_{\Delta,\ell}(z,\bar{z})=H^-_{\Delta,\ell}(z,\bar{z})+H^+_{\Delta,\ell}(z,\bar{z}),\\
 & H^-_{\Delta,\ell}(z,\bar{z}) =  \sum_{m,n} \bigg[ \bigg(\alpha^-_m(\tau|\Delta_{\phi})\alpha^-_n(\rho|\Delta_{\phi}) +\alpha^-_m(\rho|\Delta_{\phi})\alpha^-_n(\tau|\Delta_{\phi})\bigg) F_{\Delta_m}F_{\Delta_n}\\
   & +\bigg(\beta^-_m(\tau|\Delta_{\phi})\alpha^-_n(\rho|\Delta_{\phi}) +\beta^-_m(\rho|\Delta_{\phi})\alpha^-_n(\tau|\Delta_{\phi})\bigg) \partial F_{\Delta_m}F_{\Delta_n} \\
    &  +\bigg(\alpha^-_m(\tau|\Delta_{\phi})\beta^-_n(\rho|\Delta_{\phi}) +\alpha^-_m(\rho|\Delta_{\phi})\beta^-_n(\tau|\Delta_{\phi})\bigg) F_{\Delta_m}\partial F_{\Delta_n}\\
    & \bigg(\beta^-_m(\tau|\Delta_{\phi})\beta^-_n(\rho|\Delta_{\phi}) +\beta^-_m(\rho|\Delta_{\phi})\beta^-_n(\tau|\Delta_{\phi})\bigg) \partial F_{\Delta_m}
    \partial F_{\Delta_n}
    \bigg]
    \end{split}
\end{equation}

So the sum rule involving $H_{\Delta,\ell}(z,\bar{z})$ would take the following form,
\begin{equation}
    \sum_{\Delta,\ell} C^S_{\Delta,\ell} (\omega^{-} \otimes \omega^-)^{d=2} (\Delta,\ell| \Delta_{\phi})+ \sum_{\Delta,\ell} C^T_{\Delta,\ell} (\omega^{-} \otimes \omega^-)^{d=2} (\Delta,\ell| \Delta_{\phi})+ \sum_{\Delta,\ell} C^A_{\Delta,\ell} (\omega^{-} \otimes \omega^-)^{d=2} (\Delta,\ell| \Delta_{\phi})=0,
\end{equation}
where $\omega$ can be $\alpha$,$\beta$. We can also use the "+" type functionals and in that case $F_{\Delta}$ would be replaced with $H_{\Delta}$ in \eqref{2dHexp}. Using \eqref{2dFexp} we also get the sum rules arising from equations involving $F_{\Delta,\ell}(z,\bar{z})$. This way we can put bounds using our functionals.

\subsection{Mixed Correlators} \label{mixedcor}
In this section, we describe briefly the more general case involving four scalar operators with dimension $\Delta_i$ \cite{Kos:2014bka}. Therefore we have
\begin{equation}
    \langle \phi_i(x_1)\phi_j(x_2)\phi_k(x_3)\phi_l(x_4)\rangle =\frac{1}{x_{12}^{\frac{\Delta_i+\Delta_j}{2}} x_{34}^\frac{\Delta_k+\Delta_l}{2}} \big(\frac{x_{23}}{x_{14}}\big)^{\Delta_{ij}}\big(\frac{x_{14}}{x_{13}}\big)^{\Delta_{kl}} G_{ijkl}(z,\bar{z}),
\end{equation}
where $\Delta_{ij}=\Delta_i-\Delta_j$.
We can expand $G_{ijkl}(u,v)$ in conformal blocks,
\begin{equation}
    G_{ijkl}(z,\bar{z}) =\sum_{\mathcal{O}}\lambda_{i j O}\lambda_{k l O} G^{\Delta_{ij},\Delta_{kl}}_{\Delta,\ell}(z,\bar{z}).
\end{equation}
The invariance of correlator under the exchange of $(1,i)\leftrightarrow(3,k)$ gives rise to the crossing equation,
\begin{equation}
    \bigg((1-z)(1-\bar{z})\bigg)^{\frac{\Delta_j+\Delta_k} {2}}G_{ijkl}(z,\bar{z}) =(z\bar{z})^{\frac{\Delta_i+\Delta_j}{2}} G_{kjil}(1-z,1-\bar{z}).
\end{equation}

By expanding it in terms of conformal blocks and using the vectors structurally similar to (anti) crossing vectors of single correlator  the crossing equation takes the following form,
\begin{equation}
\begin{split}
   &  \sum\lambda_{ij O}\lambda_{kl O}  F^{ij,kl}_{\Delta,\ell}(z,\bar{z})+\lambda_{kjO}\lambda_{il O} F^{kj,il}_{\Delta,\ell}(z,\bar{z})=0,\\
    & 
     \sum\lambda_{ij O}\lambda_{kl O}  H^{ij,kl}_{\Delta,\ell}(z,\bar{z})-\lambda_{kjO}\lambda_{il O} H^{kj,il}_{\Delta,\ell}(z,\bar{z})=0,
    \end{split}
\end{equation}

where 
\begin{equation}
    \begin{split}
        & F^{ij,kl}_{\Delta,\ell}(z,\bar{z})= ((1-z)(1-\bar{z}))^{\frac{\Delta_k+\Delta_j}{2}} G^{\Delta_{ij},\Delta_{kl}}_{\Delta,\ell}(z,\bar{z})-(z\bar{z})^{\frac{\Delta_k+\Delta_j}{2}} G^{\Delta_{ij},\Delta_{kl}}_{\Delta,\ell}(1-z,1-\bar{z})\\
        & H^{ij,kl}_{\Delta,\ell}(z,\bar{z})= ((1-z)(1-\bar{z}))^{\frac{\Delta_k+\Delta_j}{2}} G^{\Delta_{ij},\Delta_{kl}}_{\Delta,\ell}(z,\bar{z})+(z\bar{z})^{\frac{\Delta_k+\Delta_j}{2}} G^{\Delta_{ij},\Delta_{kl}}_{\Delta,\ell}(1-z,1-\bar{z})
    \end{split}
\end{equation}
In two spacetime dimensions, we can again express the crossing vector in terms of 1d (anti) crossing vectors exploiting the factorized algebra structure of global 2d CFT ,
    \begin{equation}
    F^{d=2|ij,kl}_{\Delta,\ell}(z,\bar{z})=\frac{1}{4} (F^{ij,kl}_{\tau}(z)H^{ij,kl}_{\rho}(\bar{z})+F^{ij,kl}_{\rho}(z)H^{ij,kl}_{\tau}(\bar{z})+(z\leftrightarrow\bar{z})),
\end{equation}
where,
\begin{equation}
\begin{split}
    F^{ij,kl}_{\tau}(z)&=  F^{d=1|ij,kl}_{\tau}(z)|_{\Delta_i/2,\Delta_j/2,\Delta_k/2,\Delta_l/2}\\
   & =(1-z)^{\frac{\Delta_j+\Delta_k}{4}}G^{d=1|\Delta_{ij},\Delta_{kl}}_{\Delta}(z)-z^{\frac{\Delta_j+\Delta_k}{4}}G^{d=1|\Delta_{ij},\Delta_{kl}}_{\Delta}(1-z)
   \end{split}
\end{equation} 
 $H^{ij,kl}_{\tau}(z)$ is given by a similar equation with $``-"$ being replaced with $``+"$. We can also express $H^{ij,kl}_{\Delta,\ell}(z,\bar{z})$ in terms of one-dimensional mixed  correlator (anti) crossing symmetric vectors. The key takeaway from this section is that our framework can be extended to this general scenario if we possess knowledge of the functional actions on the one-dimensional counterparts of such (anti) crossing symmetric vectors. Recently, the Polyakov bootstrap was generalized for a mixed correlator setup in one dimension \cite{Ghosh:2023lwe}. It is feasible to determine the necessary functionals described above, and we anticipate reporting our progress in a separate work.

\section{Conclusions and Outlook}

In this work, we have presented a unique numerical framework for the bootstrap of Conformal Field Theories (CFTs) applicable in dimensions greater than one ($d>1$). This methodology capitalizes on the use of product functionals that act on the crossing vectors as the products of $1d$ functionals. We replaced the traditional approximation by polynomials with the implementation of the outer approximation method in order to conduct the conformal bootstrap using product functionals. Demonstrating remarkable numerical stability, this implementation extends beyond the results of $2d$ conformal bootstrap computed with double precision. Our findings indicate that the product functional basis demonstrates a convergence rate significantly superior to that of the conventional derivative basis. This expedited convergence has unveiled additional kinks within the scalar channel gap maximization bound.

In conclusion, let's direct our attention to several remaining challenges.

\begin{enumerate}
    \item Technical Challenges in 3D: Even though our results in three dimensions show promise, we've met with certain technical hindrances that require further consideration. Addressing these difficulties, such as optimizing the convergence rate and overcoming technical constraints, would substantially increase the effectiveness and robustness of our framework in three-dimensional CFTs.
    
    \item  Mixed correlator: We are eager to broaden our framework's scope to accommodate mixed correlators, as suggested in Section \ref{mixedcor}. Our functionals have demonstrated superior performance in one- and two-dimensional spacetime; hence, we are specifically interested in examining our approach's viability when tackling the $\mathrm{QED}_3$\cite{Albayrak:2021xtd} and long-range Ising model\cite{Behan:2018hfx}. In fact, as noted in these studies, the feasible or "island" region of the allowable parameter space continues to evolve, even after substantial derivative use. Therefore, exploring the benefits of our basis, where saturation to an optimal value occurs more rapidly, will be highly advantageous.  

    \item New kink and potential CFT candidate? We have identified new kinks in 2D CFT that represent compelling candidates for non-local CFTs. It would be intriguing to observe if we encounter any islands in a multi-correlator setup surrounding these kinks. 

    \item Extension to Higher Dimensions: Despite considerable advancement in numerically bootstrapping CFTs in two and three dimensions, transitioning our framework to operate in higher dimensions is an unexplored territory.

    \item Further examination into the hybrid category of functionals should be conducted, of which FB+FF presents an immediate example. Our preliminary explorations in this regard have yielded promising indications, suggesting a potential beneficial impact from such mixed approaches. This category could also be expanded to include the $B_v$ functionals as specified in \cite{Caron-Huot:2020adz}, and potentially even the derivative basis. 

\end{enumerate}

\section*{Acknowledgement}
We are grateful for discussions with Ant\'{o}nio Antunes, Connor Behan, Minjae Cho, Rajeev Erramilli, Yinchen He, Apratim Kaviraj, Wenliang Li, Dalimil Maz\'{a}\v{c}, Jiaxin Qiao, Junchen Rong, Miguel Paulos, Slava Rychkov, Ning Su, Pedro Vieira, and Yuan Xin. KG is thankful to Miguel Paulos for various suggestions and collaboration on related topics. Z. Zheng thanks the organizers of bootstrap 2023 meeting in  
 ICTP-SAIFR, S\~{a}o Paulo and PiTP 2023 in the 
 IAS, Princeton for providing a very stimulating environment while this work was being completed and presented. KG is supported by ANR Tremplin-ERC project FunBooTS.

\appendix

\section{Numerical Evaluation of the one-dimensional analytic functionals} \label{funcwittenrelation}
In this section, we describe the relation between functional actions and the Witten diagrams in $AdS_2$ \cite{Zhou:2018sfz,Ghosh:2021ruh}. We will utilize these relations to efficiently implement them for our numerical exploration. In our proposed method outlined in this Appendix, the most computationally demanding component is the direct evaluation of the generalized hypergeometric function, ${}_7F_6$, along with its derivative.
 The functional actions are always given by the (improved) sum of exchange Witten diagrams. Let us write down the explicit relation between the Witten diagrams and functional actions below explicitly,
\begin{equation} \label{relnfuncwitten}
\begin{split}
    & W_{\Delta,0}^{s,B}(z)+W_{\Delta,0}^{t,B}(z)+W_{\Delta,0}^{u,B}(z)+\lambda C_B(z) =\sum_{n} (\alpha^{-,B}_n(\Delta)  G^{d=1}_{\Delta^B_n}(z|\Delta_\phi)+\beta^{-,B}_n(\Delta)  \partial G^{d=1}_{\Delta^B_n}(z|\Delta_\phi)),\\
    & W_{\Delta,0}^{s,F}(z)+W_{\Delta,0}^{t,F}(z)+W_{\Delta,0}^{u,F}(z) =\sum_{n} (\alpha^{-,F}_n(\Delta)  G^{d=1}_{\Delta^F_n}(z|\Delta_\phi)+\beta^{-,F}_n(\Delta)  \partial G^{d=1}_{\Delta^F_n}(z|\Delta_\phi)),\\
    &
W_{\Delta,1}^{s,F}(z)+W_{\Delta,1}^{t,F}(z)+W_{\Delta,1}^{u,F}(z)+\lambda C_F(z) =\sum_{n} (\alpha^{+,B}_n(\Delta)  G^{d=1}_{\Delta^B_n}(z|\Delta_\phi)+\beta^{+,B}_n(\Delta)  \partial G^{d=1}_{\Delta^B_n}(z|\Delta_\phi)),\\
    &
W_{\Delta,1}^{s,B}(z)+W_{\Delta,1}^{t,B}(z)+W_{\Delta,1}^{u,B}(z) =\sum_{n} (\alpha^{+,F}_n(\Delta)  G^{d=1}_{\Delta^F_n}(z|\Delta_\phi)+\beta^{+,F}_n(\Delta)  \partial G^{d=1}_{\Delta^F_n}(z|\Delta_\phi)),
    \end{split}
\end{equation}
where $W^{i,B}_{\Delta,\ell}$ and $W^{i,F}_{\Delta,\ell}$  refers to exchange Witten diagram of dimension $\Delta$ and spin $\ell$ in the $i$-channel with bosonic and fermionic external legs respectively and $i$ could be $s,t,u$ any of these channels. We also note there are certain contact diagrams $C_B(z)$ or $C_F(z)$ in the above expansion and these contact diagrams are crucial to get a valid non-perturbative sum rule\footnote{Without the contact term, the functionals defined by the coefficients don't satisfy the swapability condition.}. The undetermined coupling $\lambda$ appearing in front of the contact diagrams can be fixed by demanding the Regge growth of the full sum is improved. As a result, we will lose one of the functionals ($\beta^{-,B}_0$,$\beta^{+,B}_0$ in our case) from the sum in the right-hand side of the above identities. For numerical implementation one of the key ingredients is to get these functional in the most efficient way. First of all, we note that fermionic exchange Witten Diagrams can be obtained from the bosonic exchange Witten diagrams \cite{Faller:2017hyt, Kaviraj:2021cvq},

\begin{equation}
\begin{split}
  W^{s,F}_{\Delta,0}(z)=W^{s,B}_{\Delta,0}(z)|_{\Delta_\phi\rightarrow\Delta_{\phi}+1/2}  \\
   W^{s,F}_{\Delta,1}(z)=W^{s,B}_{\Delta,1}(z)|_{\Delta_\phi\rightarrow\Delta_{\phi}-1/2}.
  \end{split}
\end{equation}

Therefore we just need the conformal block (CB) decomposition of Witten diagrams with bosonic legs. The exchange Witten diagram has the following CB expansion,
\begin{equation} \label{cbexpwitten}
\begin{split}
   &  W^s_{\Delta,\ell}(z)= G_{\Delta,\ell}(z)+\sum_n \bigg(\alpha^s_{n,\ell} G_{\Delta_{n,\ell}}(z)+\beta^s_{n,\ell} \partial G_{\Delta_{n,\ell}}(z)\bigg)\\
   &  W^t_{\Delta,\ell}(z)= \sum_n \bigg(\alpha^t_{n,\ell} \,G_{2\Delta_{\phi}+n}(z)+\beta^t_{n,\ell} \,\partial G_{2\Delta_{\phi}+n}(z)\bigg)\\
   &  W^u_{\Delta,\ell}(z)= \sum_n \bigg((-1)^{n+\ell} \alpha^t_{n,\ell} \,G_{2\Delta_{\phi}+n}(z)+(-1)^{n+\ell}\beta^t_{n,\ell} \,\partial G_{2\Delta_{\phi}+n}(z)\bigg),
    \end{split}
\end{equation}
where $\Delta_{n,\ell}=2\Delta_{\phi}+2n+\ell$. A generic Witten diagram is defined as,
\begin{equation}
\begin{split}
  W^{s}_{\Delta,\ell}(x_i) =  \int & \frac{d^{1+1}z_1}{z_{10}^2}\frac{d^{1+1}z_2}{z_{20}^2}  K_{\Delta_\phi}(z_1,x_1)\nabla^{\mu_1...\mu_\ell}K_{\Delta_\phi}(z_1,x_2)\pi^{\Delta}_{\mu_1...\mu_\ell,\nu_1.....\nu_\ell} (z_1,z_2) K_{\Delta_\phi}(z_2,x_3)\\
  & \nabla^{\nu_1...\nu_\ell}K_{\Delta_\phi}(z_2,x_4),
  \end{split}
\end{equation}
where $K_{\Delta_\phi}(z,x)$ and $\pi^{\Delta}_{\mu_1..\mu_\ell,\nu_1...\nu_\ell}(z_1,z_2)$ are normalized bulk to boundary and bulk to bulk propagator in $AdS_2$. Also note that the derivatives should be symmetrized (anti-symmetrized) appropriately. 

Following \cite{Zhou:2018sfz} to get these coefficients we can apply $s-$ channel Casimir\footnote{By abusing the notation slightly, we represent Witten diagram $W_{\Delta,\ell}$ as function of position or cross ratios both. But its meaning should be clear from the context.},
\begin{equation} \label{eomonsch}
    (\frac{1}{2} M^{AB}_{12}M_{12AB}+c_{\Delta,\ell}) W^{s}_{\Delta,\ell}(x_i)=\sum_{i} a_i C^{s}_i(x_i).
\end{equation}
$M^{AB}_{12}$ are the conformal generators acting on $x_1$ and $x_2$. Also $c_{\Delta,\ell}$ is the Casimir eigenvalue $\Delta(\Delta-1)$ in one dimension for $\ell=0 \,\text{and}\,  1$. The action of this operator turns exchange diagrams into a combination of contact Witten diagrams $C^{s}_i$.To see it explicitly consider scalar exchange Witten diagram,
\begin{equation}
\begin{split}
  W^{s}_{\Delta,0}(x_i) =  \int & \frac{d^{1+1}z_1}{z_{10}^2}\frac{d^{1+1}z_2}{z_{20}^2}  K_{\Delta_\phi}(z_1,x_1)K_{\Delta_\phi}(z_1,x_2)\pi^{\Delta}(z_1,z_2) K_{\Delta_\phi}(z_2,x_3)\\
  & K_{\Delta_\phi}(z_2,x_4),
  \end{split}
\end{equation}
and now using the fact that these integrals are $SO(1,2)$ invariant and the equation of motion,
\begin{equation}
      (\nabla_{z_2}^2+\Delta(\Delta-d))\pi^{\Delta}(z_1,z_2)=-\delta(z_1,z_2),
\end{equation}
we arrive at the following identity, i.e.
\begin{equation}
   (\frac{1}{2} M^{AB}_{12}M_{12AB}+c_{\Delta,0})    W^{s}_{\Delta,0}(x_i)=\int \frac{d^{1+1}z_1}{z_{10}^4} k_{\Delta_\phi}(z_1,x_1) k_{\Delta_\phi}(z_1,x_2) k_{\Delta_\phi}(z_1,x_3) k_{\Delta_\phi}(z_1,x_4).
\end{equation}
We recognize the expression appearing on the RHS is just a $\phi^4$ contact Witten diagram. This way we can always reduce the exchange Witten diagrams as a combination of finite number of contact Witten diagrams by acting with the Casimir.

Note that by commuting the operator through the trivial kinematical dependence that appears in front of a four point correlator we get $\mathcal{D}(z)$ which only depends on cross ratio. Contact Witten diagrams have CB decomposition which is well known in literature\cite{Zhou:2018sfz}, 
\begin{equation} \label{cbexpcontact}
    \sum_{i} a_i C^{s}_i(z)=\sum_n( \alpha^c_n G_{\Delta_{n,\ell}}(z)+\beta^c_n \partial G_{\Delta_{n,\ell}}(z)).
\end{equation}

Also the action of the $\mathcal{D}(z)$ on conformal block is very simple,
\begin{equation} \label{casimiractcb}
   \mathcal{D}(z) G_{\Tilde{\Delta},\ell}(z)=c_{\Tilde{\Delta},\ell} G_{\Tilde{\Delta},\ell}(z).
\end{equation}
Now acting $ \mathcal{D}(z)$ on $s-$ channel exchange Witten diagram in  \eqref{cbexpwitten} and using \eqref{cbexpcontact} we can easily find the conformal block decomposition coefficients $s-$ channel exchange Witten diagrams in terms of $\alpha^c_n$ and $\beta^c_n$.

For $t-$ channel exchange Witten diagram we apply $t-$ channel Casimir to reduce the exchange Witten diagram to contact Witten diagrams,

\begin{equation} \label{eomontch}
    (\frac{1}{2} M^{AB}_{14}M_{14AB}+c_{\Delta,\ell}) W^{t}_{\Delta,\ell}(x_i)=\sum_{k} a_k C^{t}_k(x_i).
\end{equation}

But the action of $t-$ channel operator $\mathcal{D}^t(z)$ on $s-$ channel conformal block is not diagonal and instead it can be expressed in terms of finite number of $s-$ channel conformal blocks,
\begin{equation}
  \mathcal{D}^t(z)   G_{\Tilde{\Delta}}(z)= \mu\, G_{\Tilde{\Delta}-1}(z)+\nu \,G_{\Tilde{\Delta}}(z)+\rho \,G_{\Tilde{\Delta}+1}(z),
\end{equation}
with 
\begin{equation}
\begin{split}
    & \mu=-(\Tilde{\Delta}-2\Delta_{\phi})^2\\
    & \nu=\Delta (\Delta-1)+\frac{1}{2}\Tilde{\Delta}(\Tilde{\Delta}-1)-2\Delta_{\phi}(\Delta_{\phi}-1)\\
    & \rho=-\frac{\Tilde{\Delta}^2(\Tilde{\Delta}+2\Delta_{\phi}-1)^2}{4(2\Tilde{\Delta}-1)(2\Tilde{\Delta}+1)},
 \end{split}
\end{equation}
$\Delta$ is the dimension of the exchanged operator in the Witten diagram we have considered and it appeared through the casimir eigenvalue $c_{\Delta,\ell}$.
This leads to the following recursion formulas for the CB decomposition coefficients,
\begin{equation}
\begin{split}
    & \rho_{n-1} \beta^t_{n-1}+\nu_n \beta^t_n+\mu_{n+1} \beta^t_{n+1}=R_n,\\
    & \rho_{n-1} \alpha^t_{n-1}+\nu_n \alpha^t_{n}+\mu_{n+1} \alpha^t_{n+1}+\rho'_{n-1} \beta^t_{n-1}+\nu'_n \beta^t_{n}+\mu'_{n+1} \beta^t_{n+1}= S_n,
    \end{split}
\end{equation}

$R_n$, $S_n$ are the CB decomposition coefficients of the contact diagrams in  \eqref{eomontch}. Given the leading data $\{\alpha^t_0,\beta^t_0\}$, we can use the above recursion relation to find all the sub-leading coefficients. Finally we  define two contact Witten diagrams that appeared in \eqref{relnfuncwitten} $C_B(z)$ and $C_F(z)$\footnote{$[dx]=\frac{dx}{2\pi i}$},
\begin{equation}
    \int z^{2s}(1-z)^{2t}[ds][dt] \Gamma(\Delta^i_{\phi}-s)^2 \Gamma(-t)^2 \Gamma(s+t)^2 M^i(s,t) ,
\end{equation}
where 
\begin{equation}
\begin{split}
    &\Delta^B_{\phi}=\Delta_{\phi},\,\,\,\, \Delta^{F}_{\phi}=\Delta_{\phi}+\frac{1}{2},\\
    & M^B(s,t)=1,\,\,\,\, M^F(s,t)=(s+2t)+\frac{1-z}{z}\big(\frac{1}{2}+\Delta_{\phi}-2s-t\big)+\frac{1}{z}\big(\frac{1}{2}-s+t+\Delta_{\phi}\big).
    \end{split}
\end{equation}

We can easily write these contact diagrams in terms of $D$ functions \cite{Zhou:2018sfz} and therefore the CB decomposition coefficients of these diagrams. To conclude this section, we note that the functional actions are related to exchange and contact Witten diagrams with either bosonic or fermionic external legs. Further, all exchange diagrams can be related to exchange Witten diagrams with only bosonic legs. Then following the procedure described above these exchange diagrams can be related to certain contact diagrams for which the CB decomposition coefficients are well known. So, that enables us to evaluate the functional actions numerically in an efficient manner.

\section{Relevant Numerical Packages}

In accordance with our discussion in the main text, we have employed the simplex algorithm solver and the outer approximation in the Julia package \textit{FunBoot}. The corresponding source code has been made publicly available on our \href{https://github.com/Canonical111/FunBoot}{Github} page. A comprehensive documentation detailing the source code will be uploaded in due course.

It is noteworthy that this package depends on the $1d$ functional evaluations calculated using \textit{Mathematica}. The relevant \textit{Mathematica} code is also uploaded to the \href{https://github.com/Canonical111/GenerateFunc}{Github} Page.

To elaborate further, we create an extensive table containing the functional evaluations of the various types of functionals for different values of $\Delta_\phi$. For each $\Delta_\phi$, we generate an \textit{HDF5} file that comprises all the functional data. This file then serves as the initializing element for linear programming within \textit{FunBoot} and the BSpline interpolation.

As a sample, we have provided an example \textit{HDF5} file at $\Delta_\phi=0.125$ on our \href{https://github.com/Canonical111/FunBoot}{Github} page. We also upload a $3d$ example of $\Delta_\phi=0.518$. These samples serve as a test for the package. We plan to issue further instructions on the generation of \textit{HDF5} files in the future.

\section{Summary of the properties of one-dimensional functionals}

In this subsection, we aim to consolidate the currently available results pertaining to the positivity properties of the functionals. These results have been obtained empirically, and as of yet, remain unproven.

\subsection{$\mathbf{\alpha^{+F}}$}
\begin{enumerate}
    \item $\alpha_m^{+F}$ exhibits asymptotic negativity with the lone exception of $\alpha_{0}^{+F}$, which remains consistently positive.
    \item $\alpha^{+F}_m(0)$ is invariably positive, and $\alpha^{+F}_m {}'(0)>0$.
    \item Regular zeros are typically doubly degenerate, except at maximum where they occur at:
    \begin{equation}
        2\Delta_\phi+2m+1, \quad m=0,\, 1,\, ... \, \mathrm{and} \, m\neq n\, \mathrm{for} \,\alpha_{n}
    \end{equation}
    \item For $\alpha_{n>0}$, there is an irregular singleton zero between $2\Delta_\phi+2n+1$ and $2\Delta_\phi+2n+3$.
    \item At $2\Delta_\phi$, $\alpha_{n>0}$ exhibits a local minimum.
\end{enumerate}

\subsection{$\mathbf{\alpha^{-F}}$}
\begin{enumerate}
    \item $\alpha_{m}^{-F}$ consistently approaches negative values asymptotically for $m>0$, whereas $m=0$ it is asymptotically positive.
    \item $\alpha_{m}^{-F}(0)=- a_{\mathrm{gff}}(\Delta_m)$, with $\alpha_{0}^{-F}(0)=-2\Delta_\phi$ in particular.
    \item Regular zeros occur at: 
    \begin{equation}
        2\Delta_\phi+2m+1, \quad m=0,\, 1,\, ... \mathrm{and} \quad m\neq n\, \mathrm{for} \,\alpha_{n}
    \end{equation}
    \item For $\alpha_{0}$, we observe an irregular single zero between $0$ and $2\Delta_\phi+1$, and between $2\Delta_\phi+1$ and $2\Delta_\phi+3$.
    \item For $\alpha_{n> 0}$, irregular zeros manifest between $2\Delta_\phi+2n-1$ and $2\Delta_\phi+2n+1$, and between $2\Delta_\phi+2n+1$ and $2\Delta_\phi+2n+3$, with an irregular double zero at $2\Delta_\phi$.
\end{enumerate}

\subsection{$\mathbf{\beta^{+F}}$}
\begin{enumerate}
    \item $\beta^{+F}$ trends towards positive values asymptotically.
    \item $\beta^{+F}(0)$ is always negative.
    \item Regular zeros occur at: 
    \begin{equation}
        2\Delta_\phi+2m+1, \quad m=0,\, 1,\, ... \mathrm{and} \quad m\neq n\, \mathrm{for} \,\beta_{n}
    \end{equation}
    \item No unexpected zeros have been observed.
    \item For $\beta_{n> 0}$, a local minimum appears at $\frac{1}{2}$.
\end{enumerate}

\subsection{$\mathbf{\beta^{-F}}$}
\begin{enumerate}
    \item $\beta^{-F}$ tends towards positive values asymptotically.
    \item $\beta_m^{-F}(0)$ consistently equals zero and $\beta^{-F}_m {}'(0)>0$.
    \item Regular zeros occur at: 
    \begin{equation}
        2\Delta_\phi+2m+1, \quad m=0,\, 1,\, ... \mathrm{and} \quad m\neq n\, \mathrm{for} \,\beta_{n}
    \end{equation}
    \item For $\beta_{0}$, an irregular single zero is observed between $0$ and $2\Delta_\phi+1$.
    \item For $\beta_{n> 0}$, irregular zeros manifest between $2\Delta_\phi+2n-1$ and $2\Delta_\phi+2n+1$, with an irregular double zero at $2\Delta_\phi$.
\end{enumerate}

\subsection{$\mathbf{\alpha^{+B}}$}
\begin{enumerate}
    \item $\alpha^{+B}_{m}$ trends towards negative values asymptotically, with the exception of $\alpha_0^{+B}$ and $\alpha_1^{+B}$, which tends towards positive values asymptotically. It is also noticed that the $\alpha_0^{+B}$ is always positive.
    
    \item $\alpha^{+B}_m(0)$ is always positive and $\alpha^{+B}_m {}'(0)<0$.
    
    \item Regular zeros occur at: 
    \begin{equation}
        2\Delta_\phi+2m, \quad m=0,\, 1,\, ... \mathrm{and} \quad m\neq n\,\,  \mathrm{for} \,\, \alpha_{n}
    \end{equation}
    \item For $\alpha_1$, positive for $\Delta>2\Delta_\phi$.

    \item For $\alpha_{n> 1}$, irregular zeros can be somewhat unpredictable, however no irregular zeros are found above $2\Delta_\phi+2n+2$.
\end{enumerate}

\subsection{$\mathbf{\alpha^{-B}}$}
\begin{enumerate}
    \item $\alpha_m^{-B}$ trends towards negative values asymptotically, with the exception of $\alpha_0^{-B}$, which tends towards positive values asymptotically.
    \item $\alpha_{m}^{-B}(0)=-a_{\mathrm{gff}}(\Delta_m)$, with $\alpha_{0}^{-B}(0)=-2$ in particular, and $\alpha^{-B}_m {}'(0)>0$.
    \item Regular zeros occur at: 
    \begin{equation}
        2\Delta_\phi+2m, \quad m=0,\, 1,\, ... \mathrm{and} \quad m\neq n\, \mathrm{for} \,\alpha_{n}
    \end{equation}
    \item For $\alpha_{0}$, an irregular single zero is observed between $0$ and $2\Delta_\phi$.
    \item For $\alpha_{n> 0}$, irregular zeros can be somewhat unpredictable, however, no irregular zeros are found above $2\Delta_\phi+2n+2$.
\end{enumerate}

\subsection{$\mathbf{\beta^{+B}}$}
\begin{enumerate}
    \item $\beta^{+B}$ trends towards positive values asymptotically, with $\beta_{0}^{+B}$ consistently equaling zero by our convention.
    \item $\beta^{+B}_m(0)$ is always negative and $\beta^{+B}_m {}'(0)>0$.
    \item Regular zeros occur at: 
    \begin{equation}
        2\Delta_\phi+2m, \quad m=1,\, 2,\, ... \mathrm{and} \quad m\neq n\, \mathrm{for} \,\beta_{n},\, m\neq 0
    \end{equation}
    \item For $\beta_{n> 0}$, irregular zeros can be somewhat unpredictable, however no irregular zeros are found above $2\Delta_\phi+2n$.
\end{enumerate}

\subsection{$\mathbf{\beta^{-B}}$}
\begin{enumerate}
    \item $\beta^{-B}_m$ trends towards positive values asymptotically, with $\beta_{0}^{-B}$ consistently equaling zero by our convention.
    \item $\beta^{-B}_m(0)$ is always zero with $\beta^{-B}_m {}'(0)>0$.
    \item Regular zeros occur at: 
    \begin{equation}
        2\Delta_\phi+2m, \quad m=1,\, 2,\, ... \mathrm{and} \quad m\neq n\, \mathrm{for} \,\beta_{n},\, m\neq 0
    \end{equation}
    \item For $\beta_{n> 0}$, irregular zeros can be somewhat unpredictable, however no irregular zeros are found above $2\Delta_\phi+2n$.
\end{enumerate}

We can conclude that the $\beta_m$ functional is trivial above $\Delta_m$, while the $\alpha_m$ functional is trivial above $\Delta_m+2$.

\section{Expansion coefficient of conformal blocks}
In this section, we present the formula for the expansion coefficients of the conformal block in $d$ dimensions, expressed explicitly in terms of the conformal block in $d-1$ dimensions.
\begin{equation}
     G^{d}_{\Delta,\ell}(z,\bar{z})=\sum_{n=0}^{\infty} \sum_{j=\ell,\ell-2,....,\ell\, \text{mod}\, 2} \mathcal{A}_{n, j}(\Delta,\ell)   G^{d-1}_{\Delta+2n,j}(z,\bar{z}),
\end{equation}

The formula is as follows:
\begin{equation}
    \begin{aligned}
\mathcal{A}_{n, j}(\Delta, \ell)=  & Z_{\ell}^j \frac{(1 / 2)_n}{16^n n !} \frac{((\Delta+j) / 2)_n((\tau+\ell-j+1) / 2)_n}{((\Delta+j-1) / 2)_n((\tau+\ell-j) / 2)_n} \\
& \times \frac{(\Delta-1)_{2 n}((\Delta+\ell) / 2)_n(\tau / 2)_n}{(\Delta-\nu)_n(\Delta-\nu-1 / 2+n)_n((\Delta+\ell+1) / 2)_n((\tau+1) / 2)_n}
\end{aligned}
\end{equation}
In this equation, we have used the definition $\tau=\Delta-(\ell+d-2)$. For the $Z$ coefficient (where $\nu=\frac{d-2}{2}$), we have:
\begin{equation}\label{Zeq}
    Z_{\ell}^j=\frac{(1 / 2)_p \ell !}{p ! j !} \frac{(\nu)_{j+p}(2 \nu-1)_j}{(\nu-1 / 2)_{j+p+1}(2 \nu)_{\ell}}(j+\nu-1 / 2), \quad p \equiv(\ell-j) / 2 .
\end{equation}
A notable property of the $Z$ coefficient is as follows:
\begin{equation}
    \sum _j Z_l^j=1
\end{equation}
Comparing with the original literature\cite{Hogervorst:2016hal}, we set the $c_\ell^{(d)}$ coefficient in that paper to $1$, in accordance with our notation in this study. In terms of the usual cross ratios $u$ and $v$\footnote{$u=z\bar{z},\, v=(1-z)(1-\bar{z}$).}, for $u\rightarrow 0$, our blocks in $d\geq 3$ are normalized as follows,
\begin{equation}
    G_{\Delta,\ell}(u,v)\sim \frac{\left(\frac{d-2}{2}\right)_\ell}{(d-2)_\ell} u^{\frac{\Delta-\ell}{2}} \big(-\frac{1}{2}(1-v)\big)^\ell \,_2F_1\left(\frac{\Delta+\ell}{2},\frac{\Delta+\ell}{2},\Delta+\ell,1-v\right).
\end{equation}
Readers should be particularly aware that, while these formulas are universally valid across all dimensions, special consideration is required for $3d$, where $\nu=\frac{1}{2}$ serves as a pole of the formula \eqref{Zeq}. We present the equation as follows:
\begin{equation}
\begin{small}
 Z^j_l= 
\begin{cases}
    \left(\Gamma \left(\frac{l+1}{2}\right)^2\right)/\left(\pi  \Gamma \left(\frac{l+2}{2}\right)^2\right)
     ,& \text{if } j= 0\\
    \left(2 \Gamma \left(\frac{1}{2} (-j+l+1)\right) \Gamma \left(\frac{1}{2} (j+l+1)\right)\right)\left/(\pi  \Gamma \left(\frac{1}{2} (-j+l+2)\right) \Gamma \left(\frac{1}{2} (j+l+2)\right)\right),              & \text{otherwise}
\end{cases}
\end{small}
\end{equation}
So $j=0$ is the half of the case when $j\neq 0$.

\section{Normalization of Functionals}\label{Normalization}

As discussed in the main text, the functionals, in the large $\Delta$ limit, are subject to exponential growth as per their asymptotic behavior. For the sake of maintaining numerical stability during practical numerical implementations, we introduce a factor to all functionals. This factor negates the leading exponential behavior and the leading power-law suppression.

The factors are defined by the following equations:
\begin{equation}
    n^{-}(\Delta|\Delta_\phi)=\frac{\pi ^2 \Delta  (\Delta +1)^6 \Gamma (\Delta )^2 \Gamma (2 (2 \Delta_\phi +1)-1) \Gamma (\Delta +2 \Delta_\phi )}{16 \Delta_\phi  (2 \Delta_\phi +1) \Gamma (2 \Delta ) \Gamma (4 \Delta_\phi ) \Gamma (2 \Delta_\phi +1)^2 \Gamma (\Delta -2 \Delta_\phi +5)}
\end{equation}

\begin{equation}
    n^{+}(\Delta|\Delta_\phi)=\frac{\pi ^2 \Delta  (\Delta +1)^8 \Gamma (\Delta )^2 \Gamma (2 (2 \Delta_\phi +1)-1) \Gamma (\Delta +2 \Delta_\phi )}{32 \Delta_\phi ^2 (2 \Delta_\phi +1)^2 \Gamma (2 \Delta ) \Gamma (4 \Delta_\phi ) \Gamma (2 \Delta_\phi +1)^2 \Gamma (\Delta -2 \Delta_\phi +5)}
\end{equation}
These equations exhibit a smooth behavior up to $\Delta_\phi< \frac{5}{4}$, indicating their applicability for higher dimensional bootstraps where $\Delta_\phi< \frac{5}{2}$. We define our normalized functionals as follows:

\begin{equation}\label{OneDNorm}
\Omega_N^\pm (\Delta|\Delta_\phi)= n^{\pm}(\Delta|\Delta_\phi) \times \omega^\pm (\Delta|\Delta_\phi)
\end{equation}

In essence, this step acts as a redefinition of the OPE coefficient, with the aim of absorbing the pathological behavior of the analytic functionals in the large $\Delta$ limit.

Now we proceed to establish the normalization of two-dimensional and three-dimensional functionals to achieve a uniform asymptotic behavior. We start with a recapitulation of the definition of two-dimensional functionals:

\begin{equation}
\omega_1^- \otimes \omega_2^+(\Delta,\ell| \Delta_{\phi})=\frac{1}{2} \left(\omega_1^-(\tau)\omega_2^+(\rho)+\omega_1^-(\rho)\omega_2^+(\tau)\right),
\end{equation}

This equation utilizes the shorthand substitution:

\begin{equation}
\omega\left(h \right) = \omega\left(\left. \frac{h}{2} \right\vert \frac{\Delta_\phi}{2}\right) , \quad \tau=\Delta-l,\quad \rho=\Delta+l
\end{equation}

In this context, the same notation is applicable to the normalization factor $n^\pm$ and the normalized functional $\Omega_N$. For the convenience of the reader, we would also like to note the asymptotic behavior of the factors $n^\pm(h)$ in this shorthand notation:

\begin{equation}
    n^+(h)=n^+ \left(\left. \frac{h}{2} \right\vert \frac{\Delta_\phi}{2}\right) \xrightarrow[h \to \infty]{} \frac{\pi ^{5/2} 2^{-2 \Delta _{\phi }-\frac{9}{2}} }{\Delta _{\phi } \Gamma \left(\Delta _{\phi }+2\right){}^2} 2^{-h} h^{2 \Delta _{\phi }+\frac{7}{2}}
\end{equation}

\begin{equation}
    n^-(h)=n^- \left(\left. \frac{h}{2} \right\vert \frac{\Delta_\phi}{2}\right) \xrightarrow[h \to \infty]{} \frac{\pi ^{5/2} \left(\Delta _{\phi }+1\right) 2^{-2 \Delta _{\phi }-\frac{5}{2}} }{\Gamma \left(\Delta _{\phi }+2\right){}^2}  2^{-h} h^{2 \Delta _{\phi }+\frac{3}{2}}
\end{equation}

Our numerical investigations suggest the following definition, which provides a smooth limit in the asymptotic region:

\begin{equation}\label{higherdNorm}
\Omega_N^{12}(\Delta,\ell| \Delta_{\phi})=n_\ell n^+(\tau) n^-(\tau) \left(\omega_1^- \otimes \omega_2^+(\Delta,\ell| \Delta_{\phi})\right)
\end{equation}

Here $n_\ell$ is a normalization on the large spin:
\begin{equation}
n_\ell=
\begin{cases}
    1 ,& \text{if } \ell= 0\\
    4^{-\ell} \ell^{2 \Delta_\phi+3/2}, & \text{otherwise }
\end{cases}
\end{equation}

Expressed in terms of the normalized one-dimensional functional, it can be written as:

\begin{equation}\label{eq: 2dnorm}
\Omega_N^{12}(\Delta,\ell| \Delta_{\phi})=\frac{1}{2}\left(\frac{n_\ell n^+(\tau)}{n^+(\rho)} \Omega_{N1}^- (\tau) \Omega_{N2}^+ (\rho)+\frac{n_\ell n^-(\tau)}{n^-(\rho)} \Omega_{N1}^- (\rho) \Omega_{N2}^+ (\tau)\right)
\end{equation}

For the three-dimensional case, the normalization aligns with the one for the two-dimensional case (as stated in Eq.\ref{higherdNorm}), with the explicit formula being:

\begin{equation}
\Omega_{N, \mathrm{3d}}^{12}(\Delta,\ell| \Delta_{\phi})=\sum_{n=0}^{\infty} \sum_{j} \mathcal{A}_{n, j}(\Delta,\ell) \frac{n_\ell n^+(\tau) n^-(\tau)}{n_j n^+(\tau_{n, j}) n^-(\tau_{n, j})} \Omega_N^{12}(\Delta+2n,j| \Delta_{\phi})
\end{equation}

Where $\tau_{n, j}=\Delta+2n-j$. We apologize for the minor abuse of the notation $n$.

We also briefly note the normalized OPE, both in $2d$ and $3d$\footnote{The factor $n^+(0) n^-(0)$ comes from the fact that the identity operator is also normalized.}:

\begin{equation}
a_N(\Delta,\ell)=n^+(0) n^-(0)a(\Delta, \ell)/(n_\ell n^+(\tau) n^-(\tau))
\end{equation}

Significantly, if the OPE is the one for the generalized free field, it eliminates both the exponential behavior in the large $\Delta$ limit and large $\ell$ limit.
\section{Product functional in general even spacetime dimensions} \label{prodgend}
We begin by writing down the recursion formula that relates $d$ and $d-2$ dimensional conformal blocks\cite{Dolan:2003hv},
\begin{equation}
   G^d_{\Delta,\ell}(z,\bar{z})=\frac{(z\bar{z})^2}{(z-\bar{z})^2} \left(c_{-1}G^{d-2}_{\Delta-2,\ell+2}(z,\bar{z})+c_0 G^{d-2}_{\Delta-2,\ell}(z,\bar{z})+c_1 G^{d-2}_{\Delta,\ell+2}(z,\bar{z})+c_2G^{d-2}_{\Delta,\ell}(z,\bar{z})\right)   \end{equation}
  
where,
\begin{equation}
    \begin{split}
    & c_{-1}=\frac{\mathcal{N}^d_{\Delta,\ell}}{\mathcal{N}^{d-2}_{\Delta-2,\ell+2}},\\
        & c_0=-\frac{4(\ell+d-4)(d+\ell-3)}{(d+2\ell-4)(d+2\ell-2)} \frac{\mathcal{N}^d_{\Delta,\ell}}{\mathcal{N}^{d-2}_{\Delta-2,\ell}},\\
        & c_1=-\frac{4(d-\Delta-3)(d-\Delta-2)}{(d-2\Delta-2)(d-2\Delta)}\frac{(\Delta+\ell)^2}{16(\Delta+\ell-1)(\Delta+\ell+1)}\frac{\mathcal{N}^d_{\Delta,\ell}}{\mathcal{N}^{d-2}_{\Delta,\ell+2}},\\
        & c_2=\frac{4(d-\Delta-3)(d-\Delta-2)}{(d-2\Delta-2)(d-2\Delta)}\frac{(d+\ell-4)(d+\ell-3)(d+\ell-\Delta-2)^2}{4(d+2\ell-4)(d+2\ell-2)(d+\ell-\Delta-3)(d+\ell-\Delta-1)}\frac{\mathcal{N}^d_{\Delta,\ell}}{\mathcal{N}^{d-2}_{\Delta,\ell}},
    \end{split}
\end{equation}
and  $\mathcal{N}^d_{\Delta,\ell}=\frac{\left(\frac{d}{2}-1\right)_\ell}{(d-2)_\ell}(-\frac{1}{2})^{\ell}.$
The above recursion formula also relates the (anti)crossing vector $F_{\Delta,\ell}$ ($H_{\Delta,\ell}$),
\begin{equation} \label{crossrecu}
\begin{split}
   F^d_{\Delta,\ell}(z,\bar{z}|\Delta_{\phi}) =& \frac{1}{(z-\bar{z})^2} \Big(c_{-1}F^{d-2}_{\Delta-2,\ell+2}(z,\bar{z}|\Delta_{\phi}-2)+c_0 F^{d-2}_{\Delta-2,\ell}(z,\bar{z}|\Delta_{\phi}-2)\\
   & +c_1 F^{d-2}_{\Delta,\ell+2}(z,\bar{z}|\Delta_{\phi}-2)
    +c_2 F^{d-2}_{\Delta,\ell}(z,\bar{z}|\Delta_{\phi}-2)\Big) 
\end{split}
\end{equation}
For convenience, we write the above relation using the following shorthand notation,
\begin{equation}  \label{crossrecushort}
     F^d_{\Delta,\ell}(z,\bar{z}|\Delta_{\phi})=\frac{1}{(z-\bar{z})^2} \sum_{i=0}^3 P_i F^{d-2}_{\Delta+\Delta_i,\ell+\ell_i}(z,\bar{z}|\Delta_{\phi}-2),
\end{equation}
where $P_i,\Delta_i, \ell_i$ can be read off from the equation \eqref{crossrecu}.
Then we can choose the functional kernels in the following way:
\begin{equation}
    \omega^1_\pm \times \omega^2_\pm(\Delta,\ell|\Delta_{\phi})=2 \int \frac{dz d\bar{z}}{\pi^2} h^1_\pm(z)h^2_\pm(\bar{z})(z-\bar{z})^2 \big[ \mathcal{I}_z \mathcal{I}_{\bar{z}}  F^d_{\Delta,\ell}(z,\bar{z}|\Delta_{\phi})\pm \mathcal{I}_z \mathcal{I}_{\bar{z}}  F^d_{\Delta,\ell}(z,1-\bar{z}|\Delta_{\phi}) \big]
\end{equation}
Using \eqref{crossrecu} on the RHS we can find the functional actions in $d$ dimension in terms of those in $d-2$ dimensions. But we have the prefactor $(z-\bar{z})^2$ so we have to subtract more functionals to ensure the swapping. Note that we have chosen both functionals to be either $+$ type or $-$ type. The reason is that we are assuming that $d-2$ dimensional crossing vector can be represented as a linear combination of the product of $1d$ crossing vectors ($F_\tau$) and anti-crossing vectors ($H_\tau$), with the absence of crossed terms ($F_{\tau}H_{\rho}$). In fact, we can repeatedly use the recursion to express the functional actions in $d-$ dimensions in terms of those in two spacetime dimensions. In that case we have to perform $\frac{d-2}{2}$ subtractions for both $``+$ type and $``-"$ type functionals and the external dimension would be,
\begin{equation}
 \Delta^{d=2}_{\phi}= \Delta^d_{\phi}-(d-2).
\end{equation}
To see this explicitly let us use the recursion twice to find the following relation between $d=6$ and $d=2$ crossing vector\footnote{For the sake of brevity, we omit writing down the coefficients, but obtaining them from the recursion relation \eqref{crossrecu} is a straightforward task. Also note $P_i,\Delta_i,\ell_i$ are not the same as those that appeared in \eqref{crossrecushort}.},
\begin{equation}
    F^{d=6}_{\Delta,\ell}(z,\bar{z}|\Delta_{\phi})=\frac{1}{(z-\bar{z})^4} \sum_{i=0}^{15} P_i F^{d=2}_{\Delta+\Delta_i,\ell+\ell_i}(z,\bar{z}|\Delta_{\phi}-4).
\end{equation}
Then we choose the following kernels\footnote{We are choosing $(\omega^1_+ \times \omega^2_-)$ instead of both being $+$ or $-$ type because $2d$ crossing vector has crossed terms like $F_{\tau}H_{\rho}$, see equation \eqref{2dF}.},
\begin{equation}
\begin{split}
    (\omega^1_+ \times \omega^2_-)^{d=6}(\Delta,\ell|\Delta_{\phi})=2 \sum_{i=0}^{15} P_i  \int & \frac{dz d\bar{z}}{\pi^2} h^1_+(z)h^2_-(\bar{z})(z-\bar{z})^4 \big[ \mathcal{I}_z \mathcal{I}_{\bar{z}}   F^{d=2}_{\Delta+\Delta_i,\ell+\ell_i}(z,\bar{z}|\Delta_{\phi}-4)\\
    &\pm \mathcal{I}_z \mathcal{I}_{\bar{z}}  F^{d=2}_{\Delta+\Delta_i,\ell+\ell_i}(z,1-\bar{z}|\Delta_{\phi}-4)\big].
    \end{split}
\end{equation}
So our functional action will be given by,
\begin{equation}
(\omega^1_+ \times \omega^2_-)^{d=6}(\Delta,\ell|\Delta_{\phi})=\sum_{i=0}^{15} P_i (\omega^1_+ \times \omega^2_-)^{d=2}(\Delta+\Delta_i,\ell+\ell_i|\Delta_{\phi}-4).
\end{equation}
Here the functional action on $2d$ crossing vector is defined as \eqref{eq: productdef}. But now $1d$ functionals will not act on the $6d$ crossing equation. Because there is a factor $(z-\bar{z})^4$ and it will ruin the fall off at $z=\infty$. To improve the behavior at $z=\infty$ we need to make additional subtractions,
\begin{equation}
    \begin{split}
  &  \tilde{\beta}^-_n=\beta^-_n-\sum_{i=1}^{2}c^-_n(i) \beta^-_i,\\
    &  \tilde{\alpha}^-_n=\alpha^-_n-\sum_{i=1}^{2}a^-_n(i) \beta^-_i,\\
    &     \tilde{\beta}^+_n=\beta^+_n-\sum_{i=1}^{2}c^+_n \beta^+_i,\\
    &  \tilde{\alpha}^+_n=\alpha^+_n-\sum_{i=1}^{2}a^+_n(i) \beta_i.
    \end{split}
\end{equation}
We also note that since we have used the recursion relation twice, the shift in $\Delta$ is multiple of 2. Therefore in order to ensure the presence of asymptotic zeroes, it is necessary to employ FB type functionals. Following this procedure, we can easily construct product functionals in general even spacetime dimensions. It is not immediately clear if they will be useful for giving some optimal bound, but more importantly, if they are complete. Although we didn't investigate the nature of completeness of these functionals, we found that we can construct some functionals which provide instant bound on OPE coefficients in four dimensions. We are currently investigating these exciting and intriguing functionals in detail. We must note that the positivity condition is not necessarily upheld in this context. Specifically, the coefficients $c_i$ in equation \eqref{crossrecu} do not have an assurance of being positive. In the end we also note that using dimensional reduction functional actions in all odd dimensions can be found in terms of single infinite sum as we did in the case of $d=3$.

\section{Numerical Parameters}\label{Appendix: parameter}

\begin{table}[ht]
\centering
\caption{Parameters of the SDPB}
\label{tab:SDPB}
\begin{tabular}{@{}llll@{}}
\toprule
$\Lambda$ & \# of derivatives  & $L_{\mathrm{max}}$ & kept\_pole\_order \\
\midrule
9   & 15 & 12 & 12 \\
25  & 91 & 30 & 30 \\
35  & 171& 42 & 42 \\
\bottomrule
\end{tabular}
\end{table}

In this section, we provide a detailed overview of the parameters utilized in our numerical evaluation. Our investigation involved two major computational techniques: the FunBoot and the SDPB. 

Firstly, for most of the functional bootstrap results in this study, we adopt a uniform list of spins for all the $d=2$ evaluations:
\begin{equation}
    L\in \{0, 2, 4, ..., 120\}\cup \{256\}
\end{equation}
Despite only performing a modest evaluation with $8$ functionals, we maintain this comprehensive list of spins. This is definitely much more than enough. This strategy is attributed to the efficiency of our computations, with most of them finishing instantly. As such, optimization of the list of spins has not been deemed necessary.

The exception is when we are using $144$ FF functionals. For this situation, we use a more modest spin list:
\begin{equation}
    L\in \{0, 2, 4, ..., 60\}\cup \{150\}
\end{equation}

For $3d$ bootstrap, we used an extremely modest spin list, due to the technical difficulties mentioned in the main text:
\begin{equation}
    L\in \{0, 2, 4, 6, 8, 20\}
\end{equation}

In our computations, an outer-approximation cut-off of $10^{-8}$ has been implemented. This signifies that if the operator is already optimal within this threshold, no further effort is made to improve it through outer-approximation.

Turning to the implementation in SDPB, we deployed a bisection method to pinpoint the exact location of the upper bound. The parameters relevant to this implementation are summarized in Table \ref{tab:SDPB}.

\section{The fourth kink and beyond}\label{Appendix: beyond}

\begin{figure}[h]
\centering
\includegraphics[width=12cm]{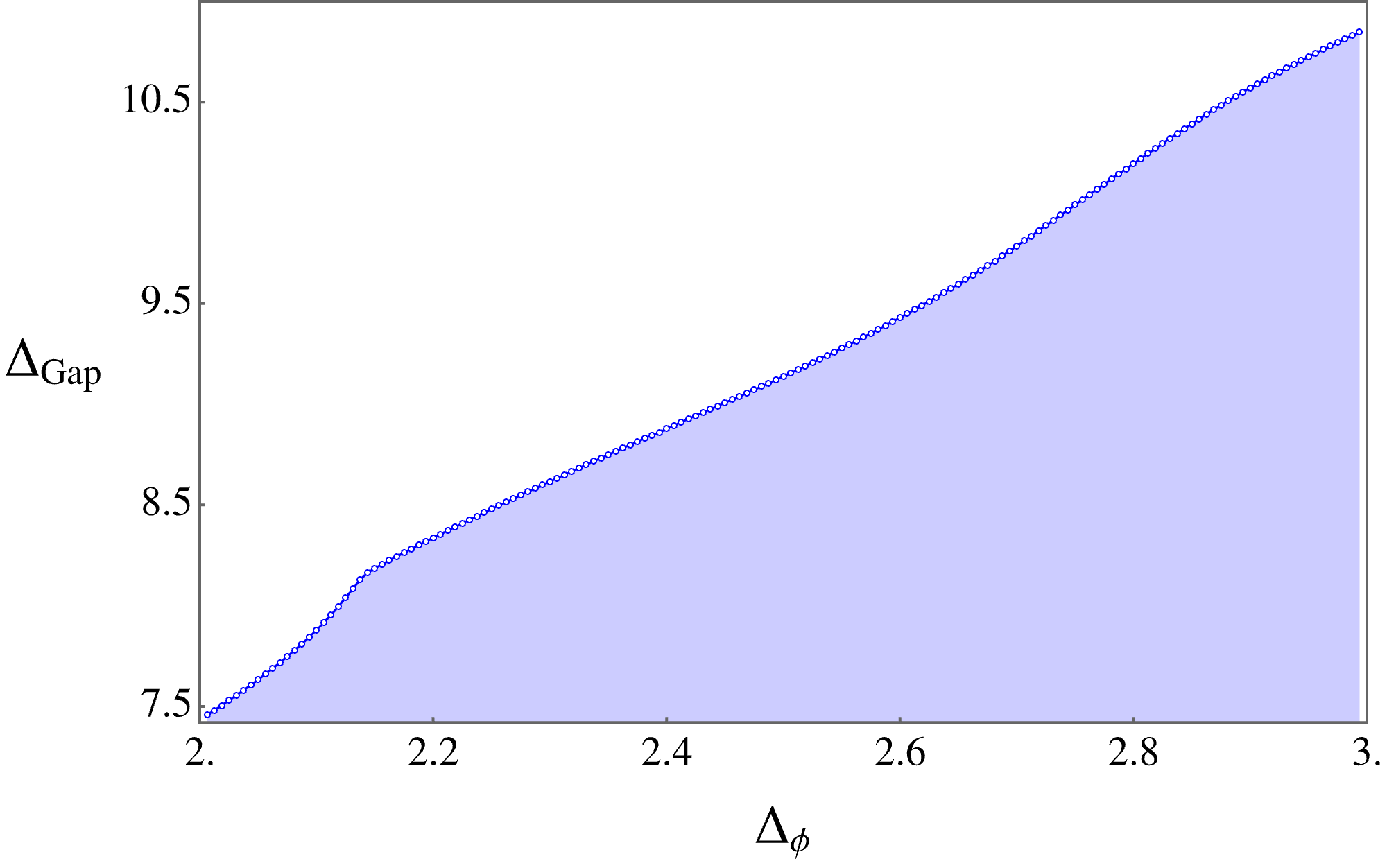}
\caption{Gap maximization for spin $\ell=0$, with $\Delta_\phi$ ranging from $2$ to $3$. This upper bound corresponds $144$ $F^+F^-$ functionals.}
\label{fig: fourthkink}
\end{figure}

In this appendix, we provide a more detailed examination of the upper bound where $\Delta_\phi$ varies from $2$ to $3$, continuing the discussion from Section.\ref{section: thirdkink}. As depicted in Fig.\ref{fig: fourthkink}, we identify a distinctive kink at $(2.14, 8.15)$, which was previously alluded to in the main body of the text. Furthermore, we suspect the existence of an additional kink around the coordinates $(2.85, 10)$, although a comprehensive exploration of this potential feature is a subject for future study. For added clarity, we list in Table.\ref{table: fourthkink} the spectrum of the optimal solution in the vicinity of the identified kink.

\begin{table}[h]
\centering
\begin{subtable}{0.49\textwidth}
\centering
\begin{tabular}{|c|c|c|}
\hline
Dimension & Spin & OPE$^2$ \\
\hline
2.1279081 & 2 & 1.505595e+01 \\
3.9247237 & 2 & 1.269176e+01 \\
4.0365950 & 4 & 2.501354e+00 \\
4.8369434 & 4 & 2.338230e+00 \\
5.7893897 & 2 & 5.237615e+00 \\
6.0000000 & 6 & 9.202021e-01 \\
6.5272048 & 6 & 4.507569e-01 \\
6.5784888 & 4 & 1.145040e+00 \\
8.0237300 & 8 & 1.558957e-01 \\
\hline
\end{tabular}
\caption{$\Delta_\phi=2.1375$}
\label{subtable:cutoff1}
\end{subtable}
\hfill
\begin{subtable}{0.49\textwidth}
\centering
\begin{tabular}{|c|c|c|}
\hline
Dim & Spin & OPE$^2$ \\
\hline
2.1486320 &   2 & 1.554380e+01 \\
3.9698081 &   2 & 1.284837e+01 \\
4.0624818 &   4 & 5.473524e+00 \\
4.8745654 &   4 & 2.255757e+00 \\
5.8507517 &   2 & 5.035630e+00 \\
6.0000000 &   6 & 9.656059e-01 \\
6.5572265 &   6 & 4.505713e-01 \\
6.5667975 &   4 & 3.355676e+00 \\
8.0141466 &   8 & 1.582781e-01 \\
\hline
\end{tabular}
\caption{$\Delta_\phi=2.14375$}
\end{subtable}
\caption{Scalar channel gap maximization with $144$ FF functionals at the kink. This table highlights the leading terms in the spectrum.}
\label{table: fourthkink}
\end{table}

\newpage

\section{Selected optimal solution Spectrum}\label{App: spectrum}

Unlike in Section \ref{Sec: convergence}, in this part of our appendix, we focus on presenting the spectra of leading operators under different cut-off values. This modification in perspective provides an enriched understanding of the underlying dynamics. As a gauge of the precision of our findings, we have benchmarked our results against a point from the minimal model $\Delta_\phi=0.25$. We also include a table for 60 derivatives at $\Delta_\phi=0.125$, the data is from the study \cite{El-Showk:2012vjm} for the comparison of Table.\ref{table:spectruml0FB0.125} in the main text.

\begin{table}[H]
\centering
\begin{subtable}{0.49\textwidth}
\centering
\begin{tabular}{|c|c|c|}
\hline
Dimension & Spin & OPE$^2$ \\
\hline
2.0451208 & 2 & 2.591169e+00 \\
3.5076379 & 2 & 4.706260e-01 \\
4.0000000 & 4 & 3.118627e-01 \\
4.5373846 & 0 & 1.164550e+00 \\
5.0052609 & 4 & 6.331897e-02 \\
6.0015942 & 6 & 2.961814e-02 \\
6.1253590 & 4 & 1.860179e-02 \\
6.1298858 & 0 & 4.344459e-02 \\
6.5483555 & 2 & 2.869116e-01 \\
\hline
\end{tabular}
\caption{60 functionals}
\label{subtable:cutoff1}
\end{subtable}
\hfill
\begin{subtable}{0.49\textwidth}
\centering
\begin{tabular}{|c|c|c|}
\hline
Dim & Spin & OPE$^2$ \\
\hline
2.0482787 &   2 & 2.506606e+00 \\
3.4378100 &   2 & 4.907859e-01 \\
4.0000000 &   4 & 2.781792e-01 \\
4.5205518 &   0 & 1.177048e+00 \\
4.8045771 &   4 & 8.200372e-02 \\
6.0000000 &   6 & 2.550358e-02 \\
6.0195505 &   4 & 2.035768e-02 \\
6.0973525 &   0 & 4.253401e-02 \\
6.5164499 &   2 & 2.943910e-01 \\
\hline
\end{tabular}
\caption{112 functionals}
\label{subtable:cutoff2}
\end{subtable}
\caption{Scalar channel gap maximization with FF functionals at $\Delta_\phi=1.2$.}
\label{table:spectruml0FF1.2}
\end{table}

\begin{table}[H]
\centering
\begin{subtable}{0.49\textwidth}
\centering
\begin{tabular}{|c|c|c|}
\hline
Dimension & Spin & OPE$^2$ \\
\hline
2.3977597 & 0 & 2.021874e+00 \\
4.0000000 & 4 & 1.090463e-03 \\
4.4129682 & 0 & 7.125925e-01 \\
4.4215296 & 2 & 9.408487e-01 \\
6.4070667 & 0 & 1.084317e-01 \\
6.4186711 & 6 & 1.514427e-04 \\
6.4415355 & 4 & 9.892015e-02 \\
6.4720431 & 2 & 1.780068e-01 \\
8.4556866 & 2 & 2.081156e-02 \\
\hline
\end{tabular}
\caption{60 functionals}
\label{subtable:cutoff1}
\end{subtable}
\hfill
\begin{subtable}{0.49\textwidth}
\centering
\begin{tabular}{|c|c|c|}
\hline
Dim & Spin & OPE$^2$ \\
\hline
2.3981660 &   0 & 2.014442e+00 \\
4.0000000 &   4 & 5.415745e-04 \\
4.4106942 &   0 & 7.163512e-01 \\
4.4116102 &   2 & 9.367668e-01 \\
6.0000000 &   6 & 1.470818e-05 \\
6.4154313 &   0 & 1.074449e-01 \\
6.4204898 &   4 & 9.828018e-02 \\
6.4410993 &   2 & 1.827097e-01 \\
6.6427085 &   6 & 7.083262e-05 \\
\hline
\end{tabular}
\caption{112 functionals}
\label{subtable:cutoff2}
\end{subtable}
\caption{$L=2$ channel gap maximization with FF functionals at $\Delta_\phi=1.2$.}
\label{table:spectruml2FF1.2}
\end{table}

\begin{table}[H]
\centering
\begin{subtable}{0.49\textwidth}
\centering
\begin{tabular}{|c|c|c|}
\hline
Dimension & Spin & OPE$^2$ \\
\hline
2.3941426 & 0 & 2.047853e+00 \\
4.0000000 & 4 & 1.555470e-03 \\
4.4354783 & 2 & 9.471885e-01 \\
4.4392920 & 0 & 7.087127e-01 \\
6.4367054 & 6 & 2.316205e-04 \\
6.4550431 & 0 & 1.031777e-01 \\
6.4576143 & 4 & 9.924106e-02 \\
6.5422600 & 2 & 1.735586e-01 \\
8.0000000 & 8 & 1.201634e-05 \\
\hline
\end{tabular}
\caption{50 functionals}
\label{subtable:cutoff1}
\end{subtable}
\hfill
\begin{subtable}{0.49\textwidth}
\centering
\begin{tabular}{|c|c|c|}
\hline
Dim & Spin & OPE$^2$ \\
\hline
2.3989922 &   0 & 2.017278e+00 \\
4.0000000 &   4 & 6.559008e-04 \\
4.4120218 &   0 & 7.128668e-01 \\
4.4173006 &   2 & 9.400492e-01 \\
6.0259888 &   6 & 6.316374e-05 \\
6.3861206 &   0 & 1.076298e-01 \\
6.4267314 &   4 & 9.860825e-02 \\
6.4746495 &   2 & 1.805964e-01 \\
8.0152016 &   8 & 5.126822e-06 \\
\hline
\end{tabular}
\caption{98 functionals}
\label{subtable:cutoff2}
\end{subtable}
\caption{$L=2$ channel gap maximization with FB functionals at $\Delta_\phi=1.2$.}
\label{table:spectruml2FB1.2}
\end{table}

\begin{table}[H]
\centering
\begin{subtable}{0.49\textwidth}
\centering
\begin{tabular}{|c|c|c|}
\hline
Dimension & Spin & OPE$^2$ \\
\hline
2.0324847 & 2 & 1.144274e+00 \\
3.1995017 & 0 & 9.253488e-01 \\
4.0105956 & 4 & 9.011871e-02 \\
4.3676758 & 0 & 1.609052e-01 \\
4.8884840 & 4 & 2.753970e-03 \\
5.1604013 & 2 & 1.838816e-01 \\
6.0000000 & 6 & 6.368615e-03 \\
6.4284822 & 2 & 2.511050e-02 \\
6.7747089 & 6 & 2.631923e-04 \\
\hline
\end{tabular}
\caption{60 functionals}
\label{subtable:cutoff1}
\end{subtable}
\hfill
\begin{subtable}{0.49\textwidth}
\centering
\begin{tabular}{|c|c|c|}
\hline
Dim & Spin & OPE$^2$ \\
\hline
2.0379323 &   2 & 1.128070e+00 \\
3.1810386 &   0 & 9.219595e-01 \\
4.0059745 &   4 & 8.597685e-02 \\
4.3289773 &   0 & 1.703759e-01 \\
4.6211210 &   4 & 5.066629e-03 \\
5.1398589 &   2 & 1.859338e-01 \\
6.0062680 &   6 & 6.282137e-03 \\
6.4120416 &   2 & 2.577807e-02 \\
7.0014879 &   6 & 2.302446e-04 \\
\hline
\end{tabular}
\caption{112 functionals}
\label{subtable:cutoff2}
\end{subtable}
\caption{Scalar channel gap maximization with FF functionals at $\Delta_\phi=0.9$.}
\label{table:spectruml0FF0.9}
\end{table}

\begin{table}[H]
\centering
\begin{subtable}{0.49\textwidth}
\centering
\begin{tabular}{|c|c|c|}
\hline
Dimension & Spin & OPE$^2$ \\
\hline
0.3676087 & 0 & 3.625326e-01 \\
1.6268165 & 0 & 9.644125e-03 \\
2.0000000 & 2 & 6.846985e-03 \\
2.1483087 & 2 & 2.616812e-03 \\
2.7035741 & 0 & 5.220688e-04 \\
4.0423120 & 4 & 1.446055e-04 \\
4.3255350 & 0 & 6.733448e-06 \\
4.4267226 & 4 & 1.294743e-05 \\
5.6301602 & 4 & 2.221984e-07 \\
\hline
\end{tabular}
\caption{32 functionals}
\label{subtable:cutoff1}
\end{subtable}
\hfill
\begin{subtable}{0.49\textwidth}
\centering
\begin{tabular}{|c|c|c|}
\hline
Dim & Spin & OPE$^2$ \\
\hline
0.3661774 &   0 & 3.654008e-01 \\
1.6366927 &   0 & 9.301289e-03 \\
2.0000000 &   2 & 6.720434e-03 \\
2.1515426 &   2 & 2.685554e-03 \\
2.6496943 &   0 & 5.750235e-04 \\
4.0383060 &   4 & 1.393874e-04 \\
4.2496797 &   0 & 8.182070e-06 \\
4.3516120 &   4 & 1.763925e-05 \\
5.3475805 &   4 & 4.441175e-07 \\
\hline
\end{tabular}
\caption{72 functionals}
\label{subtable:cutoff2}
\end{subtable}
\caption{Scalar channel gap maximization with FB functionals at $\Delta_\phi=0.075$.}
\label{table:spectruml0FB0.075}
\end{table}

\begin{table}[H]
\centering
\begin{subtable}{0.49\textwidth}
\centering
\begin{tabular}{|c|c|c|}
\hline
Dimension & Spin & OPE$^2$ \\
\hline
1.3337687 &   0 & 4.463263e-01 \\
2.0000000 &   2 & 7.821289e-02 \\
3.3380435 &   2 & 1.074416e-02 \\
4.0028314 &   4 & 1.593157e-03 \\
4.0071780 &   0 & 1.508922e-03 \\
5.3159298 &   0 & 6.647290e-05 \\
5.3456691 &   4 & 3.915309e-04 \\
6.0029565 &   6 & 5.961471e-05 \\
6.0727126 &   2 & 5.873574e-05 \\
\hline
\end{tabular}
\caption{50 functionals}
\label{subtable:cutoff1}
\end{subtable}
\hfill
\begin{subtable}{0.49\textwidth}
\centering
\begin{tabular}{|c|c|c|}
\hline
Dim & Spin & OPE$^2$ \\
\hline
1.3334859 &   0 & 4.464118e-01 \\
2.0000000 &   2 & 7.815441e-02 \\
3.3349098 &   2 & 1.077889e-02 \\
4.0008224 &   4 & 1.588745e-03 \\
4.0032582 &   0 & 1.520108e-03 \\
5.3298715 &   0 & 6.507424e-05 \\
5.3369138 &   4 & 3.957296e-04 \\
6.0000000 &   6 & 5.928866e-05 \\
6.0306677 &   2 & 6.110425e-05 \\
\hline
\end{tabular}
\caption{98 functionals}
\label{subtable:cutoff2}
\end{subtable}
\caption{Scalar channel gap maximization with FB functionals at $\Delta_\phi=0.25$.}
\label{table:spectruml0FB0.25}
\end{table}

\begin{table}[H]
\centering
\begin{subtable}{0.49\textwidth}
\centering
\begin{tabular}{|c|c|c|}
\hline
Dimension & Spin & OPE$^2$ \\
\hline
2.0000000 &   2 & 2.000021e+00 \\
3.7911913 &   2 & 3.752425e-06 \\
4.0000148 &   4 & 2.000009e-01 \\
4.0000225 &   0 & 9.999942e-01 \\
6.0000000 &   6 & 1.587351e-02 \\
6.0000617 &   2 & 1.999933e-01 \\
7.9999771 &   4 & 1.587198e-02 \\
8.0000000 &   8 & 1.160795e-03 \\
8.0004066 &   0 & 9.998061e-03 \\
\hline
\end{tabular}
\caption{50 functionals}
\label{subtable:cutoff1}
\end{subtable}
\hfill
\begin{subtable}{0.49\textwidth}
\centering
\begin{tabular}{|c|c|c|}
\hline
Dim & Spin & OPE$^2$ \\
\hline
2.0000000 &   2 & 2.000000e+00 \\
2.3917449 &   2 & 1.870680e-07 \\
4.0000001 &   0 & 1.000000e+00 \\
4.0000002 &   4 & 2.000000e-01 \\
5.9999535 &   2 & 1.929250e-01 \\
6.0000000 &   6 & 1.587301e-02 \\
6.0012797 &   2 & 7.075011e-03 \\
7.9960951 &   4 & 1.107694e-03 \\
8.0000000 &   8 & 2.141043e-04 \\
\hline
\end{tabular}
\caption{98 functionals}
\label{subtable:cutoff2}
\end{subtable}
\caption{Scalar channel gap maximization with FB functionals at $\Delta_\phi=1$.}
\label{table:spectruml0FB1}
\end{table}

\begin{table}[h]
\begin{equation*}
    \begin{array}{|c|c|c||c|c|c|c|}
\hline \text{Spin} & \Delta & \text {OPE}^2 & \Delta_{\text{num}} & \operatorname{Err}_{\Delta}(\%)& \text {OPE}_{\text{num}}^2 & \operatorname{Err}_{\mathrm{OPE}^2}(\%)     \\
\hline 0  & \,1\,  & \text{ 2.50000000e-1 } & \,1.000003  \, & \text{ 3.0e-4 } & \text{ 2.49999700e-1 } & \text{1.2e-4}\\
          & \,4\,  & \text{ 2.44140625e-4 } & \,4.0003    \, & \text{ 7.5e-3 } & \text{ 2.44112501e-4 } & \text{1.2e-2}\\
          & \,8\,  & \text{ 4.82812729e-8 } & \,8.0817    \, & \text{ 1.0 }    & \text{ 4.70903020e-8 } & \text{2.5}   \\
\hline 2  & \,2\,  & \text{ 3.12500000e-2 } & \,2.        \, & \text{ 0}       & \text{ 3.12501077e-2 } & \text{3.4e-4}\\
          & \,6\,  & \text{ 6.86644375e-6 } & \,5.9979    \, & \text{ 3.5e-2 } & \text{ 6.85151565e-6 } & \text{2.2e-1}\\
\hline 4  & \,4\,  & \text{ 4.39451562e-4 } & \,4         \, & \text{ 0}       & \text{ 4.39434791e-4 } & \text{3.8e-3}\\
          & \,5\,  & \text{ 3.05175590e-5 } & \,5.0003    \, & \text{ 6.0e-3 } & \text{ 3.05157913e-5 } & \text{5.8e-3}\\
          & \,8\,  & \text{ 2.12871504e-7 } & \,7.9920    \, & \text{ 1.0e-1 } & \text{ 2.15216199e-7 } & \text{1.1}   \\
\hline 6  & \,6\,  & \text{ 1.36239239e-5 } & \,6         \, & \text{ 0 }      & \text{ 1.36264340e-5 } & \text{1.8e-2}\\
          & \,7\,  & \text{ 1.52586728e-6 } & \,6.9978    \, & \text{ 3.1e-2 } & \text{ 1.52591668e-6 } & \text{3.2e-3}\\
\hline 8  & \,8\,  & \text{ 5.39324266e-7 } & \,8          \,& \text{ 0 }      & \text{ 5.39051108e-7 } & \text{5.1e-2}\\
\hline
\end{array}
\end{equation*}
\caption{Scalar channel gap maximization with $60$ derivatives at $\Delta_\phi=0.125$. We compare the spectrum with the exact value from the $2d$ critical Ising model. The data is from the study \cite{El-Showk:2012vjm} for the comparison of Table.\ref{table:spectruml0FB0.125} in the main text.}
\label{table:derspectruml0FB0.125}
\end{table}

\newpage

\bibliography{ref}
\bibliographystyle{jhep}



\end{document}